\documentclass[12pt,a4paper]{article}
\pdfoutput=1

\usepackage{amssymb}
\usepackage{amsmath}
\usepackage{amsfonts}
\usepackage{amsthm}
\usepackage{mathrsfs}  				
\usepackage{enumitem}
\usepackage{setspace}
\usepackage{color}
\usepackage[pdftex]{graphicx}
\usepackage{authblk}
\usepackage{subfigure}
\usepackage{float}
\usepackage[utf8]{inputenc}
\usepackage{subfigure}  			
\usepackage[font=small]{caption}	
\usepackage[compress]{cite}         
\usepackage{float}

\usepackage{datetime}

\usepackage{pgfplots}   			
	\pgfplotsset{width=10cm}
	\usepackage{bbm}
	\usepackage{tensor}
	\usepackage{physics}
	\usepackage{slashed}

\numberwithin{equation}{section}

\newcommand{\BbbR}{\mathbb{R}}
\newcommand{\BbbZ}{\mathbb{Z}}
\newcommand{\BbbC}{\mathbb{C}}

\newcommand{\sfx}{{\sf{x}}}

\DeclareMathOperator{\Realpart}{Re}

\DeclareMathOperator\arctanh{arctanh}

\theoremstyle{plain}

\usepackage{color}

\newdateformat{daymonthyear}{\THEDAY \, \monthname[\THEMONTH] \THEYEAR}
\newdateformat{monthyear}{\monthname[\THEMONTH] \THEYEAR}

\pgfplotsset{compat=1.17}

\begin{document}

\title{Detecting the massive bosonic zero-mode in expanding cosmological spacetimes}
\author{Vladimir Toussaint$^{1}$\thanks{{\tt vladimir.toussaint@nottingham.edu.cn}} }
\author{Jorma Louko$^{2}$\thanks{\tt jorma.louko@nottingham.ac.uk}}
\affil{$^{1}$School of Mathematical Sciences,\\ 
University of Nottingham Ningbo China,\\
Ningbo 315100, PR China}
\affil{$^{2}$School of Mathematical Sciences, University of Nottingham, Nottingham NG7 2RD, UK}


\date{February 2021; revised April 2021}

\maketitle

\begin{abstract}
We examine a quantised massive scalar field in $(1+1)$-dimensional spatially 
compact cosmological spacetimes in which the early time and late time expansion 
laws provide distinguished definitions of Fock ``in'' and ``out'' vacua, 
with the possible exception of the spatially constant sector, 
which may become effectively massless at early or late times. 
We show, generalising the work of Ford and Pathinayake, 
that when such a massive zero mode occurs, 
the freedom in the respective ``in'' and ``out'' states 
is a family with two real parameters. 
As an application, we consider massive untwisted and twisted scalar fields in the $(1+1)$-dimensional 
spatially compact Milne spacetime, where the untwisted field has a massive ``in'' zero mode. 
We demonstrate, by a combination of analytic and numerical methods, 
that the choice of the massive ``in'' zero mode state has a significant effect 
on the response of an inertial Unruh-DeWitt detector, 
especially in the excitation part of the spectrum. 
The detector's peculiar velocity with respect to comoving cosmological observers
has the strongest effect in the ``in'' vacuum of the untwisted field, 
where it shifts the excitation and de-excitation resonances towards 
higher values of the detector's energy gap. 
\end{abstract}

\singlespacing

\newpage

\section{Introduction\label{sec:intro}}
 
In quantum field theory on spatially compact spacetimes, 
it is well known that the wave equation may have 
normalisable zero frequency solutions, known as zero modes, 
which create an ambiguity in the choice of a vacuum state in Fock quantisation
\cite{rajaraman,dewitt83,Allen:1987tz,gavi91b,McCartor:1991ch,Kirsten:1993ug,Martin-Martinez:2014qda,Louko:2016ptn}.
It was demonstrated by Ford and Pathinayake \cite{Ford:1989mf} 
that a similar ambiguity arises in cosmological spacetimes with compact spatial sections 
when a spatially constant mode of a massive field becomes asymptotically massless in an asymptotic ``in'' or ``out'' region, 
in such a way that the adiabatic evolution criterion cannot be invoked to define a unique vacuum adapted to this ``in'' or ``out'' region. 
We refer to such field modes as massive zero modes. 

In this paper we examine how the choice of a massive zero mode quantum state affects 
the response of a localised quantum system that moves inertially in a cosmological spacetime. 
As a localised quantum system we consider an Unruh-DeWitt 
detector~\cite{Unruh:1976db,DeWitt:1979}, which provides a simplified model for the 
interaction between atomic orbitals and the quantum 
electromagnetic field when angular momentum interchange 
is negligible~\cite{MartinMartinez:2012th,Alhambra:2013uja}. 
Unruh-DeWitt detectors have a long pedigree as a device for extracting 
local information from quantum states defined by nonlocal criteria, 
including black hole spacetimes, and in recent years they have been much 
employed in entanglement extraction scenarios 
(see \cite{Ng:2018ilp,Simidzija:2018ddw,Ng:2018drz} for a sample). 
Our work fits in the context where the state of a quantum field has been 
singled out by early universe cosmological 
considerations but the state is being probed by local observers in  
the late universe~\cite{VerSteeg:2007xs,Garay:2013dya}. 
Specifically, we generalise to cosmological spacetimes previous work on 
observing zero modes in a static spacetime~\cite{Martin-Martinez:2014qda,Louko:2016ptn,Tjoa:2020riy}. 

We work in two spacetime dimensions, for the technical reason that 
this allows us to analyse the response of an 
Unruh-DeWitt detector operating for a finite time in a time-dependent geometry 
without having to smear the detector in time or in space. 
We expect similar phenomena due to the choice of the quantum state and due to the detector's motion 
to be present also in higher spacetime dimensions,
but there these phenomena will be necessarily blurred 
by choices that will need to be made for smearing 
the detector's profile in time or in space~\cite{Sachs:2017exo,Martin-Martinez:2020pss}. 

We begin in Section \ref{CompactifiedCosmol} by discussing a quantised real massive scalar 
field in a $(1+1)$-dimensional spatially homogeneous cosmological spacetime whose 
spatial sections are compactified to have the topology of a circle. 
The main points are to characterise situations in which a massive zero mode exists, 
and to characterise the freedom in the corresponding vacuum state. 
We show that the zero mode vacua form a family parametrised by two real-valued parameters, 
as observed by Ford and Pathinayake \cite{Ford:1989mf} 
for a subset of our asymptotic conditions. 
On a par with the normal scalar field, we also include in this section 
the quantisation of scalar field that is antiperiodic on traversing the spatial circle. 
As this field, called the ``twisted'' field, has no massive zero mode, 
it will provide a point of contrast for the massive zero mode effects in the later sections. 

In Section \ref{Unruh-Dewitt:ScalarDetector} 
we recall the expression for the transition probability of an Unruh-DeWitt detector~\cite{Unruh:1976db,DeWitt:1979}, 
coupled linearly to the quantum field, treated to leading order in perturbation theory, 
and operating for a finite time \cite{finitetime1,Higuchi:1993cy,Sriramkumar:1994pb,Fewster:2016ewy}. 
As we work in $(1+1)$ spacetime dimensions, we can take the detector 
to operate at constant coupling strength 
between a sharp switch-on moment and a sharp switch-off moment, 
without encountering infinities in the theory. 

In Section \ref{Compactified:Milne} we specialise to the spatially compactified Milne spacetime, 
in which the scale factor increases linearly in the cosmological time, and show that the untwisted 
scalar field has a massive zero mode in the asymptotic region near the initial singularity. 
In Section \ref{Milne:Como_NonComDetec} we write out the response of 
a detector on inertial but not necessarily comoving trajectories. 

Section \ref{sec:Minkowski} is a brief interlude in which we write out the 
response of an inertial finite time detector in Minkowski spacetime, giving numerical plots.
These plots will provide a benchmark to which the corresponding plots 
in Milne will be seen to reduce in the appropriate limits. 

Section \ref{sec:numerical} presents the core numerical results of the paper: 
the detector's response in spatially compactified Milne, on inertial but not necessarily comoving detector trajectories, 
in vacua adapted to late time dynamics and to early time dynamics, and, for the untwisted field, 
with selected choices for the early time massive zero mode state. 
When the spatial period is small, or when the trajectory is at early cosmological times, 
we see significant deviations from the Minkowski vacuum response for the de-excitation part of the spectrum, 
whereas the choice of the massive zero mode state for the 
untwisted field has a significant effect on the excitation part of the spectrum. 
The detector's peculiar velocity with respect to comoving cosmological observers
has the strongest effect in the ``in'' vacuum of the untwisted field, 
where it shifts the excitation and de-excitation resonances 
towards higher values of the detector's energy gap. 
For reasons of graphical convenience, most of the numerical plots 
for this section are delegated to an appendix. 

Section \ref{sec:conclusions} presents a summary and concluding remarks. 

We use units in which $\hbar = c =1$. 
We work in the sign convention in which $ds^2>0$ for timelike separations.

\section{Quantised scalar field in a spatially compact cosmological spacetime\label{CompactifiedCosmol}} 

In this section we quantise a real massive scalar field in a $(1+1)$-dimensional 
spatially compactified Friedmann-Lema\^itre-Robertson-Walker spacetime. 
We consider both a field that takes values in the trivial bundle, called an untwisted or periodic field, and a field that takes values 
in a nontrivial $\BbbR/\BbbZ_2$ bundle, called a twisted or antiperiodic field.

\subsection{Spacetime, field equation, and mode solutions}

We consider a spacetime with the line element 
\begin{align}\label{ExpComosMetric}
ds^2 = dt^2 - a^2(t)dx^2 = C(\eta)(d\eta^2 - dx^2) \, , 
\end{align}
where the scale factor $a(t)$ is assumed positive, 
the cosmological time $t$ and the conformal time $\eta$ are related by $d\eta/dt = 1/a(t)$, 
and $C(\eta) = a^2 (t(\eta))$. 
$C$ is by assumption positive, and we assume it to be a $C^\infty$ function of~$\eta$. 

We take $x$ to be periodic with period $L>0$, so that 
$(t, x)\sim(t, x+L)$, or $(\eta, x)\sim(\eta, x+L)$. 
The constant $\eta$ surfaces are hence topologically circles. 

We consider a real scalar field $\phi$ of mass $m>0$. 
In terms of the conformal time, the action reads 
\begin{align}
S = \int \mathcal{L} \, d\eta\,dx 
\, , 
\label{eq:action}
\end{align}
where the Lagrangian density $\mathcal{L}$ is given by 
\begin{subequations}
\label{NeutralScalarFie:FLRW}
\begin{align}\mathcal{L} 
& \, = \frac{1}{2} \left [( \partial_\eta \phi)^2 -(\partial_x \phi)^2 - \mu^2 (\eta) \phi^2\right], 
\\
\mu (\eta) & : = m \sqrt{C (\eta)} \, . 
\end{align}
\end{subequations}
Note that $\mu (\eta)$ appears as an effective time-dependent mass, and it is by assumption positive. 
The field equation is the Klein-Gordon equation, 
\begin{align}\label{scalar_fieldEqI}
\left(\frac{\partial^2}{\partial \eta^2}  - \frac{\partial^2}{\partial x^2} + \mu^2(\eta)\right) \phi(\eta, x) = 0 \, .
\end{align}
An untwisted field is periodic as $(\eta, x) \mapsto(\eta, x+L)$, 
while a twisted field is antiperiodic as $(\eta, x) \mapsto(\eta, x+L)$. 

We seek mode solutions to the field equation by the separation ansatz 
\begin{align}\label{Scalar field: discrete normal modes}
U_n(\eta,x) =  L^{-1/2}\chi_n(\eta)\exp( i k_n x)
\, ,
\end{align}
where $n\in\BbbZ$ and
\begin{align}\label{eq:cylindermomenta_ScalarField_FRW}
k_n &:= \begin{cases}
2\pi n /L & \text{for untwisted field} \, ,
\\
2\pi(n+\tfrac12)/L & \text{for twisted field} \, .
\end{cases}
\end{align}
The differential equation for $\chi_n$ is 
\begin{align}\label{ScalarFrequency_ode-discrete}
\chi''_n(\eta) + \omega_n^2(\eta)\chi_n(\eta) \,= 0 \, , 
\end{align}
where the prime denotes derivative with respect to $\eta$ and 
\begin{align}\label{ScalarFrequency_def-discrete}
\omega_n(\eta) &:= \bigl(k_n^2 +  \mu^2(\eta)  \bigr)^{1/2} \, 
\, . 
\end{align}
To make the mode solutions $U_n$ \eqref{Scalar field: discrete normal modes} 
a positive norm orthonormal set in the Klein-Gordon inner product, 
\begin{align}\label{Hilbert_scalarproduct:CompactifiedSpa}
( U_n, U_m) := i  \int_{-L/2}^{L/2} d x  \, \left( U_n^*\partial_\eta U_m - U_m\partial_\eta U_n^* \right) =\delta_{nm} 
\, ,
\end{align}
we require the mode functions $\chi_n$ to be chosen so that they satisfy the Wronskian condition 
\begin{align}\label{WronskianNorma:CompactifiedSpa}
W[\chi_n,\chi_n^*] := \chi_n  \chi^{\prime *}_n - \chi_n^*  \chi'_n = i
\, .
\end{align}
The complex conjugate mode solutions $U_n^*$ form then 
a negative norm orthonormal set in the Klein-Gordon inner product.

\subsection{Fock quantisation}

Given a choice of the mode functions~$\chi_n$, 
we can introduce a Fock quantisation by expanding the quantised field 
$\hat{\phi}$ as 
\begin{align}
\label{ScalarFieldMode_ExpanCosmo2}
\hat{\phi}(\eta, x)
&= \sum_n
\Bigl(
U_{n}(\eta,x)  \hat{a}_n
+ U^*_{n}(\eta,x)  \hat{a}^\dag_n
\Bigr) 
\ ,
\end{align}
where the nonvanishing commutators of the annihilation and creation operators are 
\begin{align}\label{anticommutatoRelat_ExpanCosmII}
\bigl[ \hat{a}_n, \hat{a}^\dag_m\bigr] 
= \delta_{n,m} \mathbbm{1} 
\ . 
\end{align}
The Fock space is built on the normalised vacuum state $|0 \rangle$, which satisfies 
\begin{align}
{\hat a}_n |0 \rangle =0 \, , 
\end{align}
and it carries a representation of 
$\hat{\phi}$ and the conjugate momentum 
$\hat{\pi}(\eta,x) = \partial_\eta \hat{\phi}(\eta,x)$, satisfying the canonical commutation relations 
\begin{align}
\label{CCRs_CosmologicalSpa}
\left[ \hat{\phi}(\eta,x), \hat{\pi}(\eta,x')    \right] = \, i \delta_{x,x'} \mathbbm{1} 
\, . 
\end{align}
The Wightman function is given by 
\begin{align}\label{Scalarfield_WightFuntII}
G (\eta,x ; \eta', x') 
&:= \langle 0 | {\hat\phi} (\eta,x ) {\hat\phi} (\eta',x' )|0 \rangle
\notag 
\\
&\,\,= 
\frac{1}{L} \sum_n  \chi_n (\eta) \chi_n (\eta') e^{ik_n(x - x')} 
\, .
\end{align}

The choice of the mode functions $\chi_n$ is however not unique: 
different choices lead to distinct vacua, 
and to unitarily inequivalent Hilbert spaces~\cite{Birrell:1982ix,Mukhanov:2007zz}. 
We shall now turn to situations in which a distinct set of mode functions, 
and a corresponding distinct vacuum, 
may be chosen by considering the behaviour of the mode functions 
in an asymptotic past or an asymptotic future.

\subsection{In and out vacua\label{subsec:in-and-out}}

To address the asymptotic past and future, 
we write the range of $\eta$ as $\eta \in (\eta_{in}, \eta_{out})$, where 
$\eta_{in}$ may be finite or $-\infty$, and $\eta_{out}$ may be finite or~$+\infty$. 
We refer to the asymptotic region $\eta \rightarrow \eta_{in}$ as the remote past, or the ``in" region, 
and to the asymptotic region $\eta \rightarrow \eta_{out}$ 
as the remote future, or the ``out" region. 
A~set of mode functions chosen by criteria set in the ``in'' region is denoted by~$\chi_n^{in}$, 
and 
a set of mode functions chosen by criteria set in the ``out'' region is denoted by~$\chi_n^{out}$. 

A situation that occurs often is that the ``in'' and ``out'' modes can be chosen to have the 
the asymptotic adiabatic form \cite{Birrell:1982ix,Mukhanov:2007zz,Parker:1974qw,Audretsch:1979uv}, 
\begin{align}\label{pos:frequency_ScalarMode_nonstatic}
\chi_n^{in/out}(\eta)  \xrightarrow[\eta \rightarrow  \eta_{in/out}]{}  
\frac{1}{ \sqrt{2\omega_n(\eta)} }\exp(-i\int^{\eta}\omega_n(\eta')d\eta') \, . 
\end{align}
A sufficient condition for this asymptotic form to exist is that 
\begin{align}\label{positiveFreq_Requirement}
\frac{d^p}{d\eta^p} \frac{C'}{C} \xrightarrow[\eta \rightarrow  \eta_{in/out}]{}  0, \quad \quad  \forall p \geq 0 \, . 
\end{align}
Technically, the asymptotic form \eqref{pos:frequency_ScalarMode_nonstatic} 
means that the modes are of locally positive frequency with respect to the 
conformal Killing vector~$\partial_\eta$. 
When the time dependence of the spacetime is slow, 
the physical interpretation is that the corresponding adiabatic vacuum state 
is perceived by local observers as a no-particle state; 
this is in particular the case if $C(\eta)$ tends to a positive constant as 
$\eta \rightarrow \eta_{in/out}$, in which case the in/out region is asymptotically static, and the condition \eqref{positiveFreq_Requirement} clearly holds. 
When the time dependence of the spacetime is not slow, 
the corresponding adiabatic vacuum state need not have a similar no-particle interpretation, 
but the state is nevertheless distinguished by the geometry of the asymptotic region; 
this criterion is often employed for choosing a state in cosmological models with 
an early universe inflationary phase~\cite{Mukhanov:2007zz}. 

Now, we wish to address the situation in which 
$C(\eta)$ 
tends to zero as $\eta \rightarrow \eta_{in/out}$, 
in such a way that modes of the asymptotic adiabatic form  
\eqref{pos:frequency_ScalarMode_nonstatic} do not exist for all~$n$. 
From 
\eqref{ScalarFrequency_ode-discrete}
and 
\eqref{ScalarFrequency_def-discrete}
we see that the mode for which the adiabatic form fails 
is the $n=0$ mode of the untwisted field. 
Following Ford and Pathinayake~\cite{Ford:1989mf}, 
we call a mode with this property a massive zero mode. 

Our main interest is in the classification of the possible choices of the mode functions 
of the massive zero mode, 
and in the consequences of these choices for the corresponding vacuum state. 
We shall briefly comment on coherent states at the end of Section~\ref{subsec:untwisted-theory}, 
but we leave the consequences for more general non-vacuum states, pure or mixed, a subject to future work. 

We assume the falloff of $C(\eta)$ as $\eta \rightarrow \eta_{in/out}$ 
to be such that the leading terms in the general solution 
for the untwisted field's spatially constant mode $\chi_0$ are 
\begin{align}\label{Zero-MomentumConstraint}
\chi_0(\eta)  \xrightarrow[\eta \rightarrow \eta_{in/out}]{} 
a_1 +a_2\eta \, , 
\quad{} \quad{} a_1, a_2 \in {\mathbb C}\, .
\end{align}
An example is when $C(\eta)$ decays exponentially as 
$\eta \to \eta^{in/out} = \mp \infty$, 
so that \eqref{positiveFreq_Requirement} holds for $p>0$ but fails for $p=0$. 
This is the out-region situation considered by Ford and Pathinayake \cite{Ford:1989mf} 
and the in-region situation that we shall encounter in Section~\ref{Compactified:Milne}. 
Another example is when $C(\eta)$ is a multiple of $\eta^2$, $0<\eta<\infty$, 
in which case $a(t)$ is a multiple of $t^{1/2}$, corresponding to a four-dimensional radiation-dominated expansion law: 
there is now a massive zero mode satisfying \eqref{Zero-MomentumConstraint} in the in-region, $\eta \to 0_+$. 
In the rest of this section we can however proceed assuming just~\eqref{Zero-MomentumConstraint}, 
leaving the details of the falloff of $C(\eta)$ unspecified. 

Given \eqref{Zero-MomentumConstraint}, 
the Wronskian condition \eqref{WronskianNorma:CompactifiedSpa} gives 
\begin{align}\label{zero-mode:constraints}
a_1 a_2^* - a_1^* a_2= i
\, ,
\end{align}
which shows that neither $a_1$ nor $a_2$ can vanish. 
We may hence fix the overall phase of $\chi_0$ uniquely by taking $a_2>0$,
and then write the general solution of \eqref{zero-mode:constraints} as 
\begin{subequations}
\label{constraint-values}
\begin{align}
a_1 &= b_1 + \frac{i}{2b_2}  \, ,
\\
a_2 &= b_2 \, , 
\end{align}
\end{subequations}
where $b_1\in\BbbR$ and $b_2>0$. 
The choices for the mode function of the massive zero mode form hence a family with two real parameters.

\subsection{Untwisted quantum theory with a massive zero mode\label{subsec:untwisted-theory}}

We now write out a quantum theory of the untwisted field, 
assuming that the $n=0$ mode is a massive zero mode satisfying \eqref{Zero-MomentumConstraint} 
in the past or in the future, 
while for all the other modes the positive norm mode functions can be 
chosen by the adiabatic criterion~\eqref{pos:frequency_ScalarMode_nonstatic}, 
respectively in the past or in the future. 
The main issue is to identify the consequences of the choice of the mode function for the massive zero mode. 

We decompose the quantum field $\hat\phi$ as 
\begin{subequations}
\label{eq:psi-osc+zm-decomposition}
\begin{align}
\hat{\phi}(\eta, x) &= \hat{\phi}_0(\eta) + \hat{\phi}_{osc}(\eta,x)
\, ,
\label{eq:psi-osc+zm-split:cosmology}
\\
\hat{\phi}_{osc}(\eta,x)
&= \sum_{n\ne0} 
\Bigl(
U_{n}(\eta, x)  \hat{a}_n
+  h.c.
\Bigr) 
\, ,
\label{Oscillatormode:ExpansionCosmologicalSpa}
\\
\hat{\phi}_0(\eta) &=\frac{1}{\sqrt{L}}
\chi_0(\eta) \hat{a}_0  +  h.c.  \, ,
\label{ModeExpan_Discrete}
\end{align}
\end{subequations}
where the mode functions have been chosen as described above, 
and we have dropped the superscripts specifying whether 
the choice of the modes refers to the ``in'' region or the ``out'' region. 
We refer to the $n\ne0$ modes as the oscillator modes. 

The Wightman function \eqref{Scalarfield_WightFuntII} decomposes as 
\begin{subequations}
\begin{align}
G(\eta,x ; \eta', x') 
&= 
G_0(\eta; \eta') 
+ 
G_{osc}(\eta,x ; \eta', x') 
\, , 
\\
G_0(\eta; \eta') 
&= 
\frac{1}{L} \chi_0(\eta)\chi_0^*(\eta')
\, , 
\label{eq:Gnought-pre}
\\
G_{osc}(\eta,x ; \eta', x') 
&= 
\frac{1}{L} \sum_{n\ne0}  \chi_n (\eta)\chi_n^*(\eta') e^{ik_n(x - x')} 
\, . 
\end{align}
\end{subequations}
The part that depends on the choice of the mode functions 
of the massive zero mode is $G_0(\eta; \eta')$~\eqref{eq:Gnought-pre}. 
In the asymptotic region, 
\eqref{Zero-MomentumConstraint}
and 
\eqref{constraint-values}
show that $G_0(\eta; \eta')$ redudes to  
\begin{align}
G_0(\eta; \eta') 
\longrightarrow  
\frac{1}{L} 
\left[ \left(b_1 + \frac{i}{2b_2} \right) + b_2 \eta \right] 
\left[ \left(b_1 - \frac{i}{2b_2} \right) + b_2 \eta' \right] 
\, . 
\end{align}

The expectation value of the stress-energy tensor requires a renormalisation, 
but the massive zero mode contribution to the 
expectation value of the energy density may be found by elementary considerations as follows. 
Let ${}_0\widehat{T}_{\mu\nu}$ denote the massive zero mode 
contribution to the stress-energy tensor operator. 
In the coordinates $(\eta,x)$, we see from \eqref{eq:action} and \eqref{NeutralScalarFie:FLRW} that 
\begin{align}
{}_0\widehat{T}_{\eta\eta}(\eta) 
= \frac{1}{2} 
\left[
\bigl( \hat \phi'_0(\eta) \bigr)^2  + m^2 C(\eta) \bigl( \hat \phi_0(\eta) \bigr)^2
\right], 
\end{align}
where we recall that the prime denotes $\frac{d}{d\eta}$. 
Using \eqref{ModeExpan_Discrete}, we have 
\begin{align}
\langle 0 | {}_0\widehat{T}_{\eta\eta}(\eta)  |0 \rangle
= \frac{1}{2L} 
\left[ \left|\chi'_0(\eta)\right|^2
+ m^2 C(\eta) \left|\chi_0(\eta)\right|^2
\right] 
\, . 
\end{align}
The contribution from the massive zero mode to the energy 
density seen by a comoving observer is hence 
\begin{align}
\rho_{0,comov} = - 
\langle 0 | {}_0\widehat{T}^{\eta}{}_\eta(\eta)  |0 \rangle
= 
\frac{1}{2L}
\left[ 
\frac{\left|\chi'_0(\eta)\right|^2}{C(\eta)}
+ m^2 \left|\chi_0(\eta)\right|^2
\right] 
\, . 
\label{eq:rhonought}
\end{align}
In the asymptotic region, 
\eqref{Zero-MomentumConstraint}
and 
\eqref{constraint-values}
give 
\begin{align}
\rho_{0,comov} \longrightarrow 
\frac{1}{2L}
\left[ 
\frac{\beta^2}{C(\eta)}
+ m^2 
\left(
(b_1 + b_2\eta)^2 + \frac{1}{4b_2^2}
\right) 
\right] 
\, . 
\end{align}
As $b_2>0$, and $C(\eta)\to0$ by assumption, $\rho_{0,comov}$ 
hence grows without bound in the asymptotic region, proportionally to~$\beta^2$. 

We end this section with two comments. 

First, we note that the contributions to the canonical commutation relations 
\eqref{CCRs_CosmologicalSpa}
from the massive zero mode and the oscillator modes decompose as 
\begin{subequations}
\begin{align}
\left[ \hat{\phi}_{0}(\eta), \hat{\pi}_{0}(\eta)    \right] &= \frac{i}{L}\mathbbm{1} 
\, ,
\label{CCRs_CosmologicalSpa_ZeroMo} 
\\
\left[ \hat{\phi}_{osc}(\eta,x), \hat{\pi}_{osc}(\eta, x')    \right] &=i \left(  \delta_{x,x'}  - \frac{1}{L} \right)  \mathbbm{1} 
\, ,
\label{CCRs_CosmologicalSpa_ZeroOsc}
\end{align}
\end{subequations}
where 
$\hat{\pi}_{0}(\eta) = \frac{d}{d\eta} \hat{\phi}_{0}(\eta)$ 
and 
$\hat{\pi}_{osc}(\eta,x) = \frac{\partial}{\partial\eta} \hat{\phi}_{osc}(\eta,x)$. 
The massive zero mode Hamiltonian is 
\begin{align}
{\hat H}_0 = \frac{L}{2} \left(  {\hat \pi}_{0}^2 + m^2 C(\eta) {\hat\phi_0}^2 \right) 
\, .  
\end{align}
The dynamics of the massive zero mode is therefore that of a nonrelativistic 
particle on the real line in a quadratic potential with a time-dependent frequency. 
From \eqref{WronskianNorma:CompactifiedSpa} and \eqref{ModeExpan_Discrete} we obtain 
\begin{align}
\chi^{\prime *}_0(\eta) \hat\phi_0(\eta) - \chi_0^*(\eta) \hat\pi_0(\eta)
= \frac{i \hat a_0}{\sqrt{L}}
\, . 
\end{align}
In a ``position'' representation, in which $\hat\pi_0 = -(i/L) \partial_{\phi_0}$ by~\eqref{CCRs_CosmologicalSpa_ZeroMo}, 
the wave function of the massive zero mode Fock vacuum is hence 
\begin{align}
\Psi_0 (\phi_0) = N \exp \! 
\left[
\frac{i L \chi^{\prime *}_0(\eta)}{2\chi_0^*(\eta_0)} 
\phi_0^2
\right]
\, ,
\end{align}
where $\eta_0$ denotes a reference moment such that 
$\phi_0$ is the position representation of $\hat\phi_0(\eta_0)$, and $N$ is a normalisation constant. 
In the asymptotic region, where the nonrelativistic particle becomes free, 
\eqref{Zero-MomentumConstraint}
and 
\eqref{constraint-values}
give 
\begin{align}
\Psi_0 (\phi_0) \longrightarrow 
\frac{{(2Lb_2^2/\pi)}^{1/4}}{\sqrt{1 + 2 i b_2^2 (\eta_0 + b_1/b_2)}}
\exp \! 
\left[
- \frac{L b_2^2}{1 + 2 i b_2^2 (\eta_0 + b_1/b_2)} 
\phi_0^2
\right]
\, ,
\end{align}
which is recognised as the Gaussian wave packet of a free nonrelativistic particle, 
with $\eta_0$ specifying the moment of time. 
This offers another physical interpretation of the parameters $b_1$ and~$b_2$. 

Second, given that the massive zero mode Fock vacuum is not unique, 
it may be of interest to consider more general states for this mode. 
As an example, following \cite{Ford:1989mf}, 
consider the coherent state $|z\rangle_0$, 
satisfying $\hat a_0 |z\rangle = z|z\rangle$, where $z \in \BbbC$ 
is a parameter. The above analysis generalises in a straightforward fashion. 
The massive zero mode contribution to the Wightman function 
generalises from \eqref{eq:Gnought-pre} to 
\begin{align}
G_0(\eta; \eta') 
= 
\frac{1}{L} 
\Bigl[
\bigl(z\chi_0(\eta) + z^*\chi^*_0(\eta)\bigr)
\bigl(z\chi_0(\eta') + z^*\chi^*_0(\eta')\bigr)
+ \chi_0(\eta)\chi_0^*(\eta')
\Bigr] 
\, , 
\end{align}
and the contribution to the 
comoving energy density 
generalises from \eqref{eq:rhonought} to 
\begin{align}
\rho_{0,comov}
&= 
\frac{1}{2L}
\Biggl\{ 
\frac{\left|\chi'_0(\eta)\right|^2
+ 
\bigl(z\chi'_0(\eta) + z^*\chi^{\prime*}_0(\eta)\bigr)^2}{C(\eta)}
\notag
\\
&\hspace{8ex}
+ m^2 
\left[ 
\left|\chi_0(\eta)\right|^2
+ 
\bigl(z\chi_0(\eta) + z^*\chi^*_0(\eta)\bigr)^2
\right] 
\Biggr\}
\, . 
\end{align}

\section{Unruh-DeWitt detector\label{Unruh-Dewitt:ScalarDetector}}

In this section we recall relevant properties of an Unruh-DeWitt 
detector coupled linearly to the scalar field \cite{Unruh:1976db,DeWitt:1979}, 
operating for a finite time \cite{finitetime1,Higuchi:1993cy,Sriramkumar:1994pb,Fewster:2016ewy}. 

The detector is a spatially pointlike two-state system, moving on the timelike worldline~$\sfx(\tau)$, 
where $\tau$ is the proper time. The detector's Hilbert space $\mathcal{H}_D$ is spanned by the orthonormal states 
$|0 \rangle_D$  and $|\omega \rangle_D$, 
satisfying $\hat{H}_D|0 \rangle_D =0$ 
and 
$\hat{H}_D|\omega \rangle_D = \omega |\omega \rangle_D$, 
where $\hat{H}_D$ is the detector's Hamiltonian and $\omega\in\BbbR$. 
$|0 \rangle_D$ is the ground state if $\omega > 0$ and
the excited state if $\omega < 0$. 

The Hilbert space of the coupled detector-field system is $\mathcal{H}_D\otimes \mathcal{H}_\phi$, 
where $\mathcal{H}_\phi$ is the Hilbert space of the field~$\phi$. 
The total Hamiltonian is 
$\hat{H} = \hat{H}_D + \hat{H}_\phi + \hat{H}_{\text{int}}$, where 
$\hat{H}_\phi$ is the Hamiltonian of the free field 
and 
\begin{equation}
\hat{H}_{\text{int}} :=  
c \chi(\tau) \hat{\mu}(\tau)\hat{\phi} \big(x(\tau)\big) 
\, , 
\label{InterHamilt}
\end{equation}
where $c\in\BbbR$ is a coupling constant, 
the real-valued switching function $\chi$ 
specifies how the interaction is turned on and off, 
and 
$\hat{\mu}(\tau)$ is the detector's 
monopole moment operator, 
evolving in the interaction picture as
\begin{align}\label{MonopoleOperator}
\hat{\mu}(\tau) = e^{i\hat{H}_{D}\tau}\hat{\mu}(0)e^{-i\hat{H}_{D}\tau}
\, . 
\end{align}

Suppose that the switching function $\chi$ has compact support, 
and the total system is initially prepared in the product state $|\Psi\rangle \otimes |0 \rangle_D $, 
where the field state $|\Psi\rangle$ 
satisfies the Hadamard condition~\cite{decanini-folacci}.
After the interaction has ceased, 
the probability for the detector to have made the transition to the state $|\omega \rangle_D$ is, 
in first-order perturbation theory, 
\begin{align}\label{TransitionProb}
P =c^2 {|{}_D\langle 0|\mu(0)| 1\rangle_D|}^2 \mathcal{F}(\omega)
\, ,
\end{align}
where 
\begin{align}
\mathcal{F}(\omega) := 
\int
d\tau \, d\tau' \, 
\chi(\tau) \chi(\tau') 
\, e^{-i\omega(\tau-\tau')} \, G(\tau,\tau')
\label{RespFunct:Scalarfiel_CB}
\end{align}
and 
\begin{align}
G (\tau, \tau')
:= 
\langle \Psi | 
\hat{\phi}\bigl( x(\tau) \bigr)
\hat{\phi}\bigl(x(\tau') \bigr)
|\Psi \rangle
\ . 
\end{align}
$\mathcal{F}$ is called the response function, and it encodes the dependence of 
$P$ on the trajectory, the switching and~$\omega$, as the prefactor 
$c^2 {|{}_D\langle 0|\mu(0)| 1\rangle_D|}^2$ in \eqref{TransitionProb} depends only on $c$ and 
the detector's internal structure. 
With minor abuse of terminology, we refer to $\mathcal{F}$ as the transition probability. 

As $|\Psi \rangle$ is by assumption Hadamard, $G (\tau, \tau')$ 
is a well-defined distribution~\cite{Hormander:1983,Fewster:1999gj}. 
If $\chi$ is smooth, $\mathcal{F}$ is hence well defined. 
In spacetime dimension $1+1$, however, the coincidence limit singularity of 
$G (\tau, \tau')$ is only logarithmic~\cite{decanini-folacci}, 
and $\mathcal{F}$ is then well defined also for less regular~$\chi$. 
We shall take $\chi$ to have a sharp switch-on and switch-off, 
\begin{align}
\chi(\tau) = \Theta (\tau - \tau_0) \Theta(\tau_1 - \tau) 
\, ,
\end{align}
where $\tau_0$ denotes the switch-on moment and $\tau_1$ denotes the switch-off moment, and we assume $\tau_0 < \tau_1$. 
The response function \eqref{RespFunct:Scalarfiel_CB} then becomes
\begin{align}\label{ResponsFunct_2D}
\mathcal{F}(\omega, \tau_1, \tau_0)
:= 
\int _{\tau_0}^{\tau_1}
d\tau \, \int _{\tau_0}^{\tau_1} d\tau' \, 
\, e^{-i\omega(\tau-\tau')} \, G(\tau,\tau')
\, .
\end{align}

\section{Detector's response in spatially compactified Milne spacetime\label{Compactified:Milne}}

In this section we specialise to the expanding Milne spacetime with compactified spatial sections. 

\begin{figure}[t]
\centering
\includegraphics[width=0.65\textwidth]{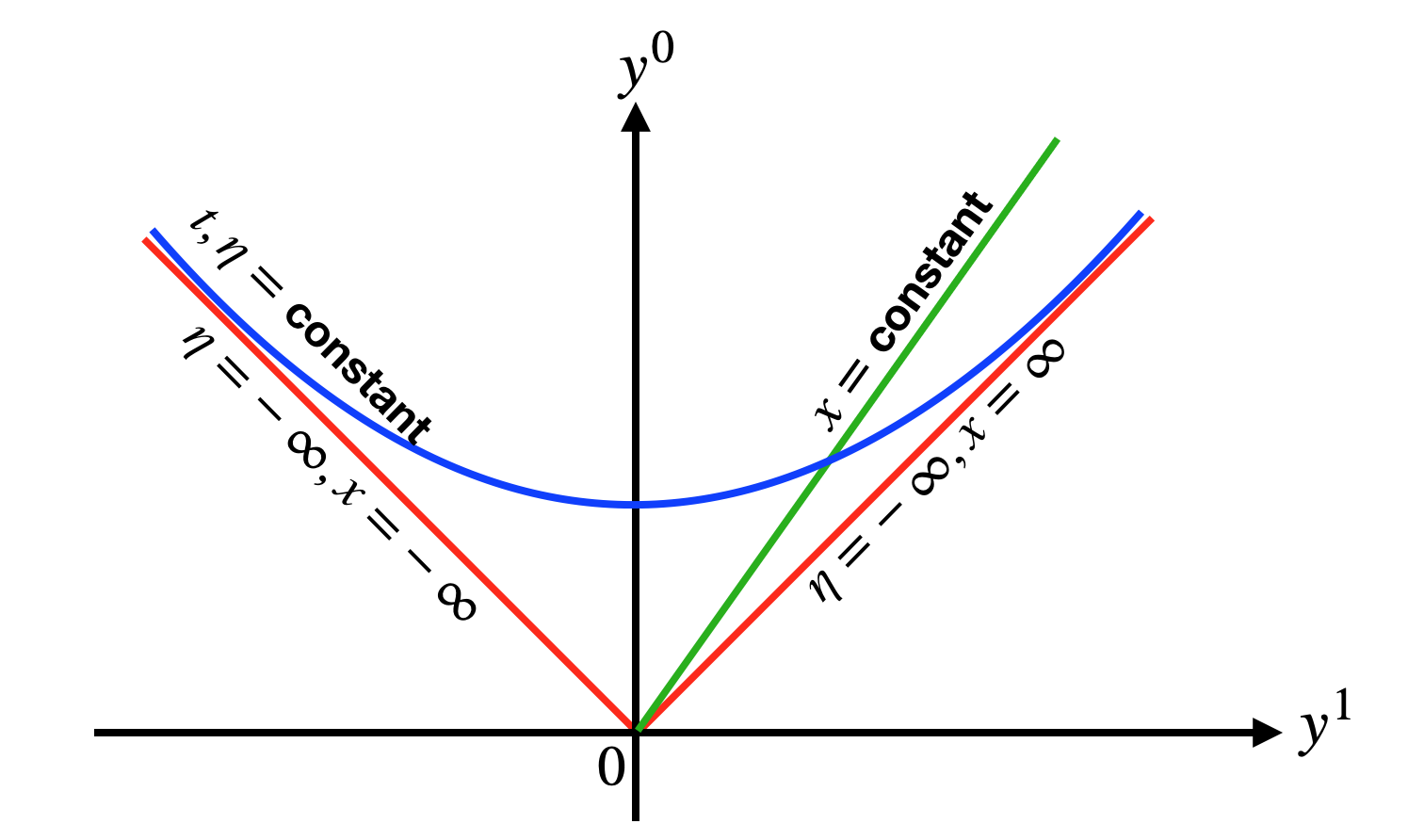} 
\caption{$(1+1)$-dimensional Milne spacetime. 
When $x$ is not compactified, the $(t, x)$ coordinates cover the future light-cone of Minkowski space, $y^0 > |y^1|$. 
The lines $x= \text{constant}$ are timelike geodesics emanating from the origin. 
Making $x$ periodic means identifying the spacetime by a boost.\label{MilneDiagram}}
\end{figure}

\subsection{Milne spacetime}
\label{Compactified:Milne_A}
The expanding $(1+1)$-dimensional Milne universe is the special case of \eqref{ExpComosMetric}
for which \cite{Birrell:1982ix}
\begin{align}\label{MilneMetric}
ds^2 &= dt^2 - a^2t^2dx^2 = e^{2a\eta}(d\eta^2 - dx^2)
\, ,  
\end{align}
where $a$ is a positive constant, $0<t < \infty$, $-\infty < \eta < \infty$ and $t =a^{-1}e^{a\eta}$. 
In the notation of~\eqref{ExpComosMetric}, 
we have $C(\eta) = e^{2a\eta}$. 
$t$, $x$ and $\eta$ have the dimension of length and $a$ has the dimension of inverse length. 

When $x$ is not compactified, this spacetime is the future quadrant, $y^0 > |y^1|$, of Minkowski spacetime, with the metric 
\begin{align}
ds^2 = {(dy^0)}^2 - {(dy^1)}^2
\, ,  
\end{align}
as seen by the coordinate transformation 
\begin{subequations}
\label{Minkowski:Time}
\begin{align}
y^0 &= a^{-1}e^{a\eta}\cosh(ax)
\, , 
\\
y^1 &= a^{-1}e^{a\eta}\sinh(ax)
\, . 
\end{align}
\end{subequations}
The cosmological initial singularity $t\to0_+$ is a coordinate singularity on the future light cone of the origin, 
$y^0 = |y^1|$, as illustrated in Figure~\ref{MilneDiagram}. 

We consider the case in which $x$ is compactified, 
by the identification $(\eta, x) \mapsto  (\eta, x +L)$, 
which is geometrically a boost of rapidity~$aL$. 
In this case the initial singularity at $t\to0_+$ is a genuine singularity, 
known as the Misner space singularity~\cite{hawking-ellis}. 

A selection of previous work on quantum field theory on the spacetime 
without spatial compactification (and on its four-dimensional counterpart) 
is \cite{Birrell:1982ix,diSessa:1974ve,Sommerfield,Gromes,Fulling:1974pu,Padmanabhan:1990fm}. 
We shall consider the spatially compactified spacetime. As the criterion 
\eqref{positiveFreq_Requirement} does not hold at $\eta \to -\infty$ for $p=0$, 
we may expect the untwisted field to have a massive zero mode, 
and we shall see that this indeed happens.

\subsection{Twisted field\label{subsec:twisted-response}}

The mode equation \eqref{ScalarFrequency_ode-discrete} is solvable in terms of Bessel functions of imaginary order~\cite{dlmf}. 
For the twisted field, the criterion 
\eqref{pos:frequency_ScalarMode_nonstatic}
selects the ``in'' and ''out'' 
positive frequency mode functions 
\begin{subequations}\label{posfreqmodes_Milne}
\begin{align}
\chi_n^{in}(\eta) &= \left[ (2a/\pi)\sinh(\pi |k_n|/a)            \right]^{-1/2}  J_{-i | k_n |/a}(me^{a\eta}/a) \, , \\
\chi_n^{out}(\eta) &= \frac{1}{2}(\pi/a)^{1/2}e^{ \pi | k_n | /(2a)} H^{(2)}_{i | k_n| /a}(me^{a\eta}/a)
\label{ModeFunctoutRebbecca}
\, ,
\end{align}
\end{subequations}
where we recall that $k_n$ is given by the twisted field expression in~\eqref{eq:cylindermomenta_ScalarField_FRW}. 
The ``in'' and ``out'' asymptotic behaviour is \cite{dlmf}
\begin{subequations}
\label{eq:asymptoticfrquency_Milne}
\begin{align}
\chi_n^{in}\left(\eta \right)  & \xrightarrow[\eta \rightarrow -\infty]{} \ 
e^{i\varphi_n}
(2 |k_n|)^{-1/2}\exp \left(-i |k_n|\eta  \right) \, ,\\
\chi_n^{out} \left(\eta  \right)  & \xrightarrow[\eta \rightarrow + \infty]{} \ 
e^{i\pi/4}(2 m \, e^{a \eta})^{-1/2} \exp \left(-i m \, e^{a\eta}/a \right)
\, ,              
\end{align}
\end{subequations}
where the real-valued phase constant $\varphi_n$ has an expression in terms of 
Euler's gamma-function. 

The Wightman functions for the ``in'' and ``out'' vacua are 
\begin{subequations}
\begin{align}
\label{Scalarfield_inWightFunt_CompMilne}
G^{in}(\eta,x, \eta',x')  &=  \frac{\pi}{2aL} \sum_{n} \frac{  J_{-i | k_n |/a}(s) J_{i |k_n|/a}(s') e^{i k_n(x -x')}}{ \sinh (\pi | k_n | /a)} \, , \\
G^{out}(\eta,x, \eta',x')  &=  \frac{\pi}{4aL}  \sum_{n} e^{ \pi | k_n| /a} H^{(2)}_{i | k_n | /a}(s)  H^{(1)}_{-i | k_n | /a}(s')  e^{i k_n(x -x')}
\label{Scalarfield_outWightFunt_CompMilne}
\, ,
\end{align}
\end{subequations}
where we have written $s := me^{a\eta}/a$ and $s' := me^{a\eta'}/a$. 
For the response of the detector in the two vacua, on the worldline $\sfx(\tau) = \bigl(t(\tau),x(\tau)\bigr)$, 
\eqref{ResponsFunct_2D} then gives 
\begin{subequations} 
\label{ResFunct_both-vacs_1}
\begin{align}
\label{ResponsFunct:TwistedField_1}
\mathcal{F}^{in}_{t}( \omega,\tau_1, \tau_0) &= \frac{\pi}{2aL}  \sum_{n=-\infty}^{\infty} \frac{\left | A_n ( \omega,\tau_1, \tau_0)  \right |^2}{ \sinh (\pi | k_n | /a)} \, , \\
\mathcal{F}^{out}_{t}( \omega,\tau_1, \tau_0)  &= \frac{\pi}{4aL}   \sum_{n=-\infty}^{\infty} e^{ \pi | k_n| /a}   \left | B_n ( \omega,\tau_1, \tau_0)  \right |^2 
\, ,
\label{ResFunct_OutVac_1}
\end{align}
\end{subequations}
where the subscript $t$ stands for ``twisted'' and 
\begin{subequations}\label{IntegralFunct:Twist}
\begin{align}
A_n ( \omega,\tau_1, \tau_0)  &:= \int_{\tau_0}^{\tau_1} d\tau \, 
J_{-i | k_n |/a}\left( m t(\tau) \right) \,  e^{-i \left(\omega \tau - k_n x(\tau) \right)} \, , 
 \\
B_n ( \omega,\tau_1, \tau_0)  &:= \int_{\tau_0}^{\tau_1} d\tau \, 
H^{(2)}_{i | k_n | /a}(m t(\tau)) \,  e^{-i \left(\omega \tau - k_n x(\tau) \right)} 
\, . 
\label{IntegralFunctB:no-twist}
\end{align}
\end{subequations}

\subsection{Untwisted field}

For the untwisted field, there are no massive zero modes at $\eta\to\infty$, 
but the spatially constant mode is a massive zero mode at $\eta\to-\infty$. 
We shall therefore consider the ``out'' and ``in'' vacua separately. 

\subsubsection{``Out'' vacuum}

For the ``out'' vacuum, we may proceed as for the twisted field in Section~\eqref{subsec:twisted-response},
with the exception that $k_n$ is now given by the untwisted field expression in~\eqref{eq:cylindermomenta_ScalarField_FRW}. 
The Wightman function is 
\begin{align}
G^{out}(\eta,x, \eta',x')
&=  \frac{\pi}{4aL}   \sum_{n = -\infty}^{\infty} e^{ \pi | k_n| /a} H^{(2)}_{i | k_n | /a}(s)  H^{(1)}_{-i | k_n | /a}(s')  e^{i k_n(x -x')}
\, ,
\label{Scalarfield_inWigCompMilneB}
\end{align}
where again $s= m e^{a\eta/a}$ and $s'= m e^{a\eta'/a}$.
For the detector's response, 
on the worldline $\sfx(\tau) = \bigl(t(\tau),x(\tau)\bigr)$, we have 
\begin{align}
\mathcal{F}^{out}_{u}( \omega,\tau_1, \tau_0)  &= \frac{\pi}{4aL}  \sum_{n=-\infty}^\infty e^{ \pi | k_n| /a}   \left | B_n ( \omega,\tau_1, \tau_0) \right |^2 
\, ,
\end{align}
where $B_n$ is as in~\eqref{IntegralFunctB:no-twist}, and the subscript $u$ stands for untwisted.

\subsubsection{``In'' vacuum}

For the ``in'' vacuum, the $n\ne0$ modes can be treated as in Section~\ref{subsec:twisted-response}. We call these modes the oscillator modes. 
Their contribution to the Wightman function is 
\begin{align}
\label{Scalarfield_inWightFunu_CompMilne}
G^{in}_{osc}(\eta,x, \eta',x')  &=  \frac{\pi}{2aL} \sum_{n\ne0} \frac{  J_{-i | k_n |/a}(s) J_{i |k_n|/a}(s') e^{i k_n(x -x')}}{ \sinh (\pi | k_n | /a)} 
\, ,
\end{align}
and their contribution to the detector's response is 
\begin{align}
\mathcal{F}^{in}_{osc}( \omega,\tau_1, \tau_0) &= \frac{\pi}{2aL}  \sum_{n\ne0} \frac{\left | A_n ( \omega,\tau_1, \tau_0)  \right |^2}{ \sinh (\pi | k_n | /a)} 
\, .
\label{eq:Fin-osc}
\end{align}

The $n=0$ mode functions are given by 
\begin{align}
\label{Comp:posfreqmodes_Milne_1}
\chi_0^{in}(\eta) &= c_1J_0(me^{a\eta}/a ) +  c_2 Y_0(me^{a\eta}/a )
\, ,
\end{align}
where
the coefficients $c_1$ and $c_2$ cannot be fixed by~\eqref{pos:frequency_ScalarMode_nonstatic}, 
but they are still constrained by the Wronskian condition~\eqref{WronskianNorma:CompactifiedSpa}, 
which reads 
\begin{align}\label{Normal:CompactZeroMode} 
c_1  c_2^* - c_1^*  c_2 = \frac{\pi i}{2a}
\, .
\end{align}
Fixing the overall phase of $\chi_0^{in}$ so that $c_2>0$, we parametrise $c_1$ and $c_2$ as 
\begin{subequations}
\label{Coeffic:ZeroMomentMode}
\begin{align}
c_1 &= \alpha + i \frac{\pi}{4a\beta}  \, , 
\\
c_2 &= \beta \, , 
\end{align}
\end{subequations}
where 
$\alpha\in\BbbR$ and $\beta>0$. 

The asymptotic early time behaviour of $\chi_0^{in}$ is 
\begin{align}
\chi_0^{in}\left(\eta \right)  & \xrightarrow[\eta \rightarrow -\infty]{} b_1 + \frac{i}{2b_2} + b_2\eta \, ,
\label{Asymptotic:ZeroMomentum-in}
\end{align}
where 
\begin{subequations}
\label{eq:b12-versus-alphabeta}
\begin{align}
b_1 &= \alpha + \frac{2\beta}{\pi} \left[  \ln(\frac{m}{2a}) +\gamma \right] 
\, , 
\\
b_2 &= \frac{2a \beta}{\pi}
\, ,
\end{align}
\end{subequations}
and $\gamma$ is the Euler-Mascheroni constant~\cite{dlmf}. 
The constants $b_1 \in \BbbR$ and $b_2>0$ in \eqref{Asymptotic:ZeroMomentum-in} 
are precisely the parameters introduced in Section 
\ref{subsec:in-and-out}
to label the massive zero mode states, and \eqref{eq:b12-versus-alphabeta}
shows how these parameters are in a one-to-one correspondence with $\alpha$ and~$\beta$. 
This is the rationale for the ``in'' label for~$\chi_0^{in}$. 

Note that in the special case
$\alpha = 0$
and $\beta = \frac12 (\pi/a)^{1/2}$, we have $c_1 = \tfrac{i}{2}(\pi/a)^{1/2}$ and $c_2 = \tfrac{1}{2}(\pi/a)^{1/2}$, 
so that $\chi_0^{in} = i \chi_0^{out}$. 
For all other values of $\alpha$ and~$\beta$, 
$\chi_0^{in}$ and $\chi_0^{out}$ are linearly independent. 

By \eqref{eq:Gnought-pre} and~\eqref{Comp:posfreqmodes_Milne_1}, 
the zero-momentum mode contribution to the Wightman function is 
\begin{align}
G_{0}^{in}(\eta, \eta') &= \frac{1}{L} \Bigl( \left | c_1 \right|^2J_0(s)J_0(s') + |c_2|^2 Y_0(s)Y_0(s') 
\notag 
\\ 
&\hspace{7ex}
+ c_1c_2^* J_0(s)Y_0(s') +c_1^*c_2 Y_0(s)J_0(s') \Bigr) 
\, , 
\end{align}
where again $s= m e^{a\eta/a}$ and $s'= m e^{a\eta'/a}$.
By~\eqref{ResponsFunct_2D}, 
the contribution to the detector's response is 
\begin{align}
\label{RespFunct_TwoReal_UntwistA}
\mathcal{F}^{in}_0(\omega,\tau_1, \tau_0) 
&= 
\frac{1}{L} \bigl\{  \left | c_1 M(\omega, \tau_1, \tau_0) \right |^2 + \left | c_2 N(\omega, \tau_1, \tau_0)\right |^2 
\notag
\\
& \hspace{7ex}
+ 2 \Realpart \left[c_1c_2^* M(\omega, \tau_1, \tau_0) N^*(\omega, \tau_1, \tau_0) \right]           
\bigr\} 
\, , 
\end{align}
where 
\begin{subequations}\label{Definitions:M&N&B}
\begin{align}
M(\omega, \tau_1,\tau_0) &:= \int_{\tau_0}^{\tau_1} d\tau \, J_0 \left(m t(\tau) \right)  e^{-i \omega \tau } \, ,  \\
N(\omega, \tau_1,\tau_0) &:= \int_{\tau_0}^{\tau_1} d\tau \, Y_0 \left(m t(\tau) \right)  e^{-i \omega \tau } \, . 
\end{align}
\end{subequations}

Collecting \eqref{eq:Fin-osc} and~\eqref{RespFunct_TwoReal_UntwistA}, 
the final formula for the response is  
\begin{align}
\mathcal{F}^{in}_{u}( \omega,\tau_1, \tau_0)  &= \mathcal{F}^{in}_{osc}( \omega,\tau_1, \tau_0)   + \mathcal{F}_0^{in}(\omega,\tau_1, \tau_0)  
\, ,
\label{FullField:ResponseFunctComMilne}
\end{align}
where the subscript $u$ stands for ``untwisted.''

\section{Inertial detector\label{Milne:Como_NonComDetec}}

In this section we specialise to detector trajectories in Milne that are inertial 
but not necessarily comoving with the cosmological expansion. 
We shall write the response as $1/m^2$ times 
a dimensionless function of dimensionless combinations of the parameters, 
suitable for numerical evaluation.

\subsection{Trajectory}
\begin{figure}[t]
\centering
\subfigure{%
\includegraphics[width=0.65\textwidth]{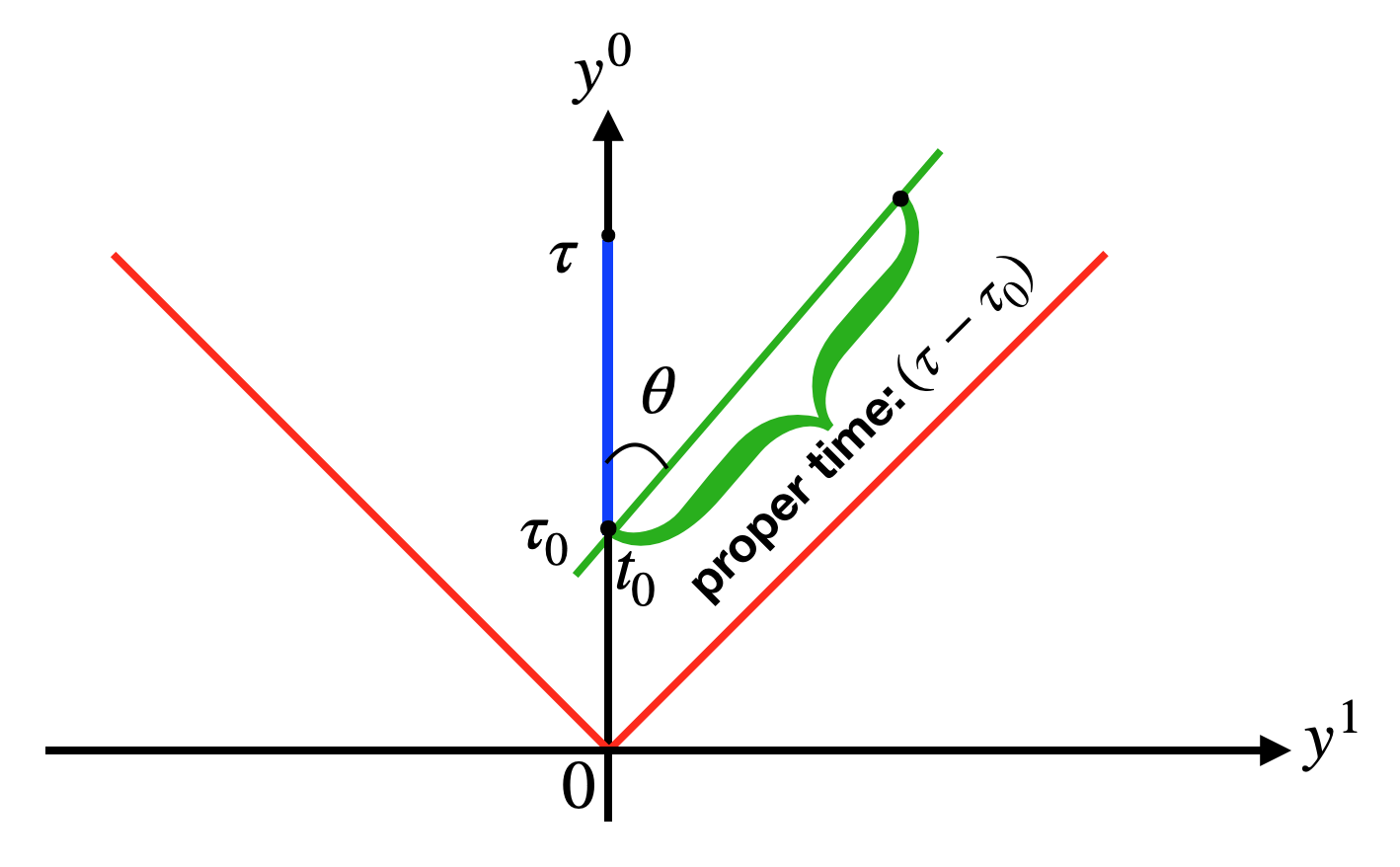}
}
\caption{Spacetime diagram of the worldline of a comoving inertial detector (on the $y^0$-axis, blue)
and a non-comoving inertial detector (green). The switch-on event is at $(y^0,y^1) = (t_0,0)$, or $(t,x) = (t_0,0)$, 
at which event the rapidity of the non-comoving detector with respect to the comoving detector is $\theta\in\BbbR$. 
The proper time parameter $\tau$ on the trajectory is chosen to have the value $\tau_0 := t_0$ 
at the switch-on event.\label{Non:ComovingWordline}}
\end{figure}

We parametrise the detector's worldline as   
\begin{subequations}\label{Non-ComovingWordlines}
\begin{align}
t(\tau) &= \sqrt{\bigl[t_0 + (\tau - t_0)\cosh\theta \bigr]^2 -\bigl[ (\tau -t_0 )\sinh\theta \bigr]^2}, \\
x(\tau) &= \frac{1}{a} \arctanh \! \left( {\frac{(\tau -t_0 )\sinh\theta}{t_0+(\tau - t_0)\cosh\theta}} \right)
\, ,
\end{align}
\end{subequations}
where $t_0 > 0$ is the moment of cosmological time 
$t$ at which the detector is switched on, $\theta\in\BbbR$ is the rapidity of the detector 
with the respect to the comoving worldline at the switch-on moment, 
and the range of the proper time $\tau$ is chosen such that $\tau= \tau_0 := t_0$ at the switch-on moment. 
The comoving worldline is obtained as the special case~$\theta=0$. 
Figure \ref{Non:ComovingWordline} shows a spacetime diagram 
with a comoving trajectory and a non-comoving trajectory with the same value of~$t_0$.

\subsection{Twisted field}

For the twisted field, \eqref{ResFunct_both-vacs_1}
and 
\eqref{Non-ComovingWordlines} 
give 
\begin{subequations} 
\begin{align}
\mathcal{F}^{in}_{t}( \omega,\tau_1, \tau_0) &= m^{-2} \, \Pi^{in}_{t}(\omega/m, m\tau_1, m\tau_0) \, , \\
\mathcal{F}^{out}_{t}( \omega,\tau_1, \tau_0) &= m^{-2} \, \Pi^{out}_{t}(\omega/m, m\tau_1, m\tau_0) 
\, ,
\end{align}
\end{subequations}
where 
\begin{subequations} 
\label{ResponseFunct_TwistCom_Milne}
\begin{align}
\label{ResponsFunct:TwistedField_Com_in}
\Pi^{in}_{t}(\mu, \tilde\tau_1, \tilde \tau_0) 
&=
\frac{\pi}{2aL} \sum_{n=-\infty}^{\infty} 
\frac{\bigl| \tilde A_{n}  (\mu, \tilde\tau_1, \tilde\tau_0) \bigr |^2}{ \sinh \! \left(\frac{2\pi^2}{aL} \! \left| n+\frac{1}{2}\right|\right)}  \, ,\\
\Pi^{out}_{t}(\mu, \tilde\tau_1, \tilde \tau_0) 
&=
\frac{\pi}{4aL} \sum_{n=-\infty}^{\infty} \exp \! \left({\frac{2\pi^2}{aL} \! \left|n + \frac{1}{2} \right|} \right)
\bigl | \tilde B_{n} \left(\mu, \tilde\tau_1, \tilde\tau_0 \right) \bigr |^2 
\label{ResponsFunct:TwistedField_Com_out}
\, ,
\end{align}
\end{subequations} 
with 
\begin{subequations}\label{A_n&B_nFunct:Twist}
\begin{align}
\tilde A_n ( \mu, \tilde\tau_1, \tilde\tau_0) 
&:= 
\int_{\tilde\tau_0}^{\tilde\tau_1} d u\, J_{-i\frac{2\pi}{aL}  \left| n+\frac{1}{2} \right|} \bigl(f_{\theta,\tilde \tau_0}(u) \bigr)
\exp[-i \mu u +i\frac{2\pi}{aL} \! \left( n+\frac{1}{2} \right) \! g_{\theta, \tilde \tau_0} (u) ] \, , \\
\tilde B_n ( \mu, \tilde\tau, \tilde\tau_0) 
&:= 
\int_{\tilde\tau_0}^{\tilde\tau_1} d u\, H^{(2)}_{i\frac{2\pi}{aL} \left| n+\frac{1}{2} \right|} \bigl(f_{\theta, \tilde \tau_0}(u) \bigr) 
\exp[-i \mu u +i\frac{2\pi}{aL} \! \left(n+\frac{1}{2} \right) \! g_{\theta,\tilde \tau_0} (u) ]
\, ,
\end{align}
\end{subequations}
and 
\begin{subequations}\label{AuxiliaryFunct_Comov}
\begin{align}
\label{AuxiliaryFunct_ComovA1}
f_{\theta,\tilde \tau_0}(u) &:= \sqrt{\bigl[ \tilde \tau_0 + \left(u-\tilde \tau_0 \right)\cosh\theta \bigr]^2 - \bigl[\left(u-\tilde \tau_0\right)\sinh\theta \bigr]^2}   \, , \\
g_{\theta,\tilde \tau_0} (u) &:= \arctanh \! \left(  \frac{(u-\tilde \tau_0)\sinh\theta }{\tilde \tau_0 + (u-\tilde \tau_0)\cosh\theta} \right)
\label{AuxiliaryFunct_ComovB1}
\, .
\end{align}
\end{subequations}
Apart from the overall dimensionful factor~$1/m^2$, 
the response hence depends on the parameters only via the dimensionless combinations $\mu = \omega/m$, 
$\tilde \tau_1 = m\tau_1$, $\tilde \tau_0 = m\tau_0$ and~$aL$. 
The value of the cosmological time $t$ at which the detector starts to operate is~$\tau_0$, 
and the detector operates for the total proper time $\tau_1-\tau_0$.

\subsection{Untwisted field}

For the untwisted field, we need to consider separately the ``in'' and ``out'' vacua. 

\subsubsection{``Out'' vacuum}

For the ``out'' vacuum, we may proceed as with the twisted field. We find 
\begin{align}
\mathcal{F}^{out}_{u}( \omega,\tau_1, \tau_0) &= m^{-2} \, \Pi^{out}_{u}(\omega/m, m\tau_1, m\tau_0) 
\, ,
\end{align}
where 
\begin{align}
\Pi^{out}_{u}(\mu, \tilde\tau_1, \tilde \tau_0) 
&=
\frac{\pi}{4aL} \sum_{n = -\infty}^\infty \exp \! \left({\frac{2\pi^2 |n| }{aL}} \right)
\left | \tilde{\tilde B}_{n} \left(\mu, \tilde\tau_1, \tilde\tau_0 \right) \right |^2 
\, ,
\end{align}
with
\begin{align}
\tilde{\tilde B}_n ( \mu, \tilde\tau_1, \tilde\tau_0) 
&:= 
\int_{\tilde\tau_0}^{\tilde\tau_1} d u\, H^{(2)}_{i\frac{2\pi |n|}{aL}} \bigl(f_{\theta, \tilde \tau_0}(u) \bigr) 
\exp[-i \mu u +i\frac{2\pi n}{aL} g_{\theta,\tilde \tau_0} (u) ]
\, .  
\end{align}

\subsubsection{``In'' vacuum}

For the ``in'' vacuum, the oscillator modes may be treated as for the twisted field. 
We find that their contribution is 
\begin{align}
\mathcal{F}^{in}_{osc}( \omega,\tau_1, \tau_0) &= m^{-2} \, \Pi^{in}_{osc}(\omega/m, m\tau_1, m\tau_0) 
\, ,
\end{align}
where 
\begin{align}
\label{ResponsFunct:TwistedField_Com_in-osc}
\Pi^{in}_{osc}(\mu, \tilde\tau_1, \tilde \tau_0) 
=
\frac{\pi}{2aL} \sum_{n \ne 0} 
\frac{\left| \tilde{\tilde A}_{n}  (\mu, \tilde\tau_1, \tilde\tau_0) \right |^2}{ \sinh \! \left(\frac{2\pi^2 |n|}{aL} \right)}  
\, ,
\end{align}
with
\begin{align}
\tilde{\tilde A}_n ( \mu, \tilde\tau_1, \tilde\tau_0) 
&:= 
\int_{\tilde\tau_0}^{\tilde\tau_1} d u\, J_{-i\frac{2\pi |n|}{aL}} \bigl(f_{\theta,\tilde \tau_0}(u) \bigr)
\exp[-i \mu u +i\frac{2\pi n}{aL} g_{\theta, \tilde \tau_0} (u) ] 
\, .  
\end{align}

For the zero momentum mode, we have, using 
\eqref{Coeffic:ZeroMomentMode}, 
\eqref{RespFunct_TwoReal_UntwistA}
and~\eqref{Definitions:M&N&B}, 
\begin{align}
\mathcal{F}^{in}_{0}( \omega,\tau_1, \tau_0) &= m^{-2} \, \Pi^{in}_{0}(\omega/m, m\tau_1, m\tau_0)
\, ,
\end{align}
where 
\begin{align}
\Pi_{0}^{in}(\mu, \tilde\tau_1, \tilde\tau_0) 
&= 
\frac{\pi}{4aL} \biggl\{   \Bigl({\tilde\alpha}^2 + {\tilde\beta}^{-2}\Bigr) 
\left| \tilde{\tilde M}(\mu, \tilde\tau_1, \tilde\tau_0) \right|^2  
+ {\tilde\beta}^2 \left| \tilde{\tilde N}(\mu, \tilde\tau_1, \tilde\tau_0) \right|^2
\notag 
\\
&\hspace{9ex} 
+ 2 \Realpart \! \left[ \Bigl( \tilde\alpha \tilde\beta + i \Bigr) 
\tilde{\tilde M}(\mu, \tilde\tau_1, \tilde\tau_0)
\tilde{\tilde N}^*(\mu, \tilde\tau_1, \tilde\tau_0)
\right]
\biggr\}
\, , 
\label{ResponsFunct:TwistedField_Com_in-zero}
\end{align}
with
\begin{subequations}
\label{AuxiliaryFunNocomXZ_2}
\begin{align}
\tilde{\tilde M}(\mu, \tilde\tau_1, \tilde\tau_0) 
&:= 
\int_{\tilde\tau_0}^{\tilde\tau_1} d u \, J_0\bigl(f_{\theta,\tilde\tau_0}(u) \bigr)   
\exp(-i\mu u) \, , 
\\
\tilde{\tilde N}(\mu, \tilde\tau_1, \tilde\tau_0) 
&:= 
\int_{\tilde\tau_0}^{\tilde\tau_1} d u \, Y_0\bigl(f_{\theta,\tilde\tau_0}(u) \bigr)   
\exp(-i\mu u) 
\, ,  
\end{align}
\end{subequations}
and the dimensionless parameters $\tilde\alpha \in \BbbR$ and $\tilde\beta>0$ 
are related to $\alpha$ and $\beta$ in \eqref{Coeffic:ZeroMomentMode} by 
\begin{subequations}
\begin{align}
\alpha &= \tfrac12 {(\pi/a)}^{1/2} \, \tilde\alpha
\, , 
\\
\beta &= \tfrac12 {(\pi/a)}^{1/2} \, \tilde\beta
\, . 
\end{align}
\end{subequations}

Collecting 
\eqref{ResponsFunct:TwistedField_Com_in-osc}
and~\eqref{ResponsFunct:TwistedField_Com_in-zero}, 
we have 
\begin{align}
\mathcal{F}^{in}_{u}( \omega,\tau_1, \tau_0)  &= m^{-2} \, \Pi^{in}_{u}(\omega/m, m\tau_1, m\tau_0) 
\, ,
\end{align}
where 
\begin{align}
\Pi^{in}_{u}(\mu, \tilde\tau_1, \tilde\tau_0) 
&= 
\Pi^{in}_{osc}(\mu, \tilde\tau_1, \tilde\tau_0) 
+ 
\Pi^{in}_{0}(\mu, \tilde\tau_1, \tilde\tau_0) 
\, .
\end{align}

\section{Interlude: inertial detector in Minkowski vacuum\label{sec:Minkowski}}

In Milne spacetime with noncompactified spatial sections, 
the ``out'' vacuum defined by the adiabatic criterion 
\eqref{pos:frequency_ScalarMode_nonstatic}
coincides with the Minkowski vacuum~\cite{Birrell:1982ix,Fulling:1974pu}. 
In this section we give results for the response of a finite-time inertial detector in Minkowski vacuum. 
We shall see in Section \ref{sec:numerical} below that the response 
in spatially compactified Milne will duly reduce to the Minkowski vacuum response in appropriate limits. 

Recall that in two-dimensional Minkowski spacetime, 
the pull-back of the Minkowski vacuum Wightman function 
of a massive scalar field to an inertial worldline is \cite{Takagi:1986kn}
\begin{align}
G(\tau, \tau')
= \frac{1}{2\pi}
K_0
\bigl(
m [\epsilon + i (\tau-\tau')] \bigr)
\, , 
\label{eq:MinkWightman-pullback}
\end{align}
where $K_0$ is the modified Bessel function of the second kind
and the limit $\epsilon \to 0_+$ is understood. 
While in higher spacetime dimensions the $\epsilon \to 0_+$ 
limit is distributional, 
in two dimensions the coincidence singularity is so weak that 
the limit can be represented by an integrable function, as \cite{dlmf}
\begin{align}
\label{eq:scalar-wightman_Massive}
G(\tau, \tau') = 
\begin{cases}
\displaystyle{
\vphantom{\frac{A^A}{A}}
-\frac{i}{4} H_0^{(2)}\bigl( m  (\tau -\tau') \bigr)    } \, ,
&
\text{for $\tau > \tau' $}, 
\\[3ex]
\displaystyle{
\vphantom{\frac{A}{A_A}}
\frac{i}{4} H_0^{(1)}\bigl(  m  (\tau' -\tau) \bigr)    \, ,}
&
\text{for $\tau' > \tau $} \, ,
\end{cases}    
\end{align}
where $H_0^{(1)}$ and $H_0^{(2)}$ are the Hankel functions. 
For a detector operating for the total proper time~$\Delta\tau$, 
formula \eqref{ResponsFunct_2D} hence gives the response function 
\begin{align}
\mathcal{F}_{\text{Mink}}(\omega, \Delta \tau) = m^{-2} \, \Pi_{\text{Mink}}(\omega/m, m\Delta \tau)
\, , 
\label{eq:Minkvac-Fcal}
\end{align}
where 
\begin{align}
\Pi_{\text{Mink}}(\mu, \Delta \tilde\tau)
= 
-\frac{1}{2}\int_{0}^{\Delta \tilde\tau} du \, (\Delta\tilde\tau - u) 
\bigl[ 
J_0 (u) \sin(\mu u)  
+ Y_0 (u) \cos( \mu u)
\bigr ]        
\, .  
\label{eq:Minkvac-Pi}
\end{align}

\begin{figure}[t]
\centering
\subfigure[$\Pi_{\text{Mink}}(\mu, \Delta \tilde\tau)$]{%
\includegraphics[width=0.51\textwidth]{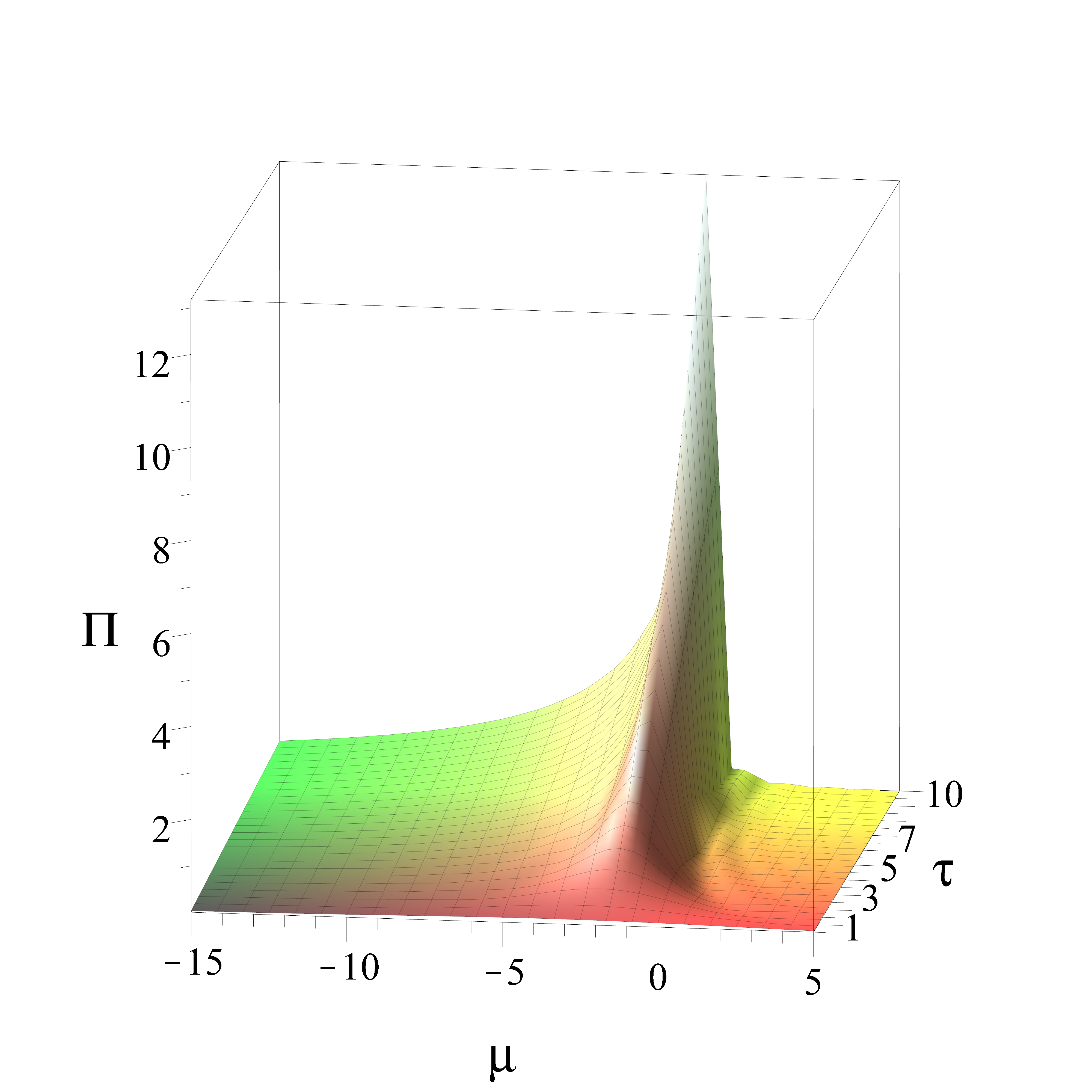} 
\label{Pers:MassiveResFun3D}}
\subfigure[$\Pi_{\text{Mink}}(\mu, 10)$]{%
\includegraphics[width=0.46\textwidth]{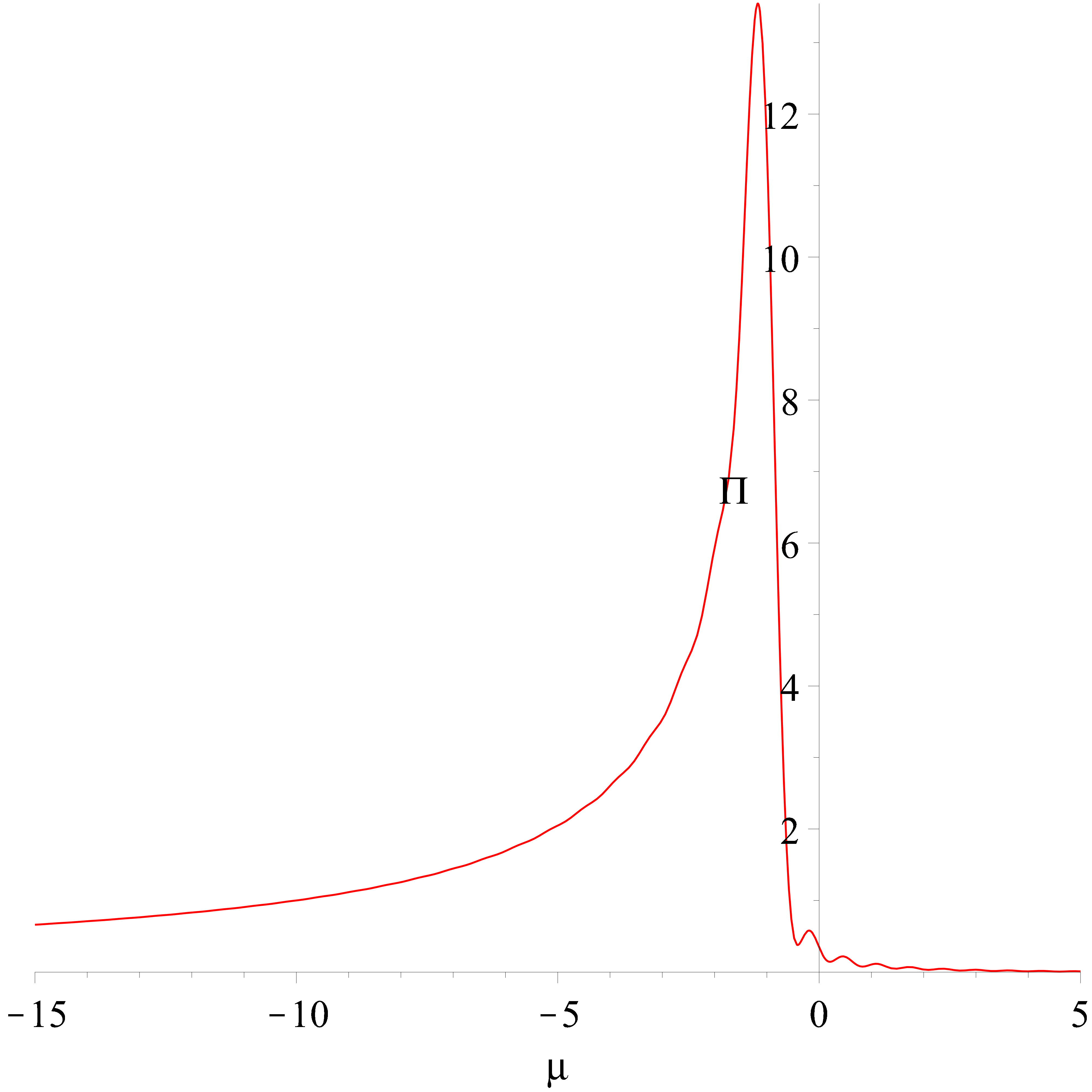} 
\label{CrossSection:MassiveResFun2D}}
\caption{Inertial detector's response in Minkowski vacuum, evaluated from~\eqref{eq:Minkvac-Pi}. 
Part (a) is a perspective plot, with the axis label $\tau$ in the plot  
denoting $\Delta \tilde\tau$ in~\eqref{eq:Minkvac-Pi}. 
Part (b) is the cross-section at $\Delta \tilde\tau = 10$.\label{MassiveResFunc_AB}}
\end{figure}

Numerical plots of $\Pi_{\text{Mink}}(\mu, \Delta \tilde\tau)$ 
\eqref{eq:Minkvac-Pi} are shown in 
Figure~\ref{MassiveResFunc_AB}. At large $\Delta\tilde\tau$, 
the prominent feature is a de-excitation peak near $\mu \approx -1$, 
corresponding to $\omega \approx -m$. 

\section{Numerical results in spatially compactified Milne\label{sec:numerical}}

In this section we discuss the core numerical results of the paper: 
the detector's response on selected  
inertial trajectories in spatially compactified Milne, 
in the ``in'' and ``out'' vacua. 
The key aim is to see how the Milne response differs from 
the Minkowski vacuum response of Section~\ref{sec:Minkowski}. 
For graphical convenience, we give in this section a verbal discussion of the results 
while delegating the numerical plots to Figures \ref{Tw_CM_RF}--\ref{CM_T_in_M} in the appendix.

\subsection{Comoving detector}

Consider first the comoving detector. 

For the twisted field, the results in Figure 
\ref{Tw_CM_RF} 
show that 
the prominent feature of the spectrum at late times or at large $aL$ 
is still the de-excitation peak near $\omega = -m$, 
for both the ``in'' vacuum and the ``out'' vacuum, 
in close agreement with the Minkowski vacuum results of Section~\ref{sec:Minkowski}, 
as was to be expected. 
When $aL$ is small, or when the detector operates at early times, 
the de-excitation spectrum develops more structure, 
but we see no evidence of significant excitations in the parameter range probed. 

For the untwisted field, the results shown in Figure \ref{Untw_CM_RF_inVac} are very 
similar, for both the ``in'' vacuum and the ``out'' vacuum, provided the massive zero mode ``in'' vacuum 
is chosen to agree with the massive zero mode ``out'' vacuum, 
although there is now more quantitative structure in the de-excitation spectrum 
when $aL$ is small or when the detector operates at early times. 

For the ``in'' vacuum of the untwisted field, 
effects of varying the parameters $\tilde\beta$ and $\tilde\alpha$ of the state 
are shown in respectively Figures \ref{CM_T_in_I} and~\ref{CM_T_in_II}. 
The prominent de-excitation peak survives, but it is now 
accompanied by an excitation peak, near $\omega \approx m$. 
This is consistent with the intuitive picture that 
changing the massive zero mode state puts in the field a `particle'  
that can be absorbed by the detector.

\subsection{Non-comoving detector}

Consider then a non-comoving detector. 

Figures \ref{NC_UT_out_IA}--\ref{CM_T_in_M}
show results for the twisted and untwisted fields in the ``in'' and ``out'' vacua, 
with large and small values of~$aL$, with 
the detector operating at early and late times, 
and with a selection of detector rapidities with respect to a comoving observer at the turn-on moment.  
The ``in'' vacuum of the spatially constant mode of the untwisted field is chosen 
to agree with that of the ``out'' vacuum, except in Figure~\ref{CM_T_in_M}, 
where the $\tilde\alpha$ and $\tilde\beta$ parameters of this mode are varied. 

The results show that in most cases within the parameter range probed, 
the effect of the rapidity is significant only when the detector operates at early times 
and $aL$ is small: 
a representative example is the twisted field in the ``out'' vacuum, shown in 
Figure~\ref{NC_UT_out_IA}, 
where a significant effect appears only in Figure~\ref{Pers:MassiveResFunC}, 
as additional structure in the de-excitation probabilities. 

The exception to this pattern is the ``in'' vacuum of the untwisted field, 
for which results are shown in 
Figures \ref{UTCMD_inVACC} and~\ref{CM_T_in_M}. 
When the parameters 
$\tilde\alpha$ and $\tilde\beta$ of the spatially constant mode
are chosen so that this mode agrees with the spatially constant mode of the ``out'' vacuum, 
Figure \ref{UTCMD_inVACC} shows a net shift of the de-excitation peaks to more negative values of $\omega$ 
as the rapidity increases, and this shift persists even at late times and for large~$aL$, 
within the parameter range probed. 
When the parameters $\tilde\alpha$ and $\tilde\beta$ are varied, 
there appears also an excitation peak, 
and this excitation peak becomes shifted to more positive values of 
$\omega$ as the rapidity increases, as shown in Figure~\ref{CM_T_in_M}. 

\begin{figure}[t]
\centering
\includegraphics[width=0.45\textwidth]{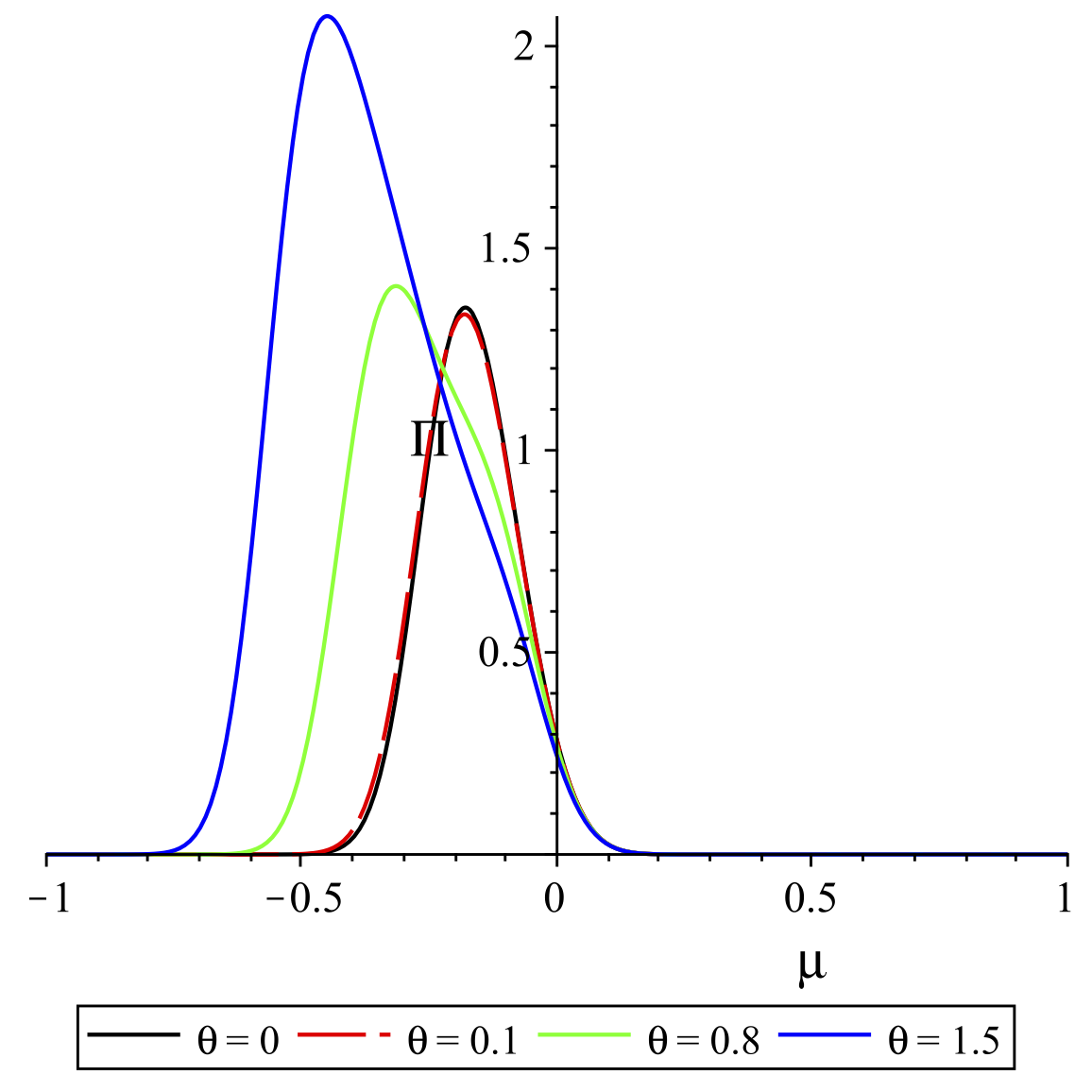}
\caption{Response of an inertial detector on a static $(1+1)$-dimensional cylinder, as a function of the detector's excitation energy, for selected values of the detector's rapidity $\theta$ with respect to static observers, evaluated from (IV.12a) in~\cite{Martin-Martinez:2014qda}. The detector is coupled linearly to the proper time derivative of an untwisted massless scalar field, the field is prepared in the Minkowski-like vacuum, except for the zero mode, whose state is chosen so that the contribution to the response, from (IV.12b) in~\cite{Martin-Martinez:2014qda}, is negligible. The spatial circumference, the duration of the interaction and the normalisation of the horizontal scale 
are chosen to be comparable to those in Figure~\ref{CrossSection:MassiveResFun4H}, adjusted for the absence of a mass parameter. 
Note the shift of the de-excitation peak towards negative $\mu$ as the rapidity increases, in qualitative agreement with the shift in Figure~\ref{UTCMD_inVACC}, but note also the increase in the height of the peak, in contrast with the decrease of the height in Figure~\ref{UTCMD_inVACC}.\label{fig:static-cylinder}}
\end{figure}

While we do not have an analytic explanation of why the 
spectral shift due to the detector's rapidity is most persistent for the 
``in'' vacuum of the untwisted field, 
we have verified that a similar 
shift appears for an untwisted field even in a spatially 
periodic static spacetime, 
in a Minkowski-like vacuum~\cite{Martin-Martinez:2014qda}. 
Taking the field on the static cylinder to be massless, 
taking the detector to be coupled to the proper time derivative of the field 
(rather than to the value of the field), and taking the switching function to be Gaussian (rather than sharp), 
we may evaluate the detector's response from formulas 
(IV.12) in~\cite{Martin-Martinez:2014qda}. 
Choosing the state of the zero mode (which does not have a Fock vacuum) 
such that its contribution to the response is negligible, we may use just 
(IV.12a) in~\cite{Martin-Martinez:2014qda}. 
The results are shown in Figure~\ref{fig:static-cylinder}, 
with parameter choices that are comparable to those in Figure~\ref{CrossSection:MassiveResFun4H}, 
adjusted for the absence of a mass parameter. 
The plot shows a clear shift of the de-excitation peak towards negative $\mu$ as the rapidity increases, 
in qualitative agreement with the shift seen in Figure~\ref{UTCMD_inVACC}. 
Note, however, that the heights of the peaks in Figure \ref{fig:static-cylinder} increase as the peaks shift, 
whereas the heights of the peaks in Figure \ref{UTCMD_inVACC} decrease as the peaks shift.

\section{Conclusions\label{sec:conclusions}}

We have investigated quantum field theory in spatially compact cosmological spacetimes where the early time or late time asymptotic behaviour makes a massive field asymptotically massless, without a distinguished Fock vacuum that could be singled out by adiabatic considerations. Focusing on a massive scalar field in $1+1$ spacetime dimensions, we showed that the freedom in the choice of the vacuum state is a family with two real-valued parameters, in agreement with the observations of Ford and Pathinayake \cite{Ford:1989mf} for a subset of our asymptotic conditions. Specialising to the expanding Milne spacetime with compactified spatial sections, where the ambiguity arises in the early time vacuum, we examined how the ambiguity affects the response of an inertial Unruh-DeWitt detector coupled to the quantum field. 
In parallel, for contrast, we analysed the Unruh-DeWitt's detector's 
response to a $\BbbZ_2$-twisted scalar field, 
for which adiabatic considerations do single out unique ``in'' and ``out'' vacua. 

We found that the choice of the massive ``in'' zero mode state has a significant effect 
on the response of an inertial Unruh-DeWitt detector, 
especially in the excitation part of the spectrum. 
We also found that the inertial detector's peculiar velocity with 
respect to the comoving cosmological observers affects the detector's 
response mainly at early times in spacetimes with a small spatial circumference, 
as could perhaps have been expected, but with one notable exception: 
for an untwisted field in the ``in'' vacuum, the peculiar velocity effect 
survives even for large circumferences and late times, 
within the parameter range of our numerical simulations, 
and produces a shift of the de-excitation and excitation resonances 
to larger detector energy gaps as the detector's peculiar velocity increases. 
We verified that a qualitatively similar resonance shift occurs also for 
a static spacetime with compact spatial sections, 
but we do not have a quantitative explanation of why in Milne this effect 
is specific to the ``in'' vacuum of the untwisted field. 

While Milne spacetime is flat, we expect similar phenomena 
to arise also in curved spacetimes with compact spatial sections, 
including locally de~Sitter and locally anti-de~Sitter spacetimes. 
We leave investigation of these spacetimes subject to future work. 

Finally, we recall that our quantum scalar field was free, with a quadratic action. 
In an interacting theory, new phenomena could be expected to arise: 
there exist situations in which 
a vacuum associated with linearly-growing field modes is highly sensitive to loop corrections, 
so that loop corrections to correlation functions 
have secular growth and become quickly comparable to the 
tree-level values~\cite{Krotov:2010ma,Akhmedov:2014hfa,Akhmedov:2014doa,Akhmedov:2019rvx}. 
It would be interesting to examine whether similar secular growth occurs in our setting 
when the scalar field Lagrangian density \eqref{NeutralScalarFie:FLRW}
is generalised to include an interaction term, such as~$\phi^4$, and if so, 
how this growth depends on the vacuum state, and how the growth affects the response of the detector.

\section*{Acknowledgments}

We thank Larry Ford and Bei Lok Hu for helpful comments and discussions about 
the dominant behaviour of oscillator modes versus zero mode at early times. 
We also thank Chris Fewster and Atsushi Higuchi for useful discussions. 
JL~thanks Adam Magee for discussions on twisted and untwisted massive fields 
in Milne spacetime~\cite{Magee-GPD}. 
We thank an anonymous referee for helpful comments, 
including bringing the work in 
\cite{Krotov:2010ma,Akhmedov:2014hfa,Akhmedov:2014doa,Akhmedov:2019rvx} to our attention.
JL~acknowledges partial support by 
United Kingdom Research and Innovation (UKRI) 
Science and Technology Facilities Council (STFC) grant ST/S002227/1 
``Quantum Sensors for Fundamental Physics'' 
and Theory Consolidated Grant ST/P000703/1.

\appendix 

\section*{Appendix: Figures for Section~\ref{sec:numerical}}

\begin{figure}[p]
\centering
\subfigure[$\Pi_t^{out}(\mu, 10,20)$ for selected $aL$.]{%
\includegraphics[width=0.48\textwidth]{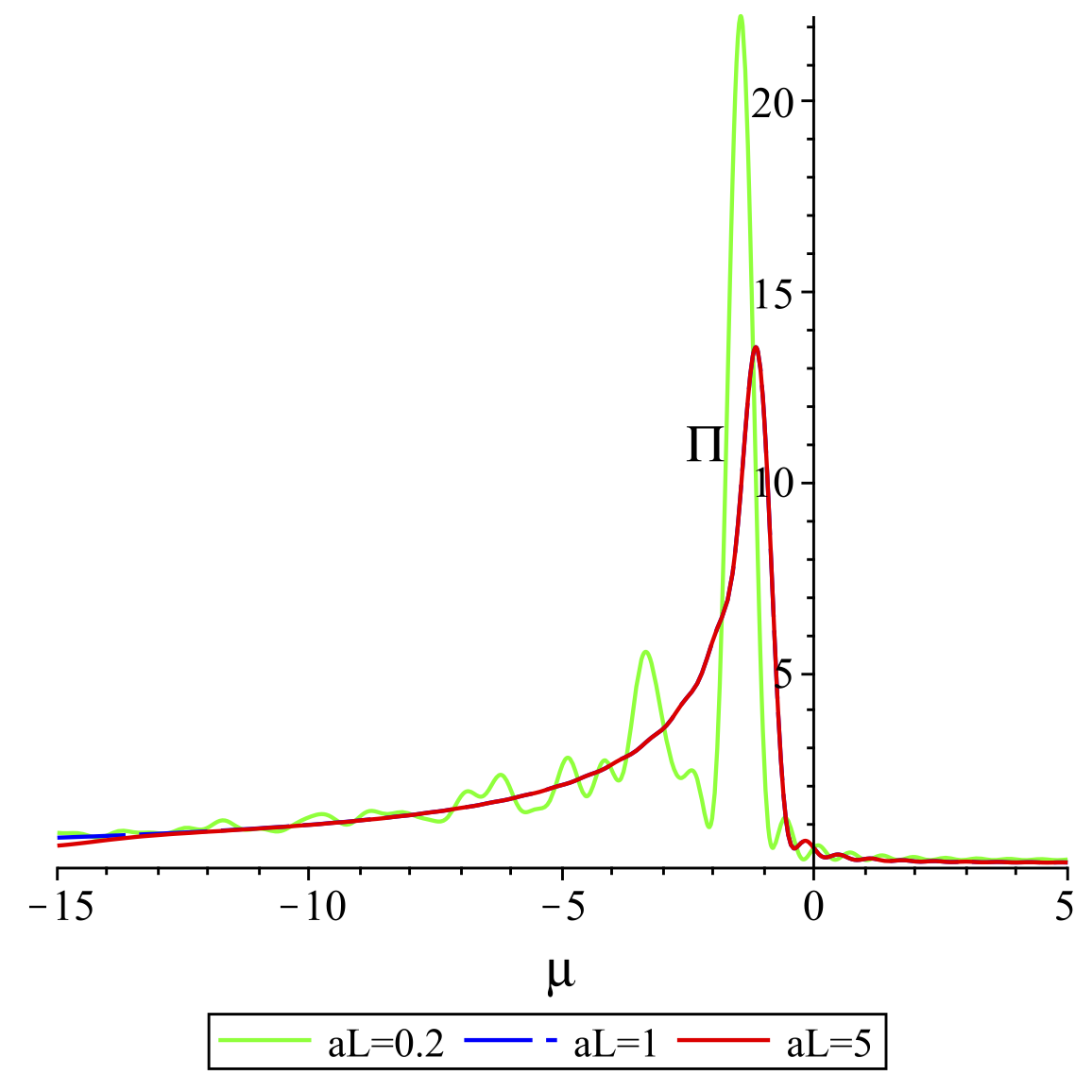} 
\label{Tw_CM_RF_outVac_1a}}
\subfigure[$\Pi_t^{out}(\mu, m\tau_0, m\tau_1)$ for $aL=0.2$ and selected $m\tau_0, m\tau_1$.]{%
\includegraphics[width=0.48\textwidth]{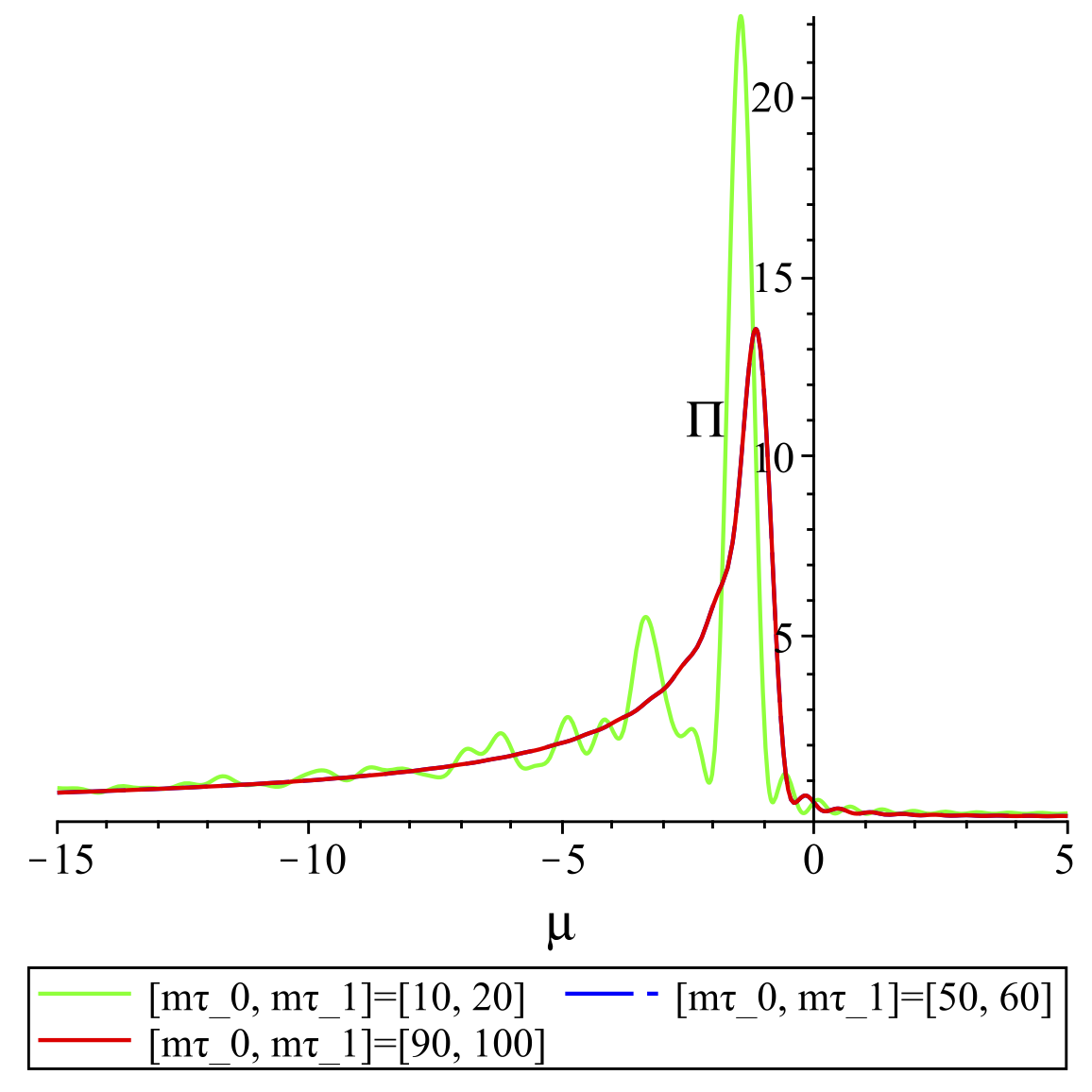} 
\label{Tw_CM_RF_outVac_1b}}
\subfigure[$\Pi_t^{in}(\mu, 10,20)$ for selected $aL$.]{%
\includegraphics[width=0.48\textwidth]{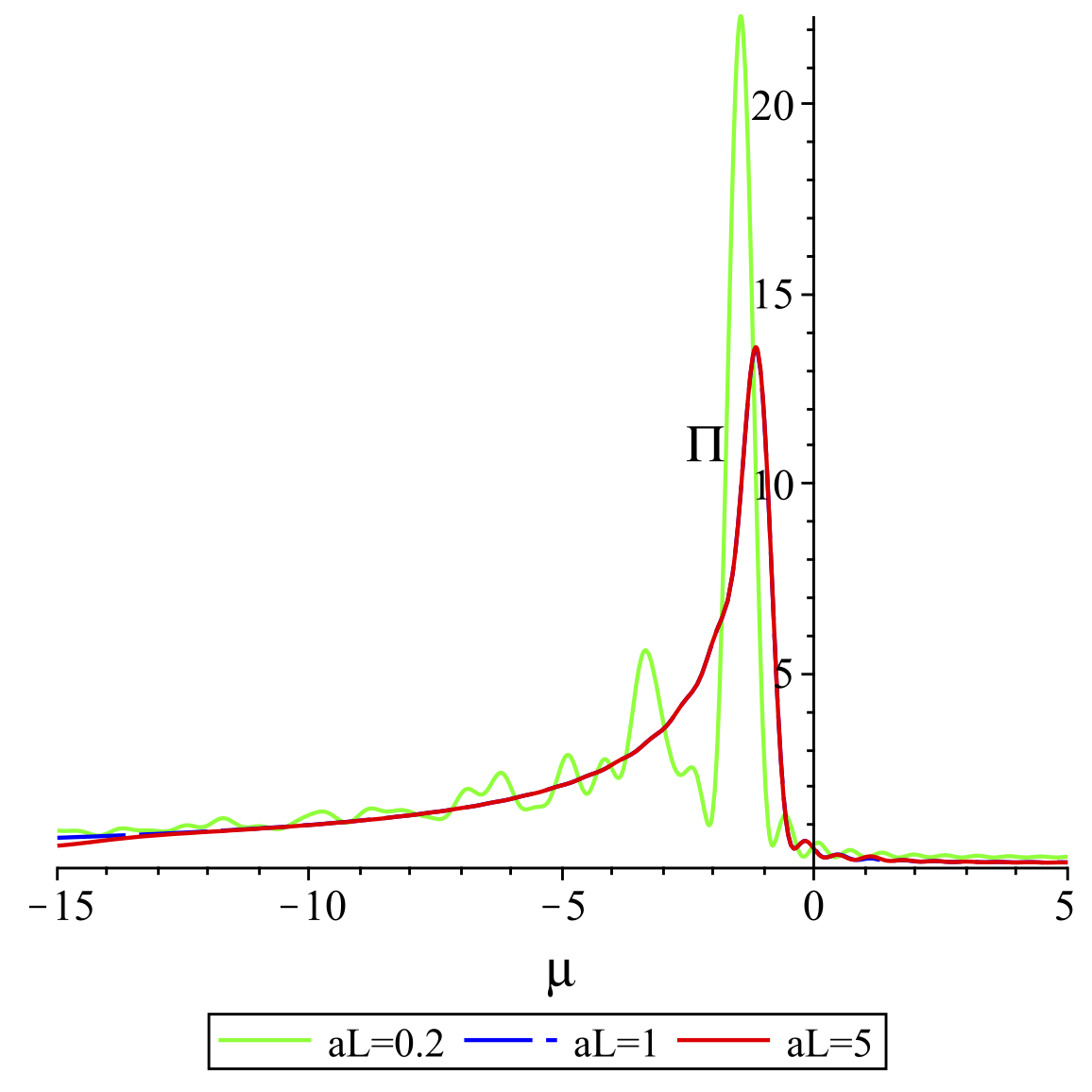} 
\label{Tw_CM_RF_intVac_1a}}
\subfigure[$\Pi_t^{in}(\mu, m\tau_0, m\tau_1)$ for $aL=0.2$ and selected $m\tau_0, m\tau_1$.]{%
\includegraphics[width=0.48\textwidth]{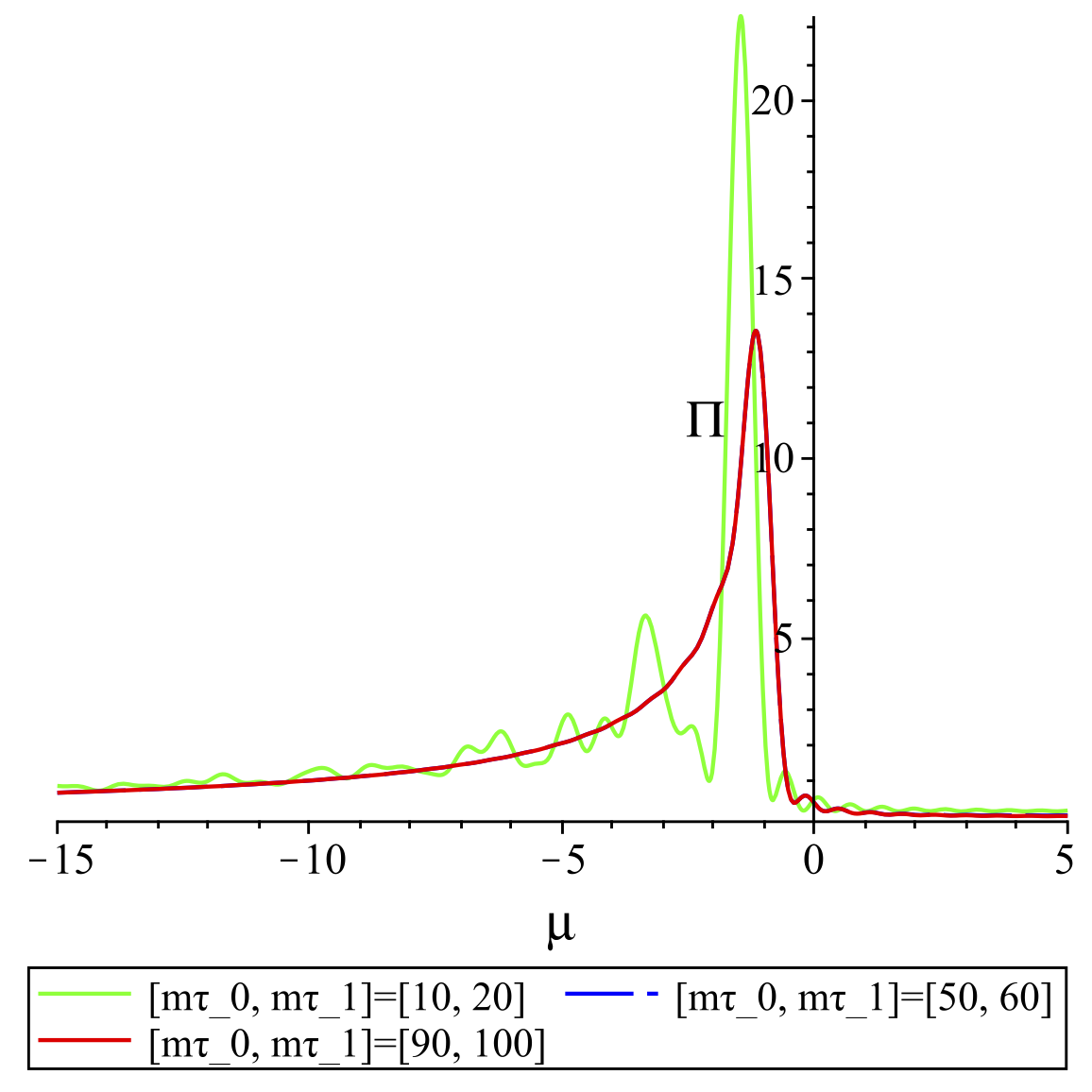} 
\label{Tw_CM_RF_intVac_1b}}
\caption{Comoving detector's response for the twisted field in Milne, 
as a function of $\mu = \omega/m$, 
for the ``in'' and ``out'' vacua, and parameter values as indicated. 
The red curve masks the blue curve fully or almost fully.\label{Tw_CM_RF}}
\end{figure}

\begin{figure}[p]
\centering
\subfigure[$\Pi_u^{out}(\mu, 10,20)$ for selected $aL$.]{%
\includegraphics[width=0.48\textwidth]{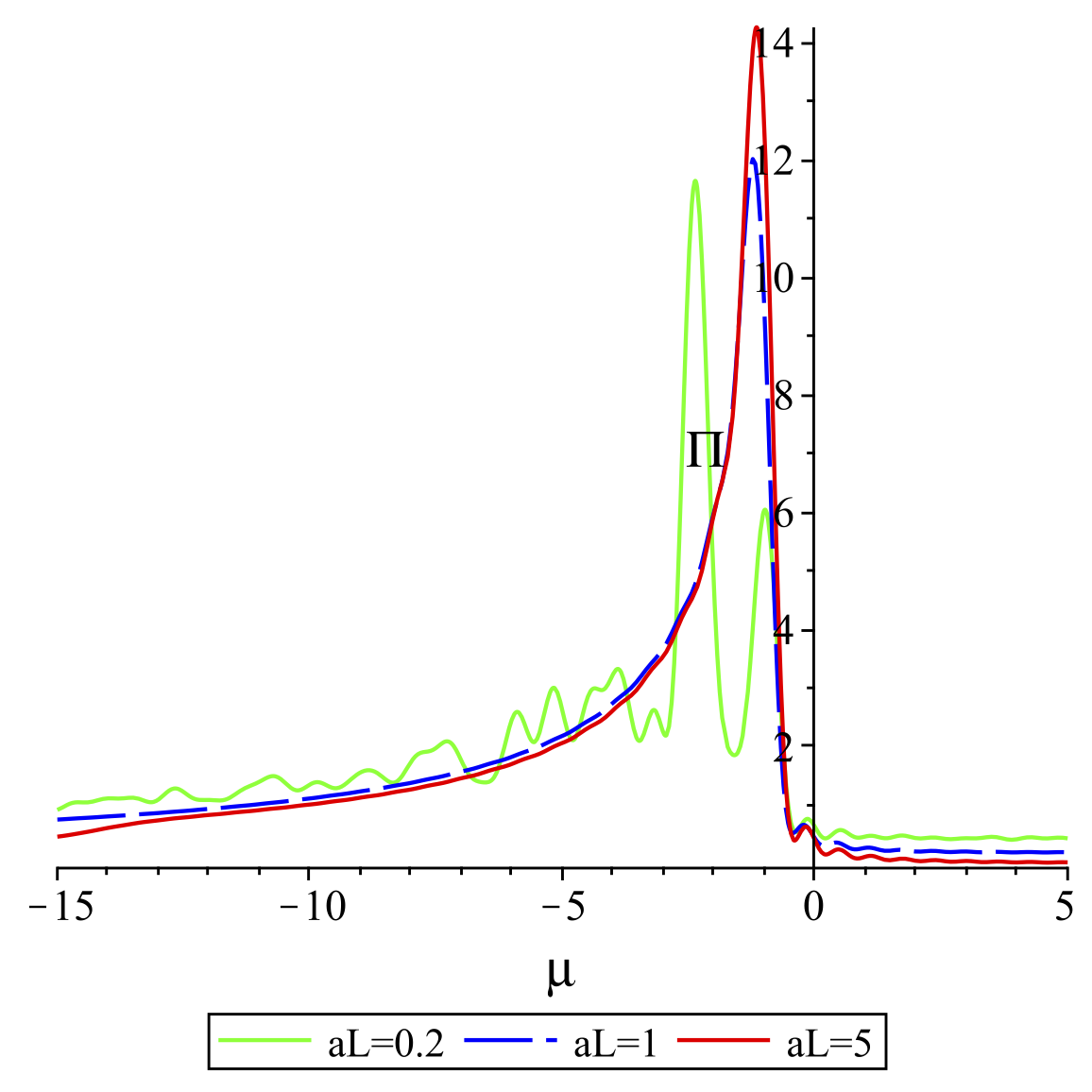} 
\label{Untw_CM_RF_outVac_1a}}
\subfigure[$\Pi_u^{out}(\mu, m\tau_0, m\tau_1)$ for $aL=0.2$ and selected $m\tau_0, m\tau_1$.]{%
\includegraphics[width=0.48\textwidth]{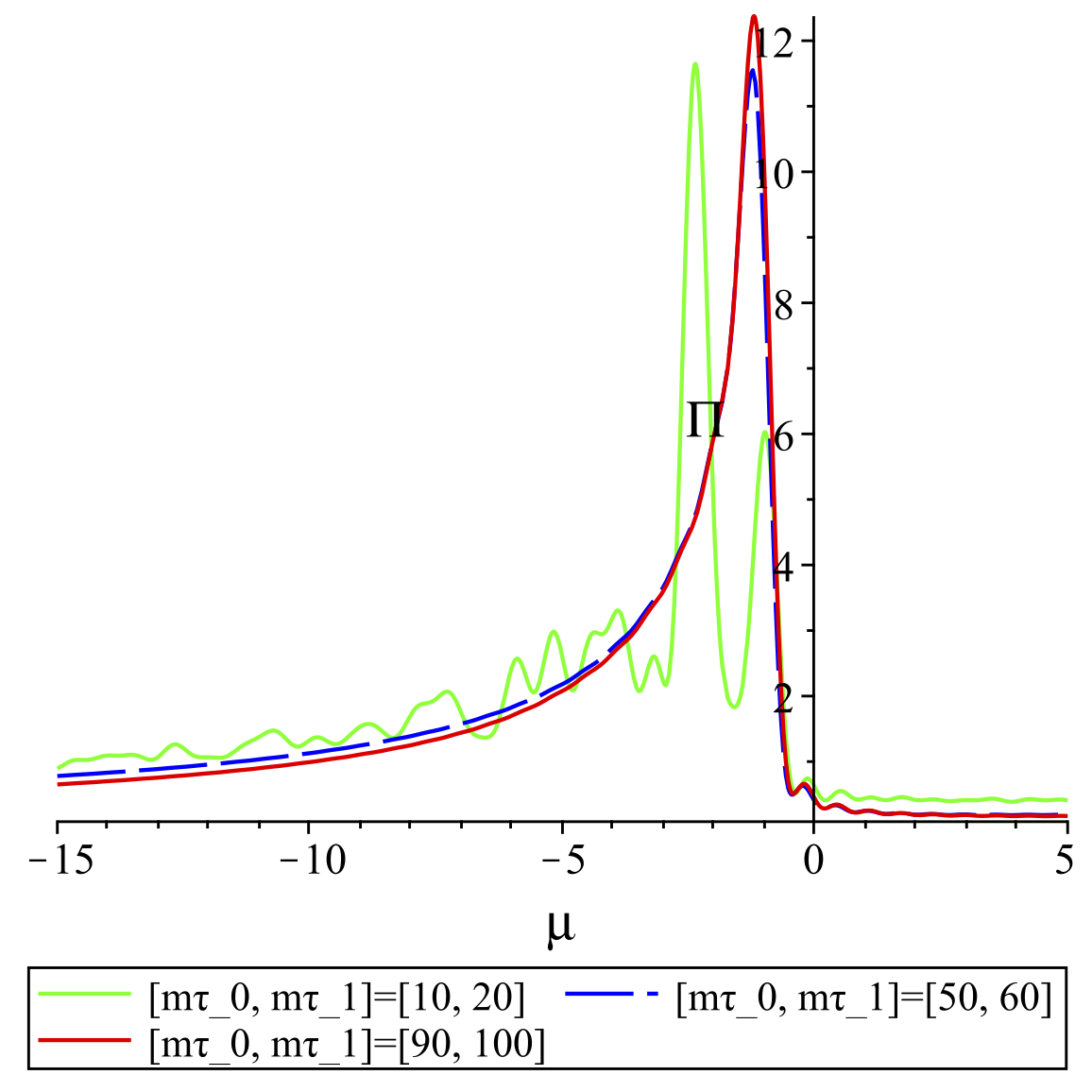} 
\label{Untw_CM_RF_outVac_1b}}
\subfigure[$\Pi_u^{in}(\mu, 10,20)$ for selected $aL$, with $\tilde\alpha =0$ and $\tilde\beta=1$.]{%
\includegraphics[width=0.48\textwidth]{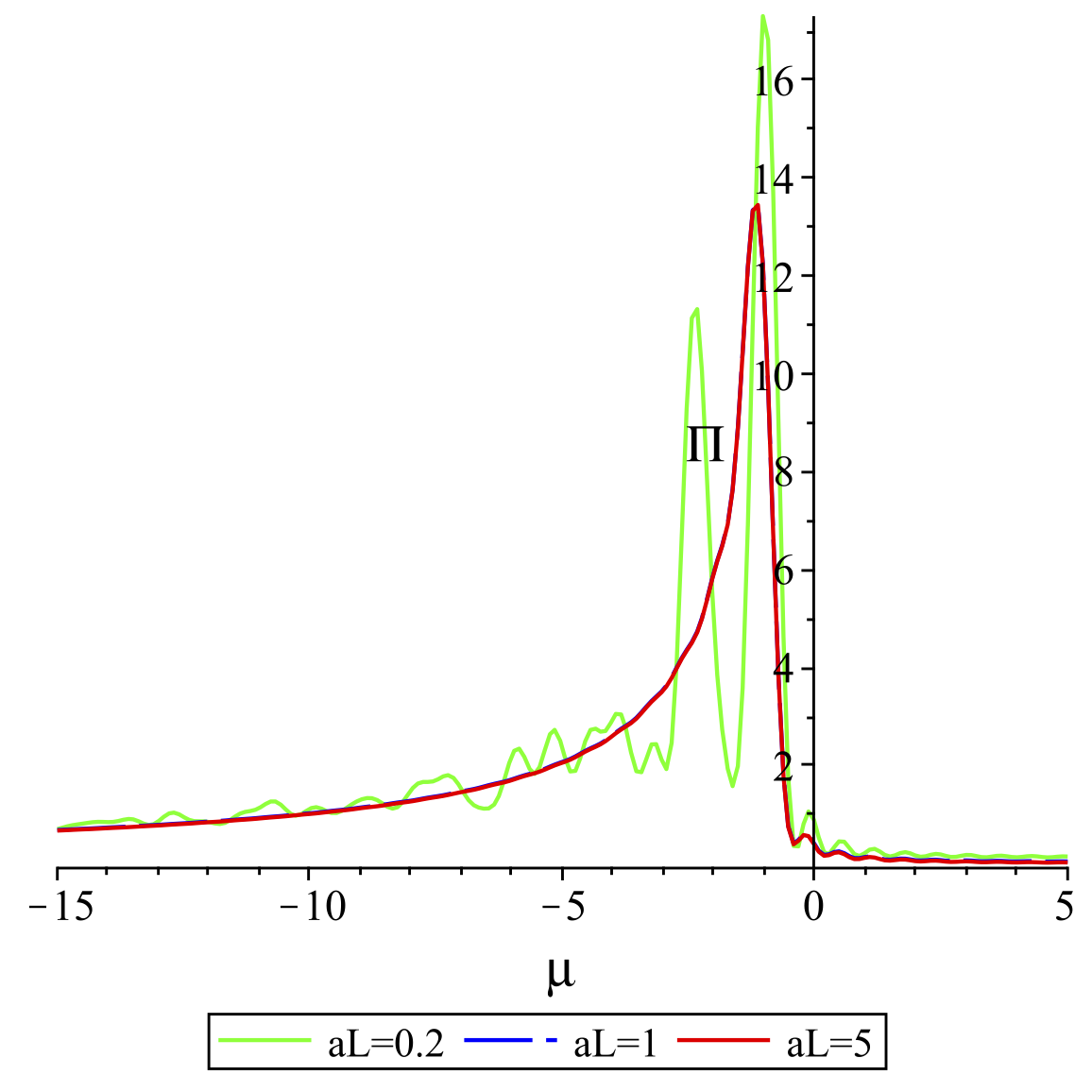} 
\label{Untw_CM_RF_inVac_1a}}
\subfigure[$\Pi_u^{in}(\mu, m\tau_0, m\tau_1)$ for $aL=0.2$ and selected $m\tau_0, m\tau_1$, 
with $\tilde\alpha =0$ and $\tilde\beta=1$.]{%
\includegraphics[width=0.48\textwidth]{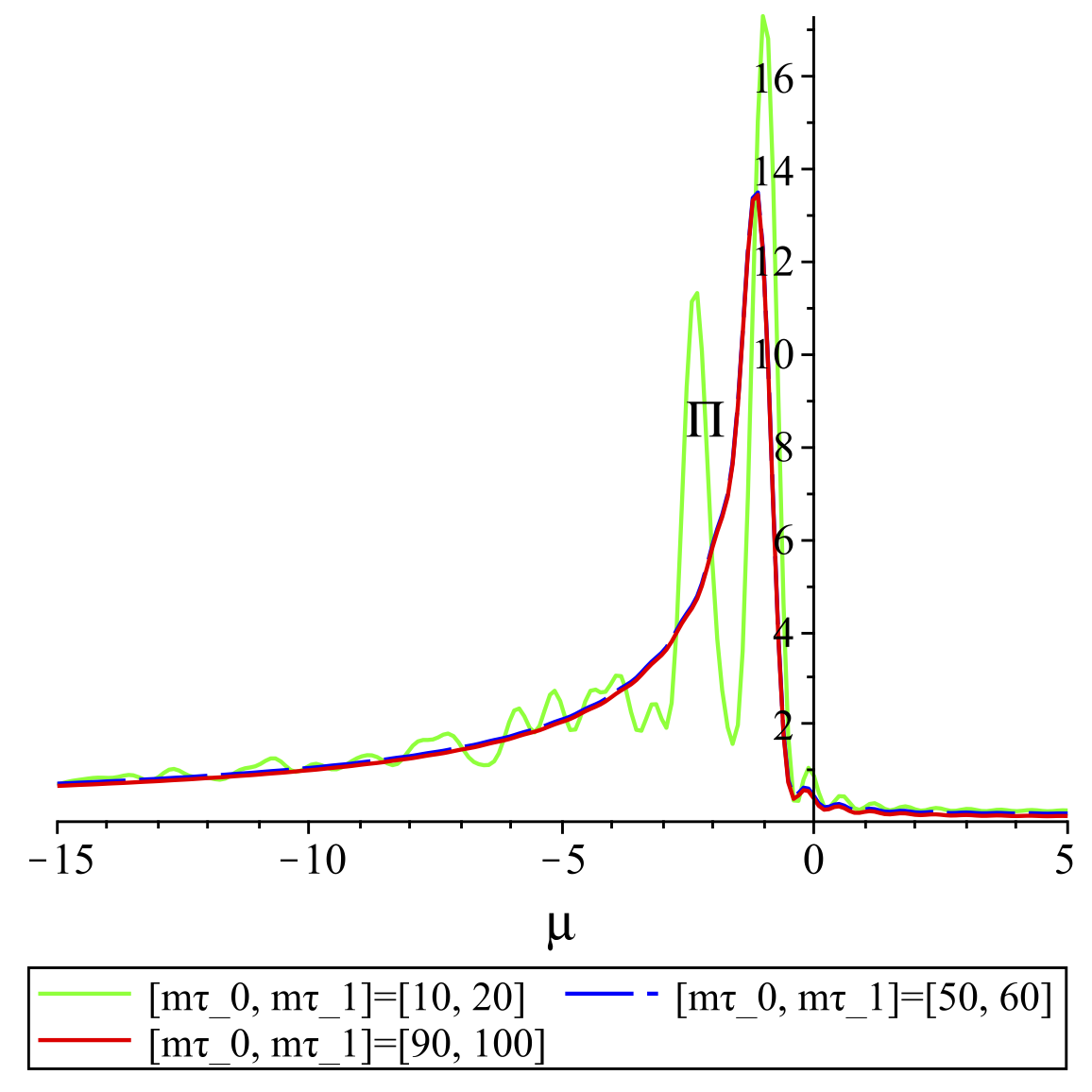} 
\label{Untw_CM_RF_inVac_1b}}
\caption{Comoving detector's response for the untwisted field in Milne as a function of $\mu = \omega/m$, 
for the ``in'' and ``out'' vacua, and parameter values as indicated. 
The ``in'' vacuum spatially constant mode parameters are 
$\tilde\alpha=0$ and $\tilde\beta=1$, 
so that this mode coincides with that of the ``out'' vacuum.\label{Untw_CM_RF_inVac}}
\end{figure}

\begin{figure}[p]
\centering
\subfigure[$\tilde\beta = 2/\sqrt{\pi}$]{%
\includegraphics[width=0.48\textwidth]{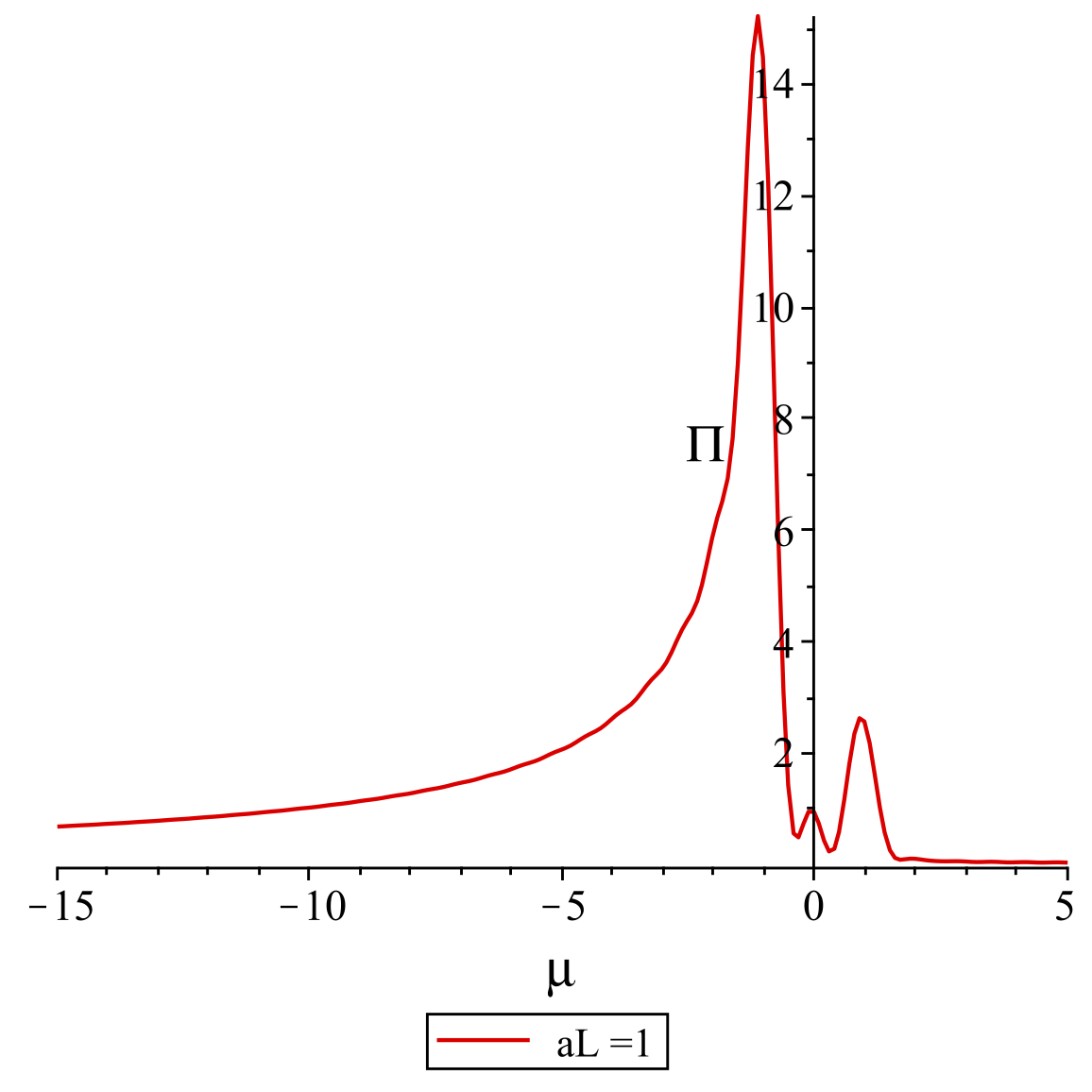} 
\label{Pers:MassiveResFunA}}
\subfigure[$\tilde\beta = 10/\sqrt{\pi}$]{%
\includegraphics[width=0.48\textwidth]{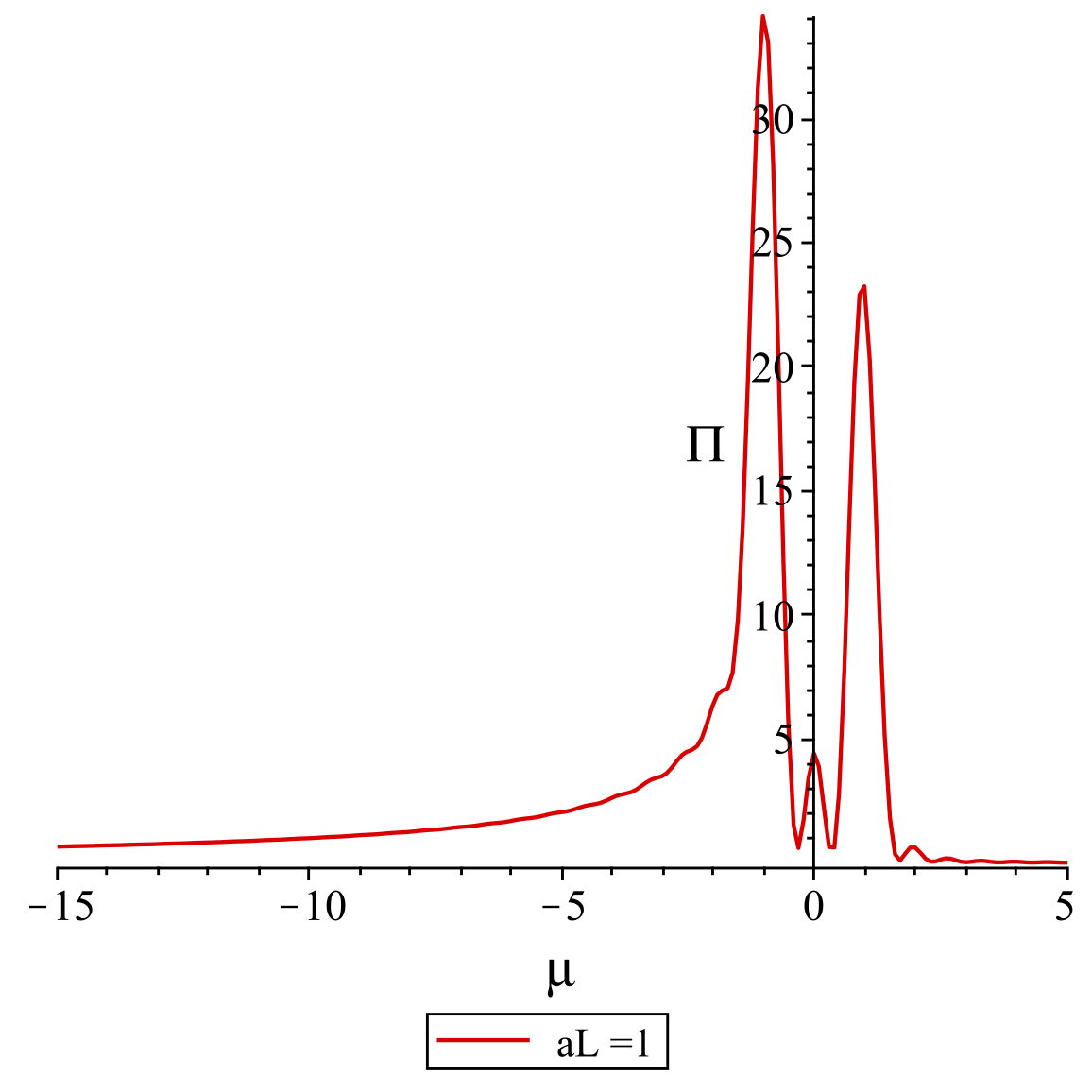} 
\label{CrossSection:MassiveResFun2A}}
\subfigure[$\tilde\beta = 20/\sqrt{\pi}$]{%
\includegraphics[width=0.48\textwidth]{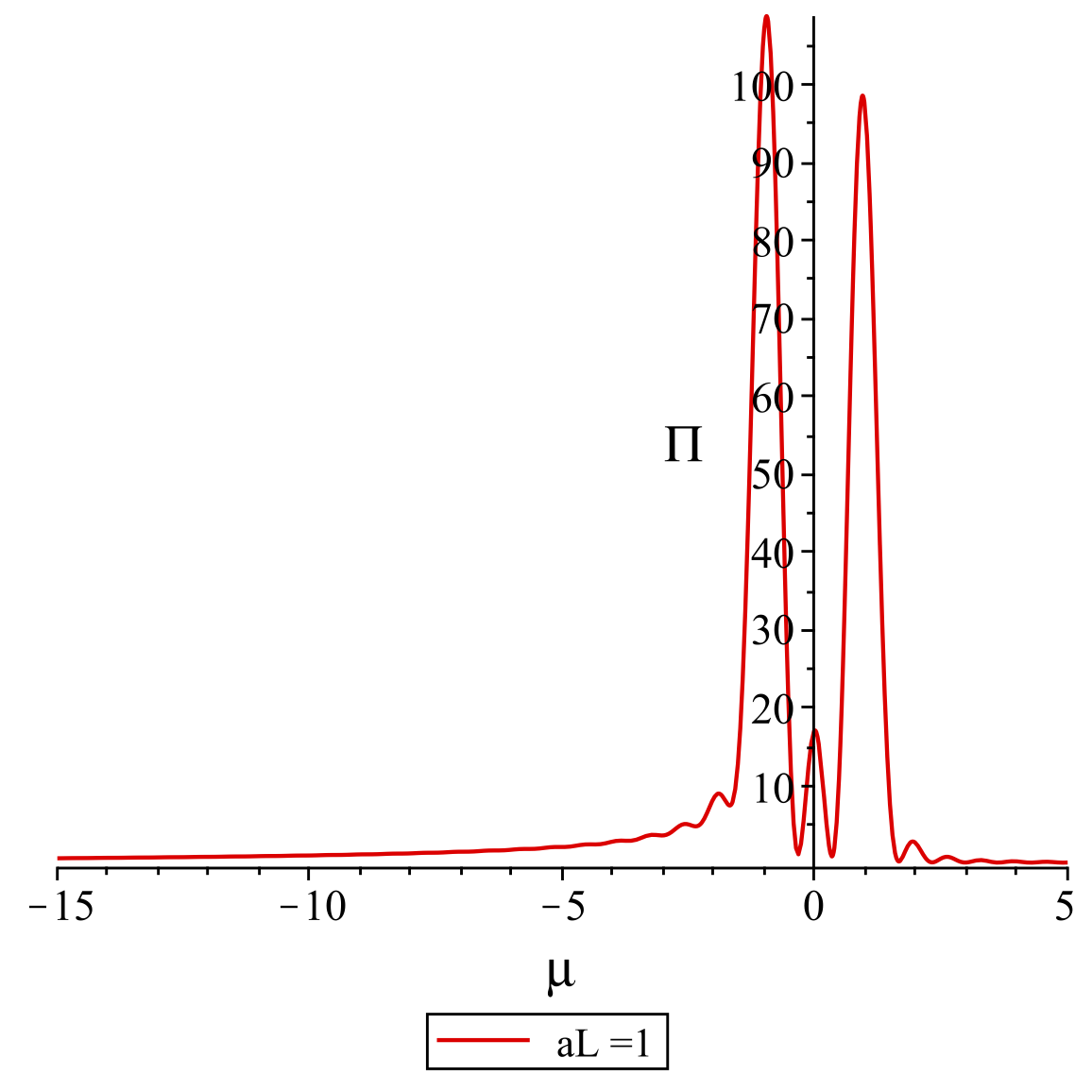} 
\label{CrossSection:MassiveResFun3A}}
\subfigure[$\tilde\beta = 30/\sqrt{\pi}$]{%
\includegraphics[width=0.48\textwidth]{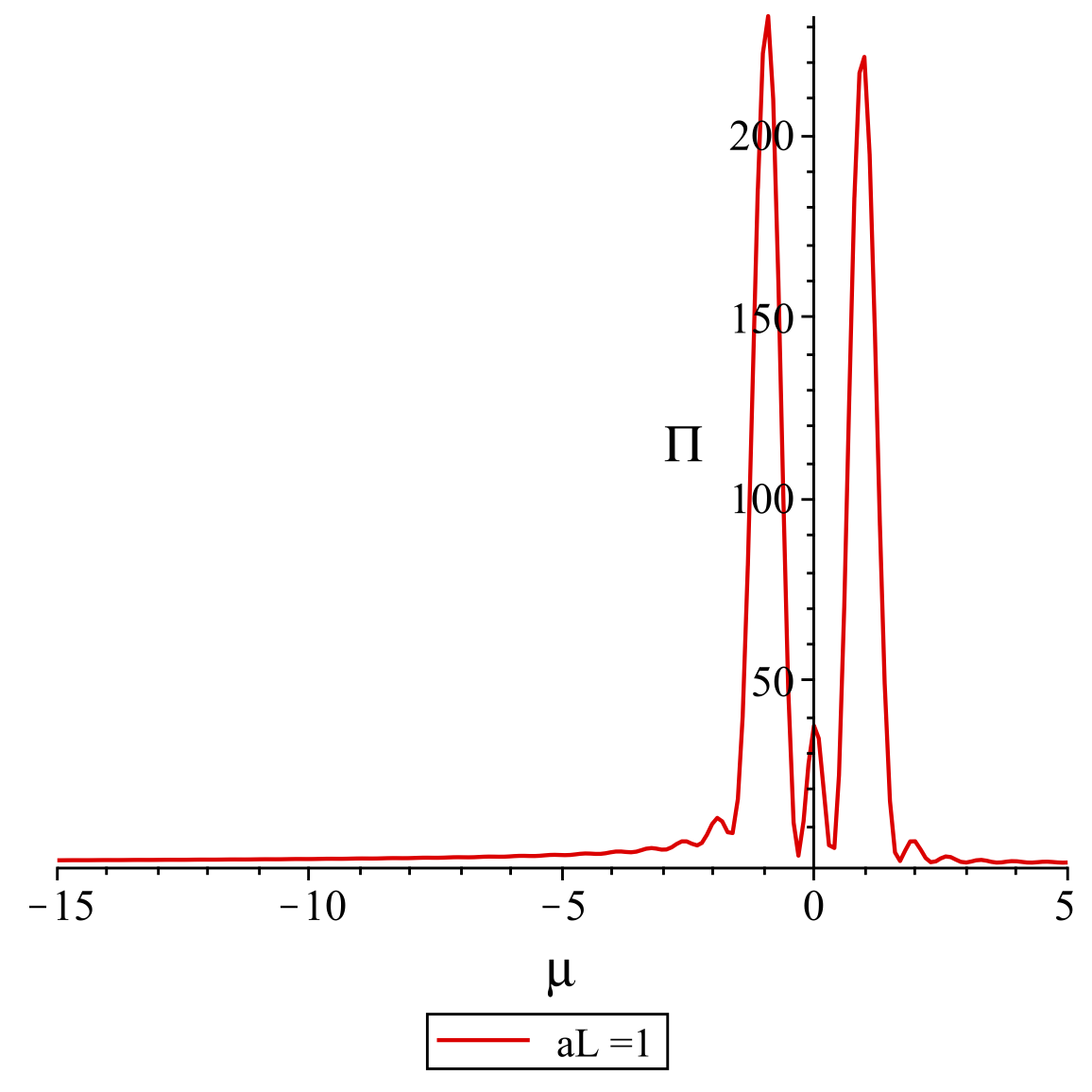} 
\label{CrossSection:MassiveResFun4A}}
\caption{Comoving detector's response for the untwisted field in Milne with $aL=1$ 
in the ``in'' vacuum, $\Pi_u^{in}(\mu, 10, 20)$, with $\tilde\alpha=0$ but 
varying the parameter $\tilde\beta$ as indicated.\label{CM_T_in_I}}
\end{figure}

\begin{figure}[p]
\centering
\subfigure[$\tilde\alpha=4/\sqrt{\pi}$]{%
\includegraphics[width=0.48\textwidth]{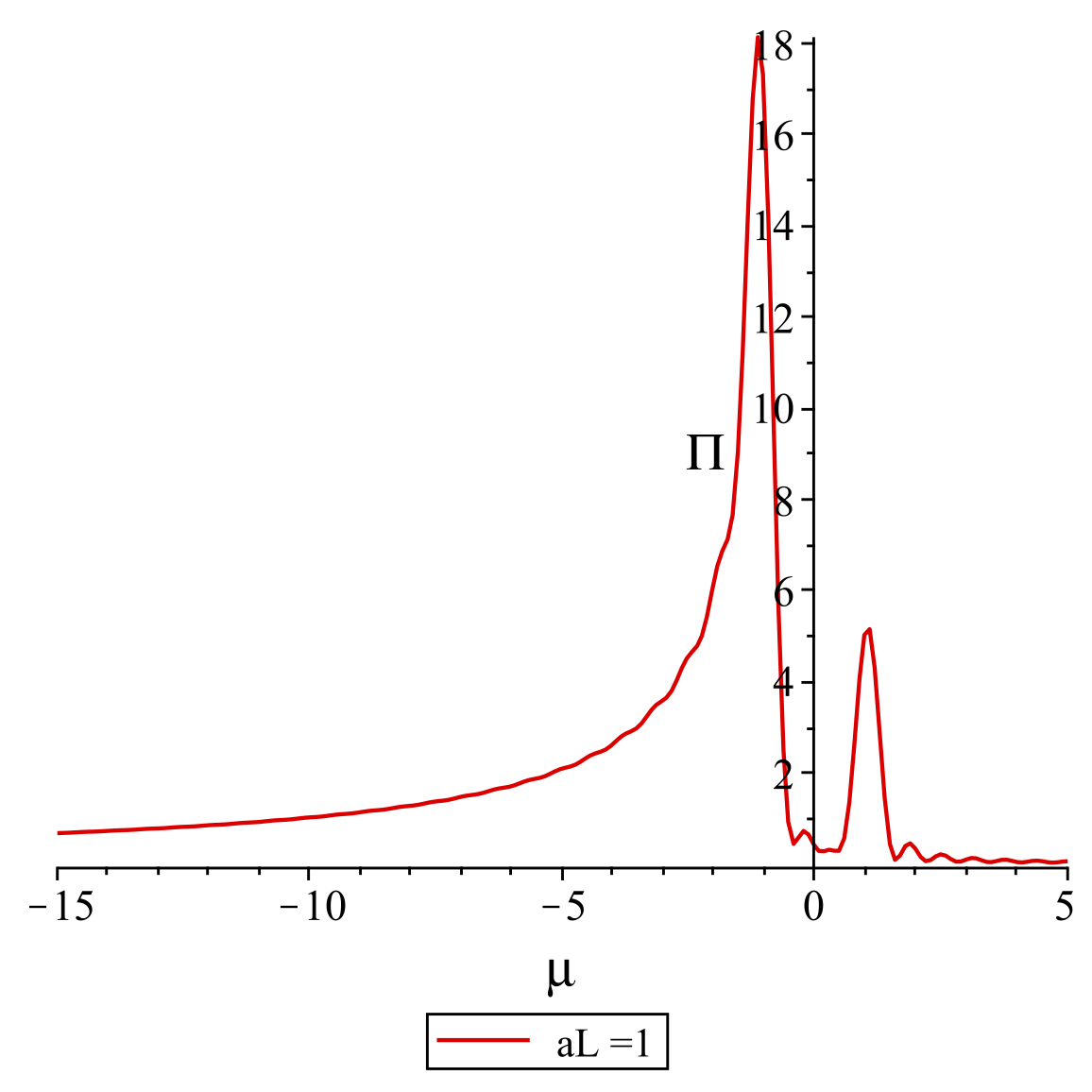} 
\label{Pers:MassiveResFunB}}
\subfigure[$\tilde\alpha=6/\sqrt{\pi}$]{%
\includegraphics[width=0.48\textwidth]{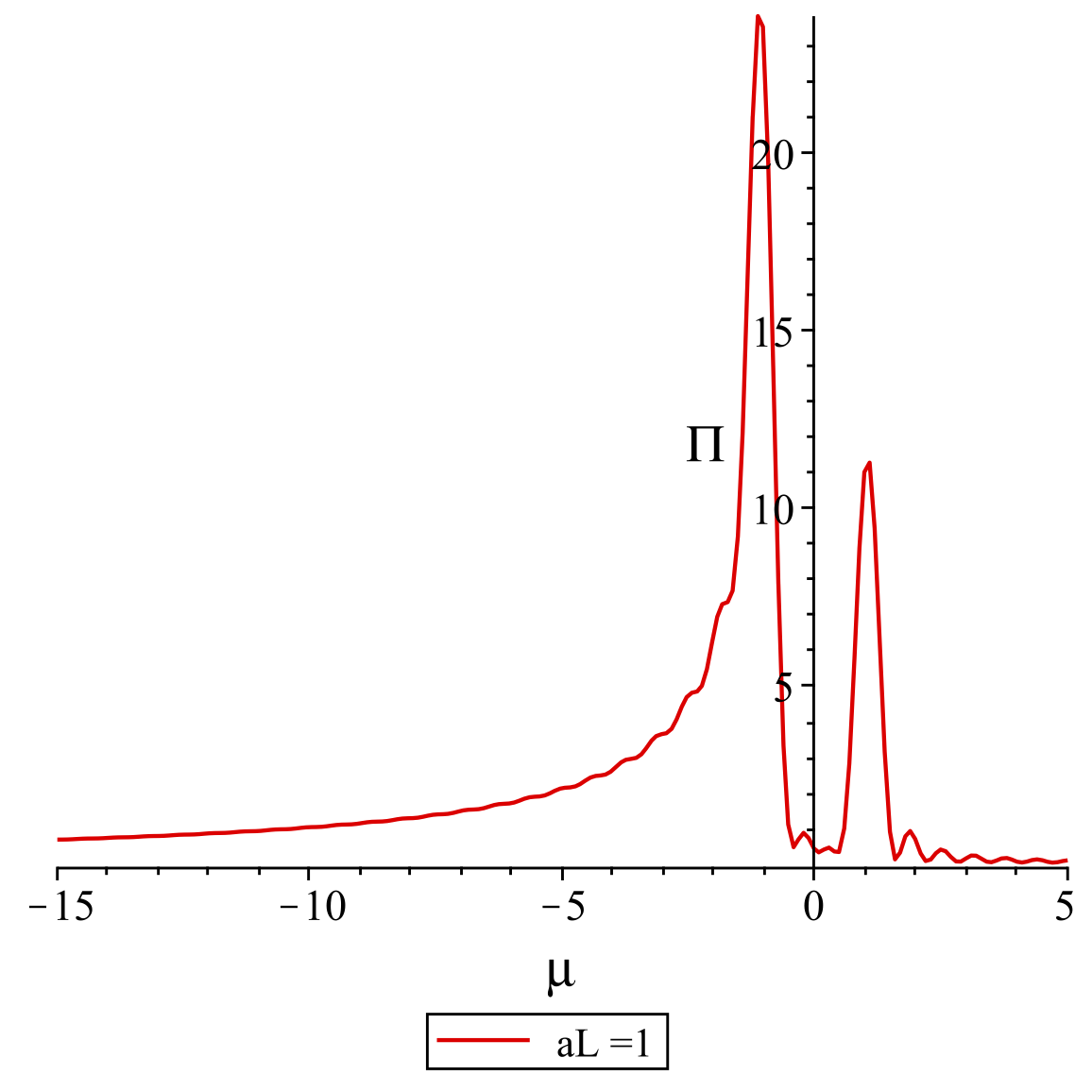} 
\label{CrossSection:MassiveResFun2B}}
\subfigure[$\tilde\alpha=-4/\sqrt{\pi}$]{%
\includegraphics[width=0.48\textwidth]{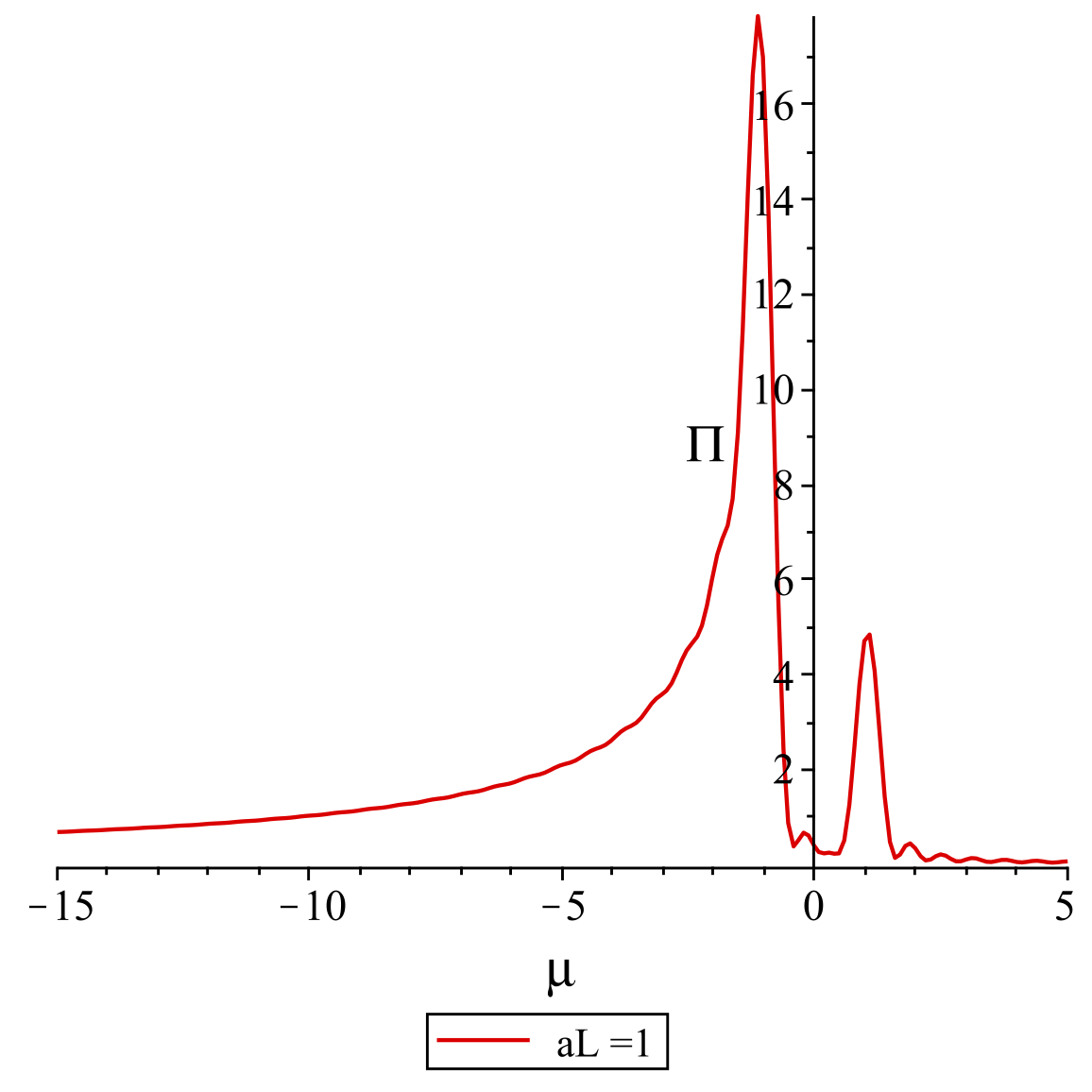} 
\label{CrossSection:MassiveResFun3B}}
\subfigure[$\tilde\alpha=-6/\sqrt{\pi}$]{%
\includegraphics[width=0.48\textwidth]{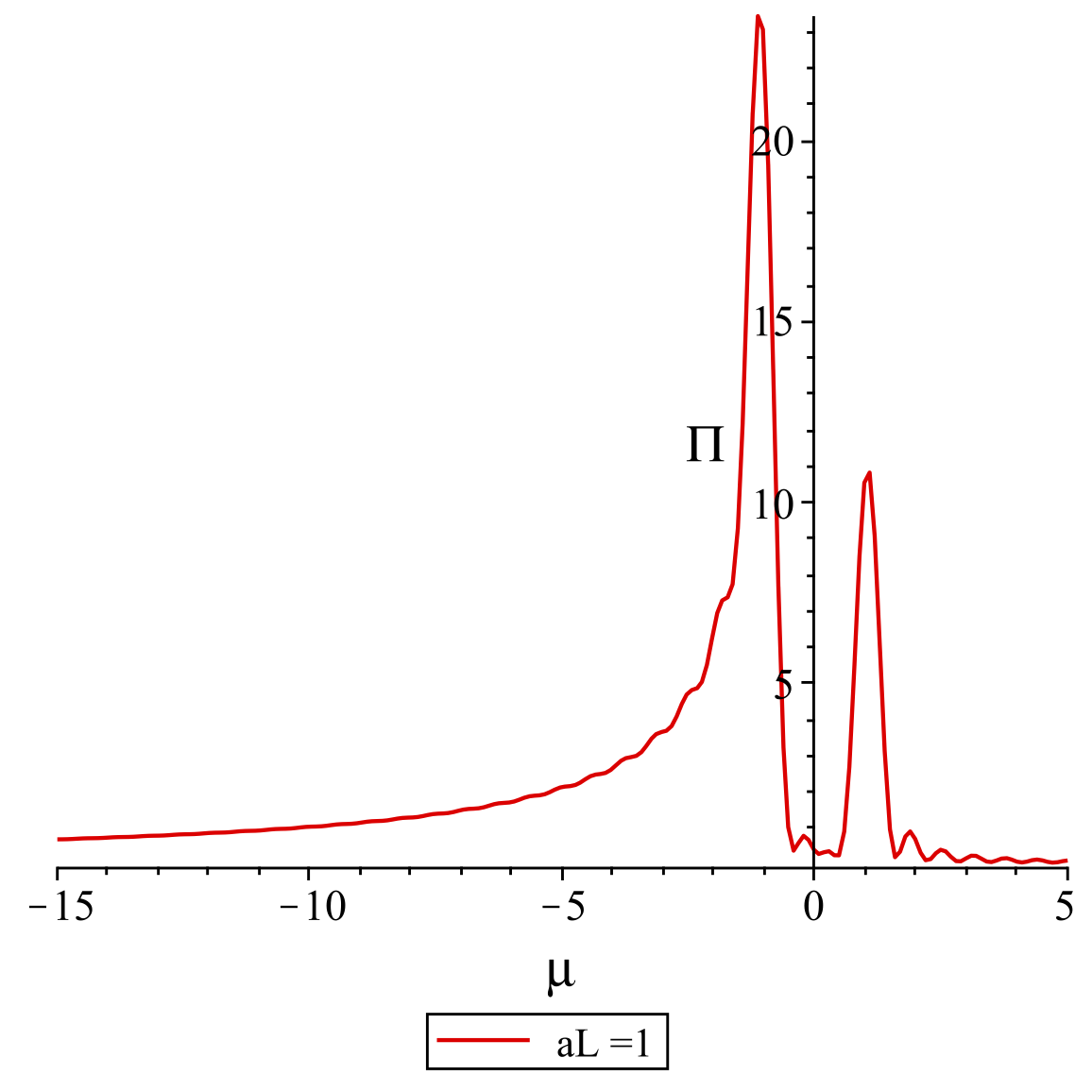} 
\label{CrossSection:MassiveResFun4B}}
\caption{Comoving detector's response for the untwisted field in Milne with $aL=1$ 
in the ``in'' vacuum, $\Pi_u^{in}(\mu, 10, 20)$, with $\tilde\beta=1$ but 
varying the parameter $\tilde\alpha$ as indicated.\label{CM_T_in_II}}
\end{figure}

\begin{figure}[p]
\centering
\subfigure[$\Pi_t^{out}(\mu, 10,20)$ for $aL = 0.2$ and selected $\theta$.]{%
\includegraphics[width=0.48\textwidth]{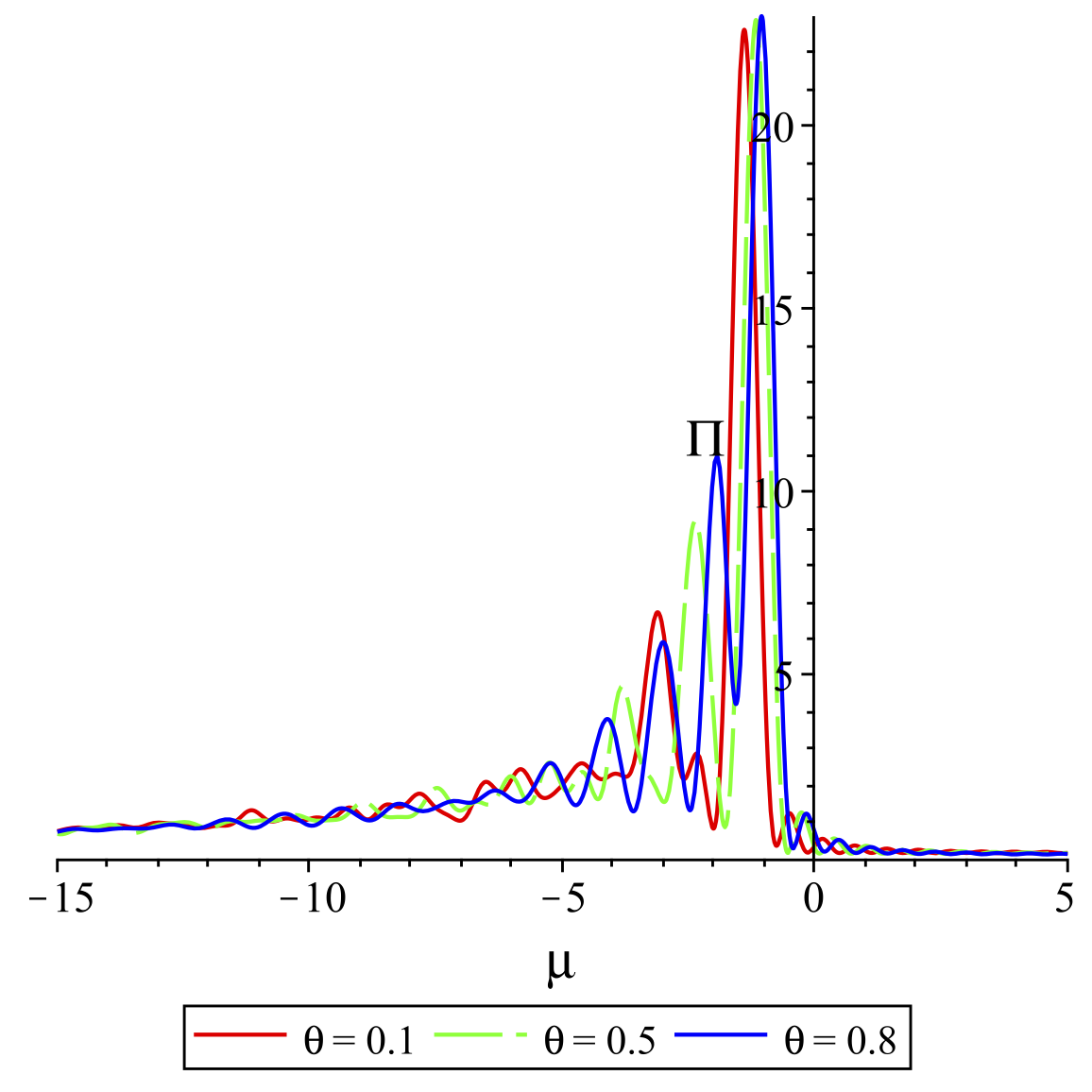} 
\label{Pers:MassiveResFunC}}
\subfigure[$\Pi_t^{out}(\mu, 90,100)$ for $aL = 0.2$ and selected $\theta$.]{%
\includegraphics[width=0.48\textwidth]{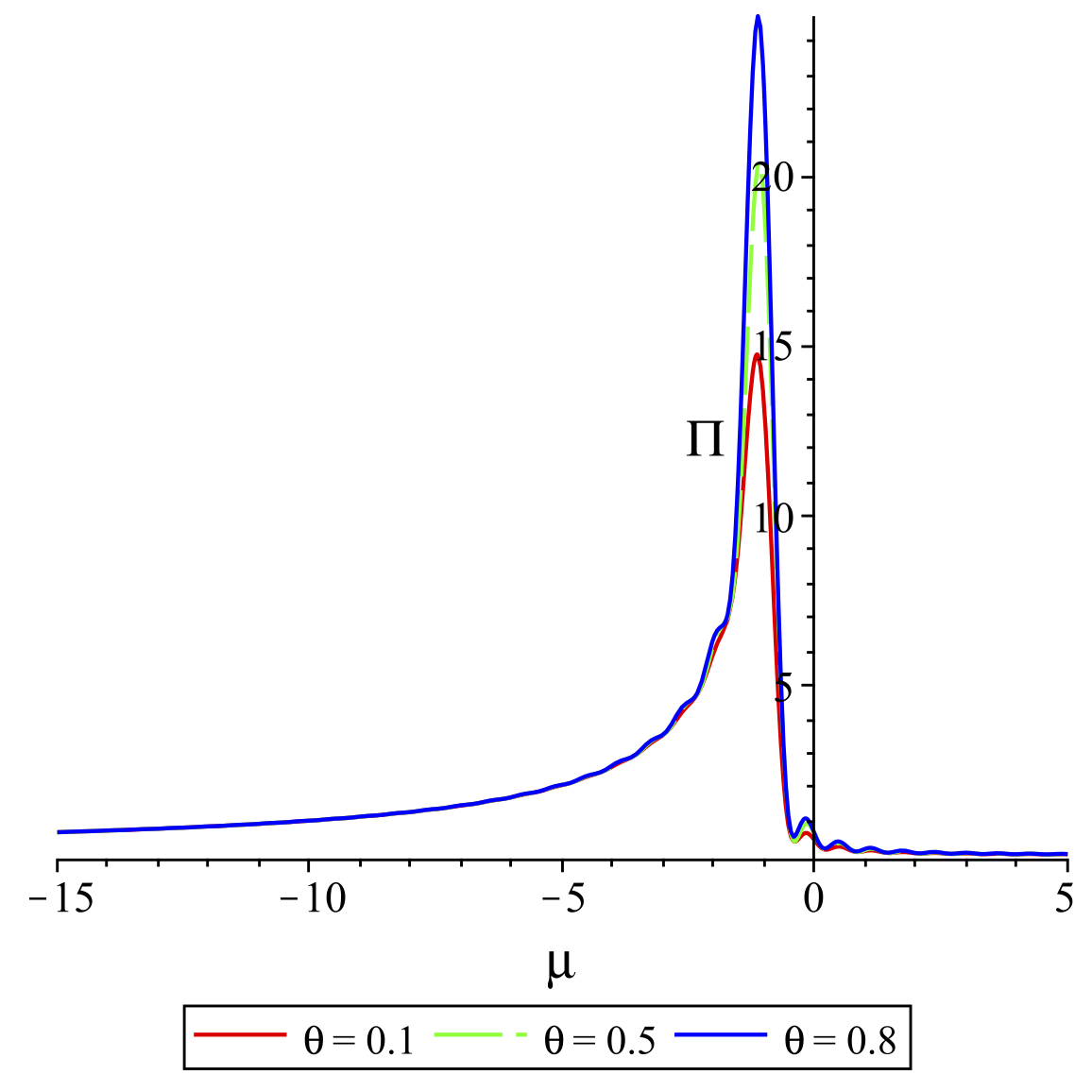} 
\label{CrossSection:MassiveResFun2C}}
\subfigure[$\Pi_t^{out}(\mu, 10,20)$ for $aL = 1$ and selected $\theta$.]{%
\includegraphics[width=0.48\textwidth]{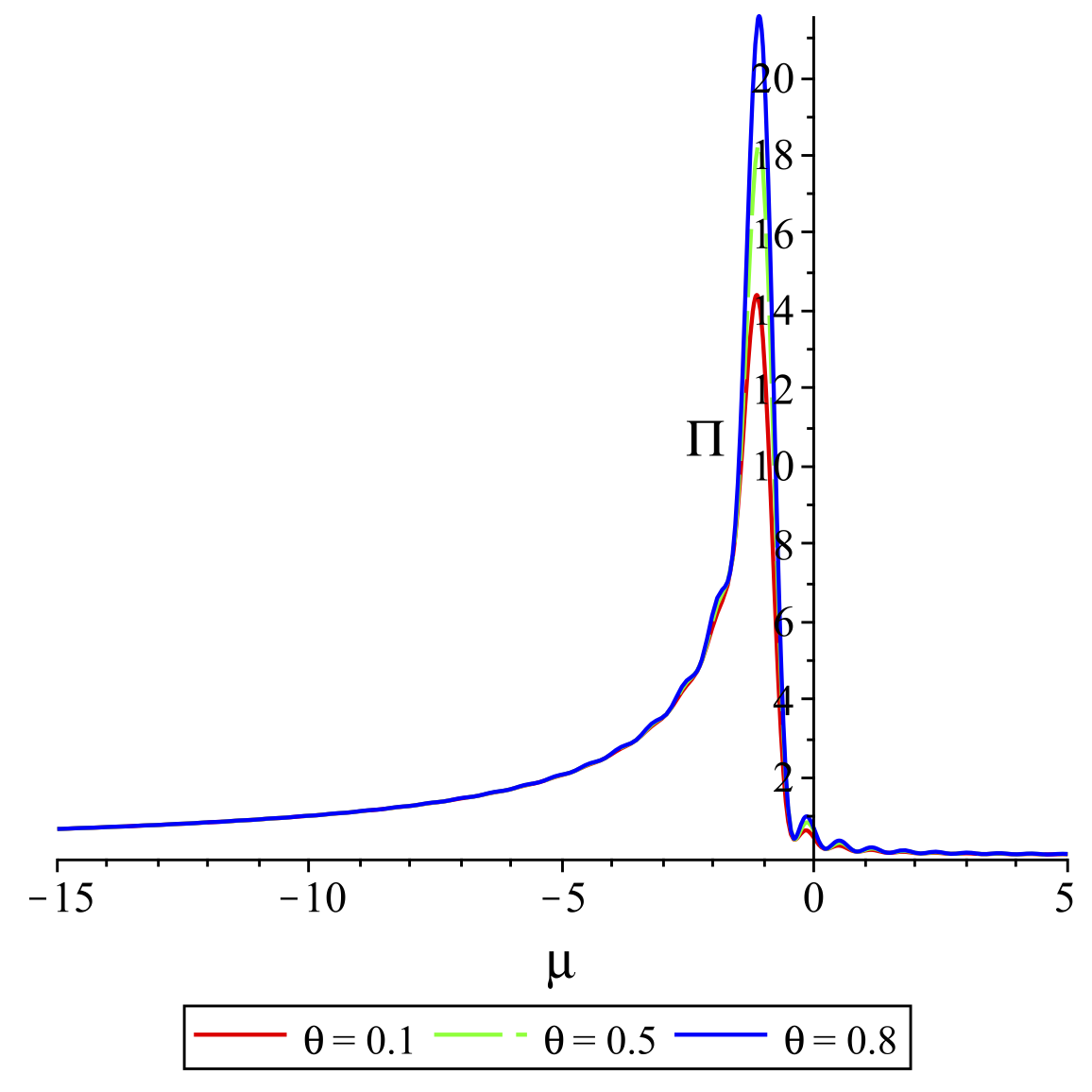} 
\label{CrossSection:MassiveResFun3C}}
\subfigure[$\Pi_t^{out}(\mu, 90,100)$ for $aL = 1$ and selected $\theta$.]{%
\includegraphics[width=0.48\textwidth]{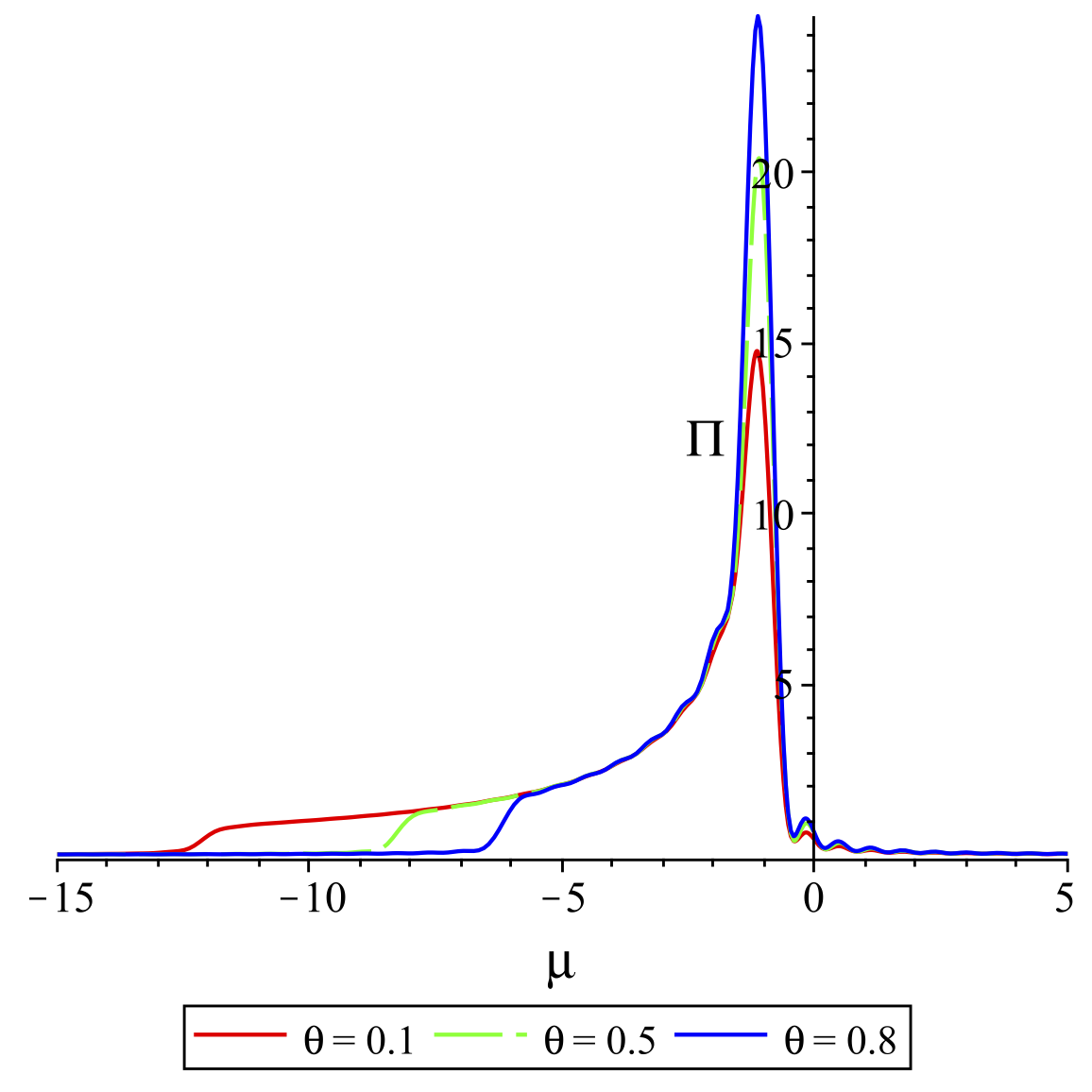} 
\label{CrossSection:MassiveResFun4C}}
\caption{Non-comoving detector's response for the twisted field in Milne, 
as a function of $\mu = \omega/m$, 
for the ``out'' vacuum, and parameter values as indicated. The sudden drop in Figure \ref{CrossSection:MassiveResFun4C} at large negative $\mu$ 
appears to be a numerical artefact.\label{NC_UT_out_IA}} 
\end{figure}

\begin{figure}[p]
\centering
\subfigure[$\Pi_t^{in}(\mu, 10,20)$ for $aL = 0.2$ and selected $\theta$.]{%
\includegraphics[width=0.48\textwidth]{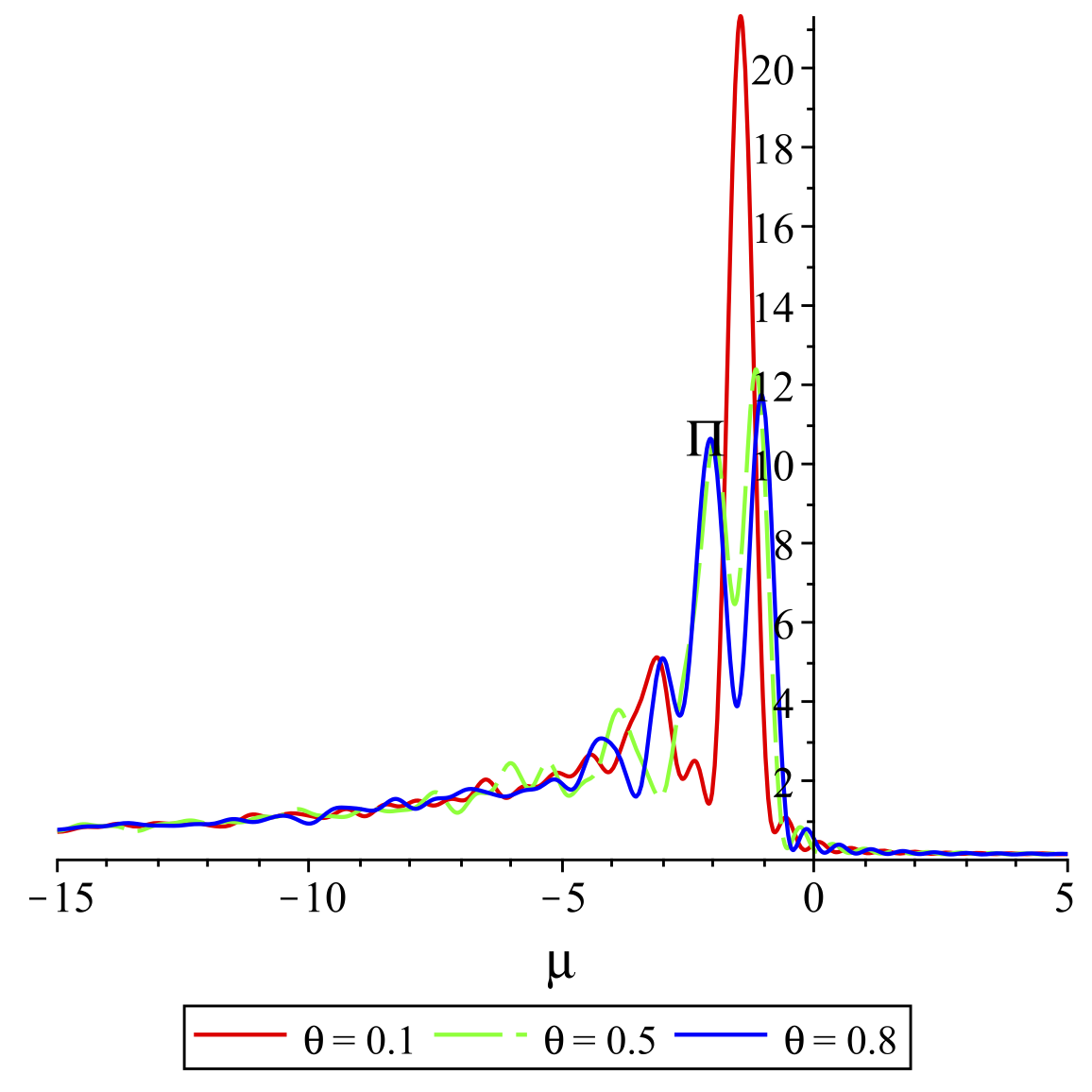} 
\label{Pers:MassiveResFunF}}
\subfigure[$\Pi_t^{in}(\mu, 90,100)$ for $aL = 0.2$ and selected $\theta$.]{%
\includegraphics[width=0.48\textwidth]{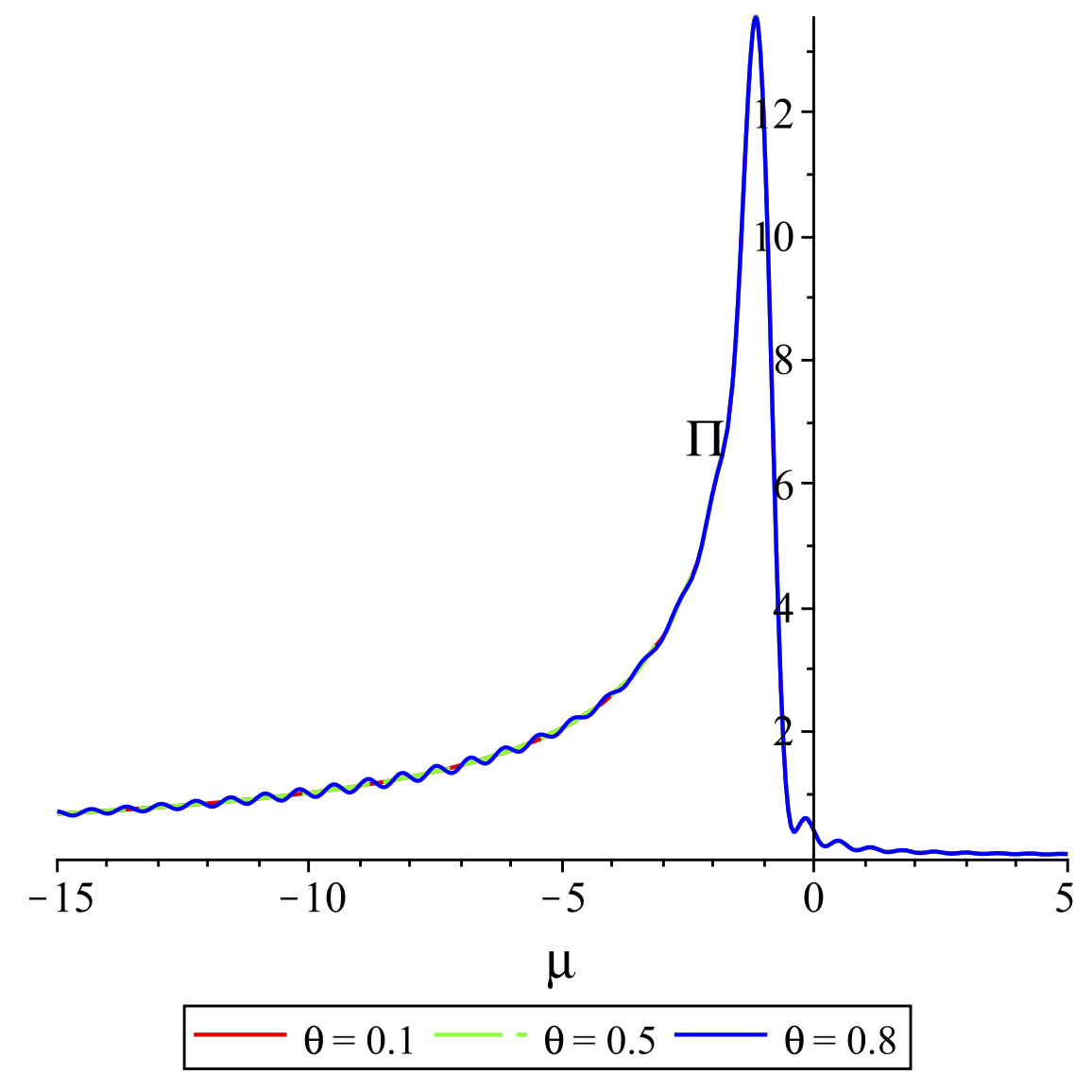} 
\label{CrossSection:MassiveResFun2F}}
\subfigure[$\Pi_t^{in}(\mu, 10,20)$ for $aL = 1$ and selected $\theta$.]{%
\includegraphics[width=0.48\textwidth]{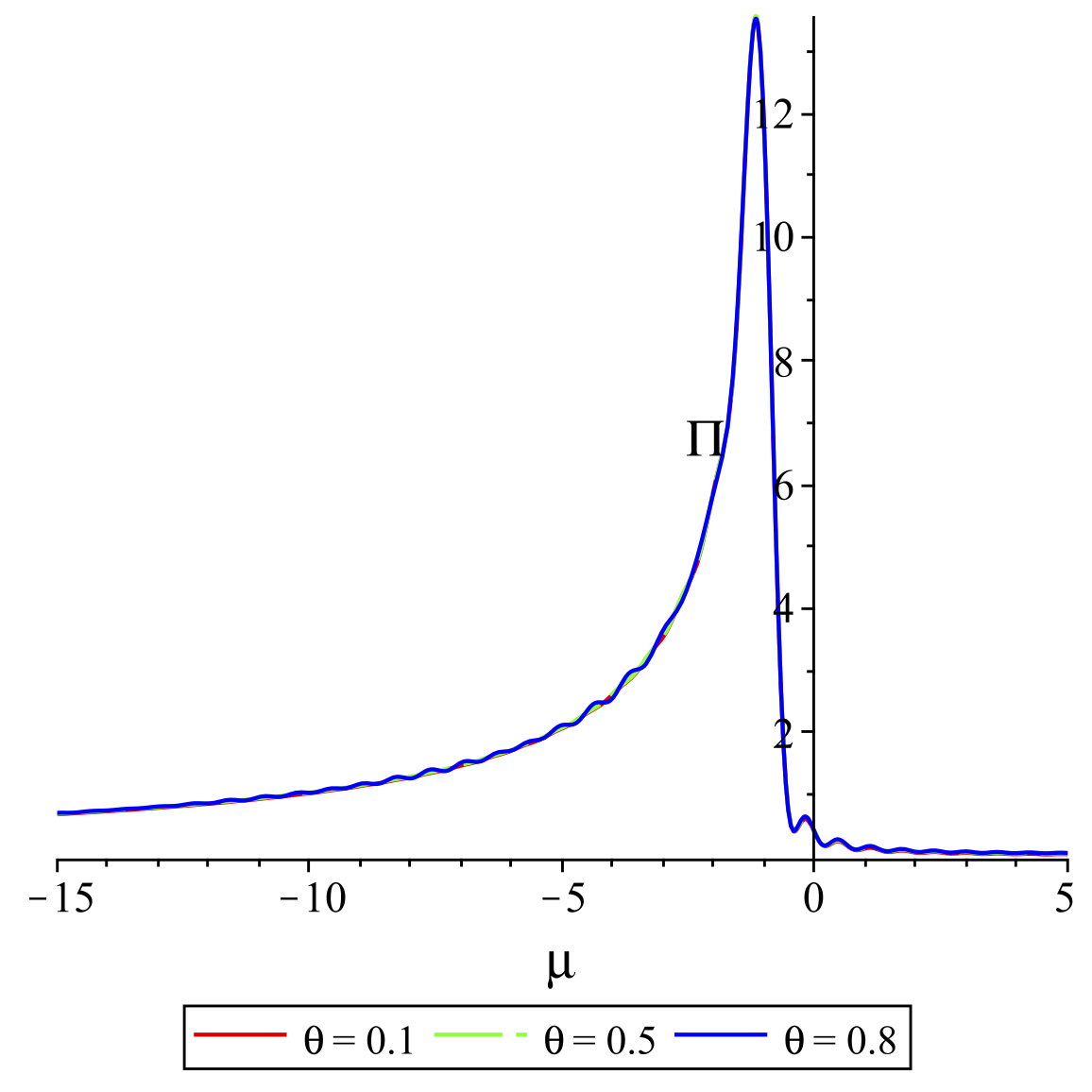} 
\label{CrossSection:MassiveResFun3F}}
\subfigure[$\Pi_t^{in}(\mu, 90,100)$ for $aL = 1$ and selected $\theta$.]{%
\includegraphics[width=0.48\textwidth]{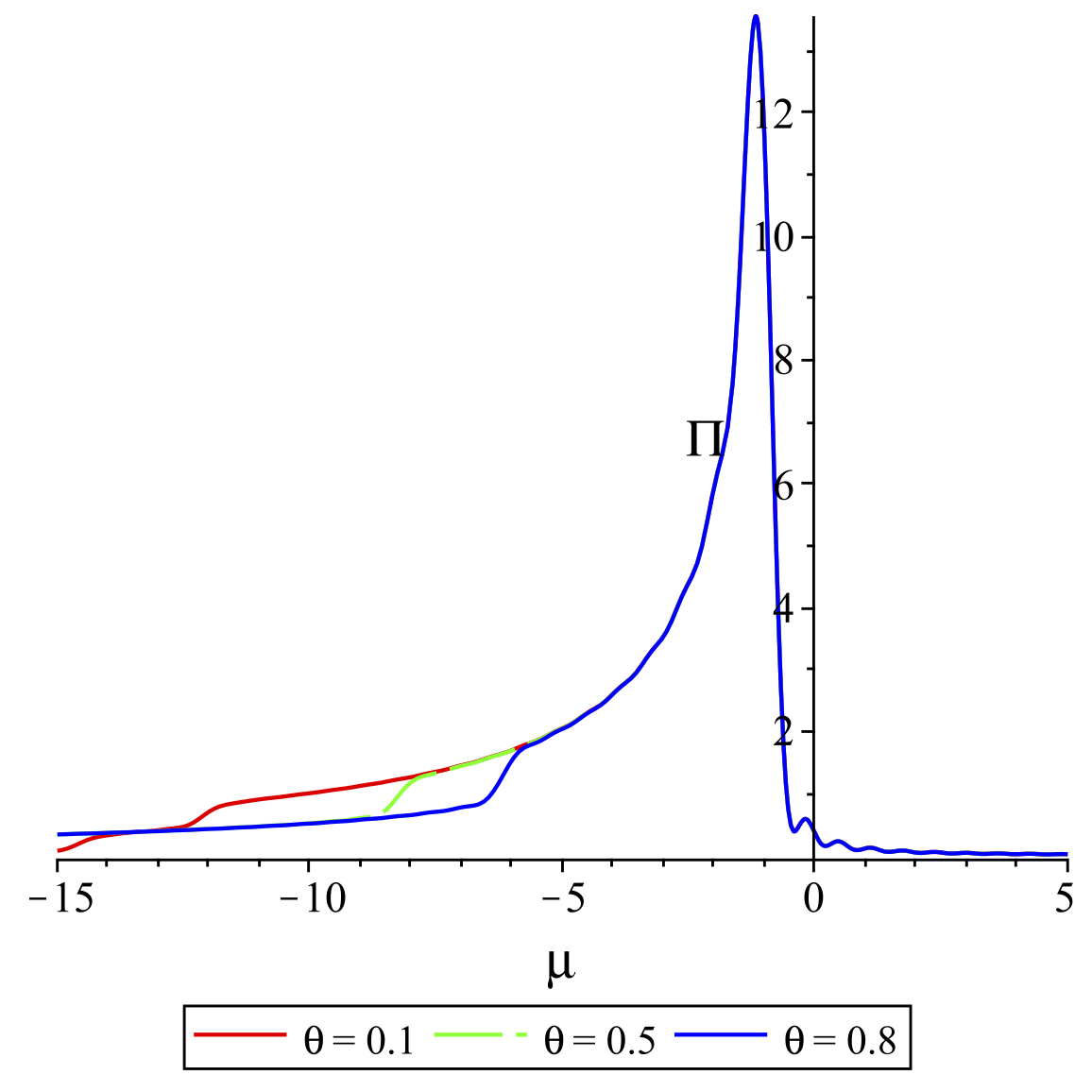} 
\label{CrossSection:MassiveResFun4F}}
\caption{Non-comoving detector's response for the twisted field in Milne, 
as a function of $\mu = \omega/m$, 
for the ``in'' vacuum, and parameter values as indicated. 
The sudden drop in Figure \ref{CrossSection:MassiveResFun4F} at large negative $\mu$ 
appears to be a numerical artefact.\label{NC_UT_out_IB}} 
\end{figure}

\begin{figure}[p]
\centering
\subfigure[$\Pi_t^{out}(\mu, 10,20)$ for $aL = 0.2$ and selected $\theta$.]{%
\includegraphics[width=0.48\textwidth]{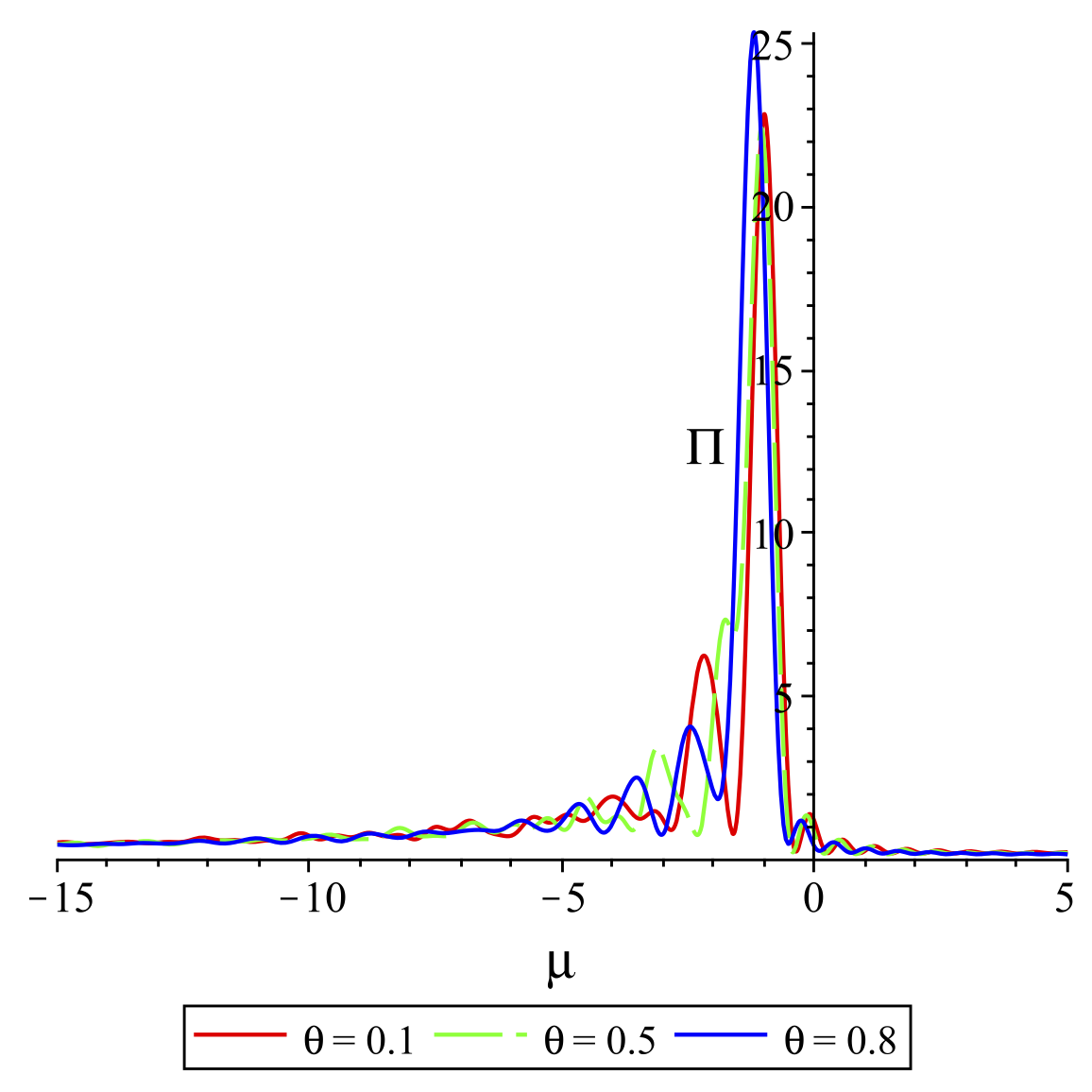} 
\label{Pers:MassiveResFunG}}
\subfigure[$\Pi_u^{out}(\mu, 90,100)$ for $aL = 0.2$ and selected $\theta$.]{%
\includegraphics[width=0.48\textwidth]{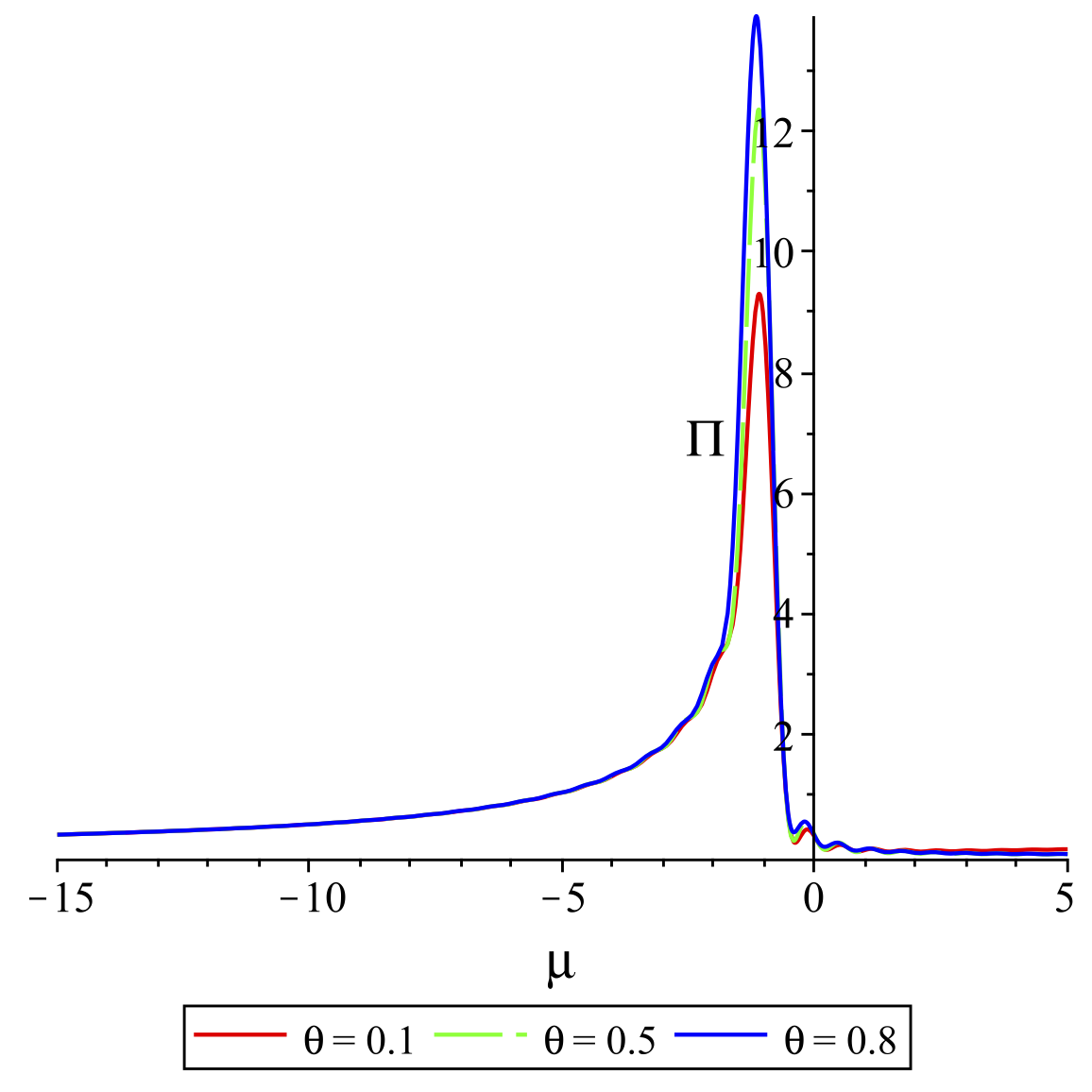} 
\label{CrossSection:MassiveResFun2G}}
\subfigure[$\Pi_u^{out}(\mu, 10,20)$ for $aL = 1$ and selected $\theta$.]{%
\includegraphics[width=0.48\textwidth]{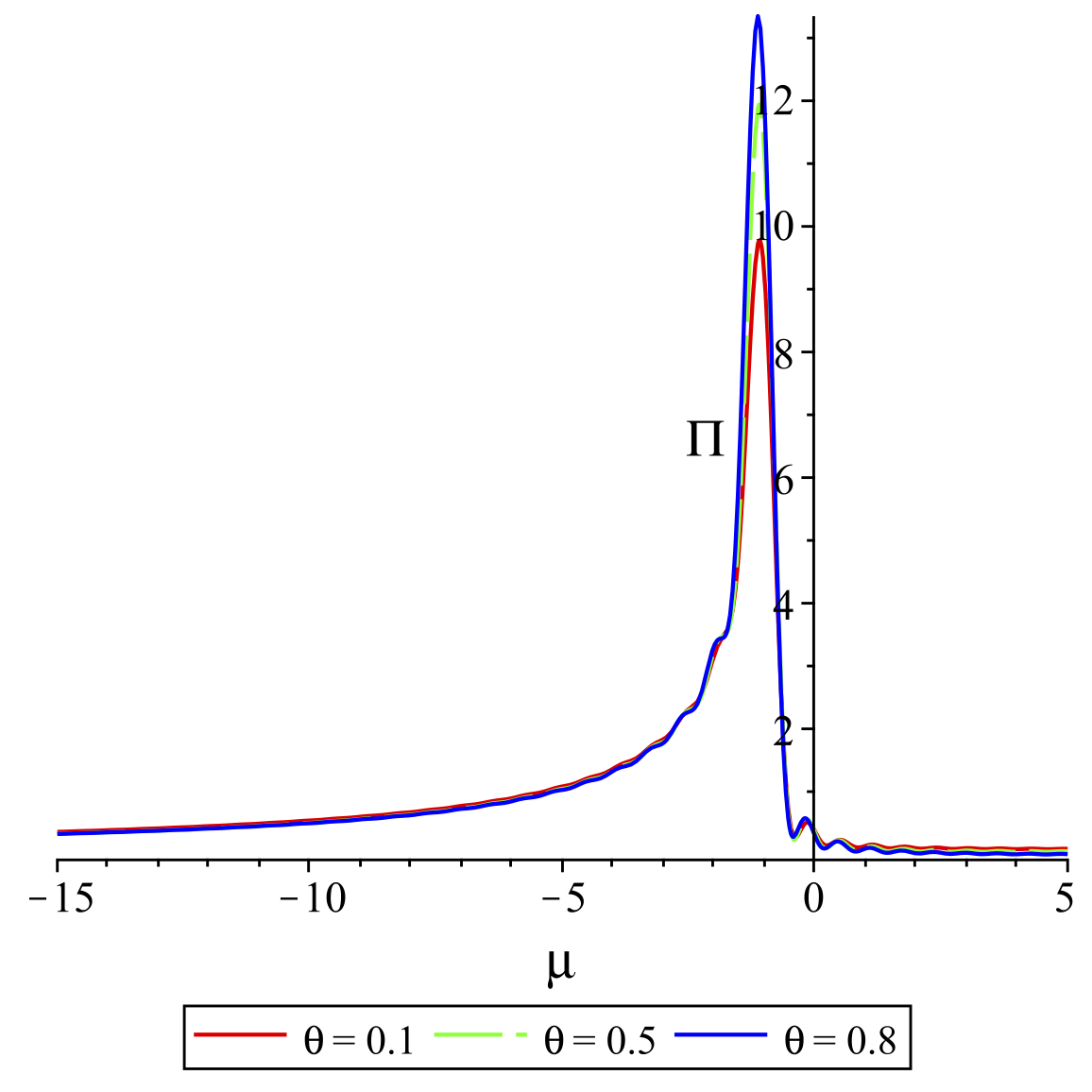} 
\label{CrossSection:MassiveResFun3G}}
\subfigure[$\Pi_u^{out}(\mu, 90,100)$ for $aL = 1$ and selected $\theta$.]{%
\includegraphics[width=0.48\textwidth]{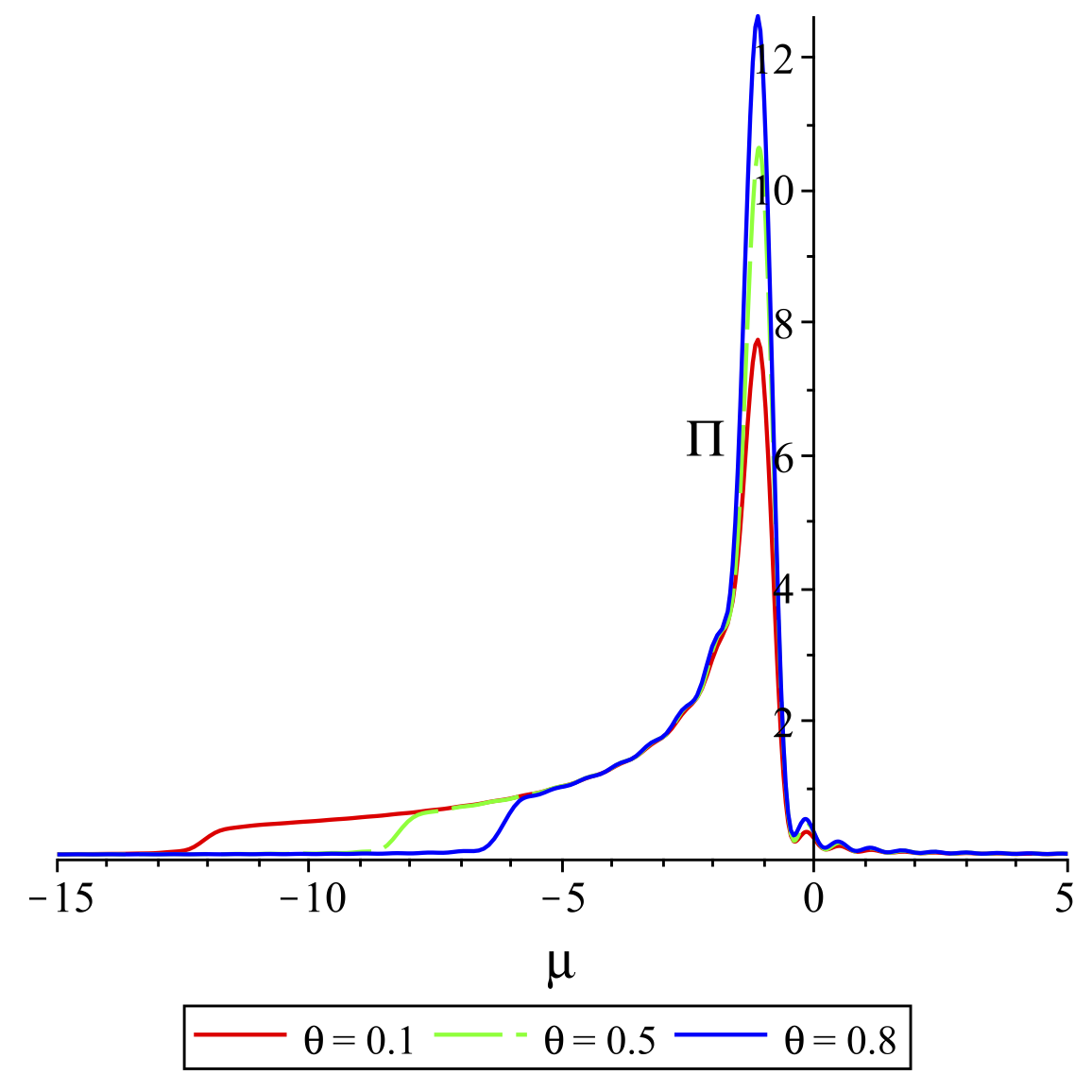} 
\label{CrossSection:MassiveResFun4G}}
\caption{Non-comoving detector's response for the untwisted field in Milne, 
as a function of $\mu = \omega/m$, 
for the ``out'' vacuum, and parameter values as indicated. 
The sudden drop in Figure \ref{CrossSection:MassiveResFun4G} at large negative $\mu$ 
appears to be a numerical artefact.\label{NC_UT_out_IC}}
\end{figure}

\begin{figure}[p]
\centering
\subfigure[$\Pi_u^{in}(\mu, 10,20)$ for $aL = 0.2$ and selected $\theta$.]{%
\includegraphics[width=0.48\textwidth]{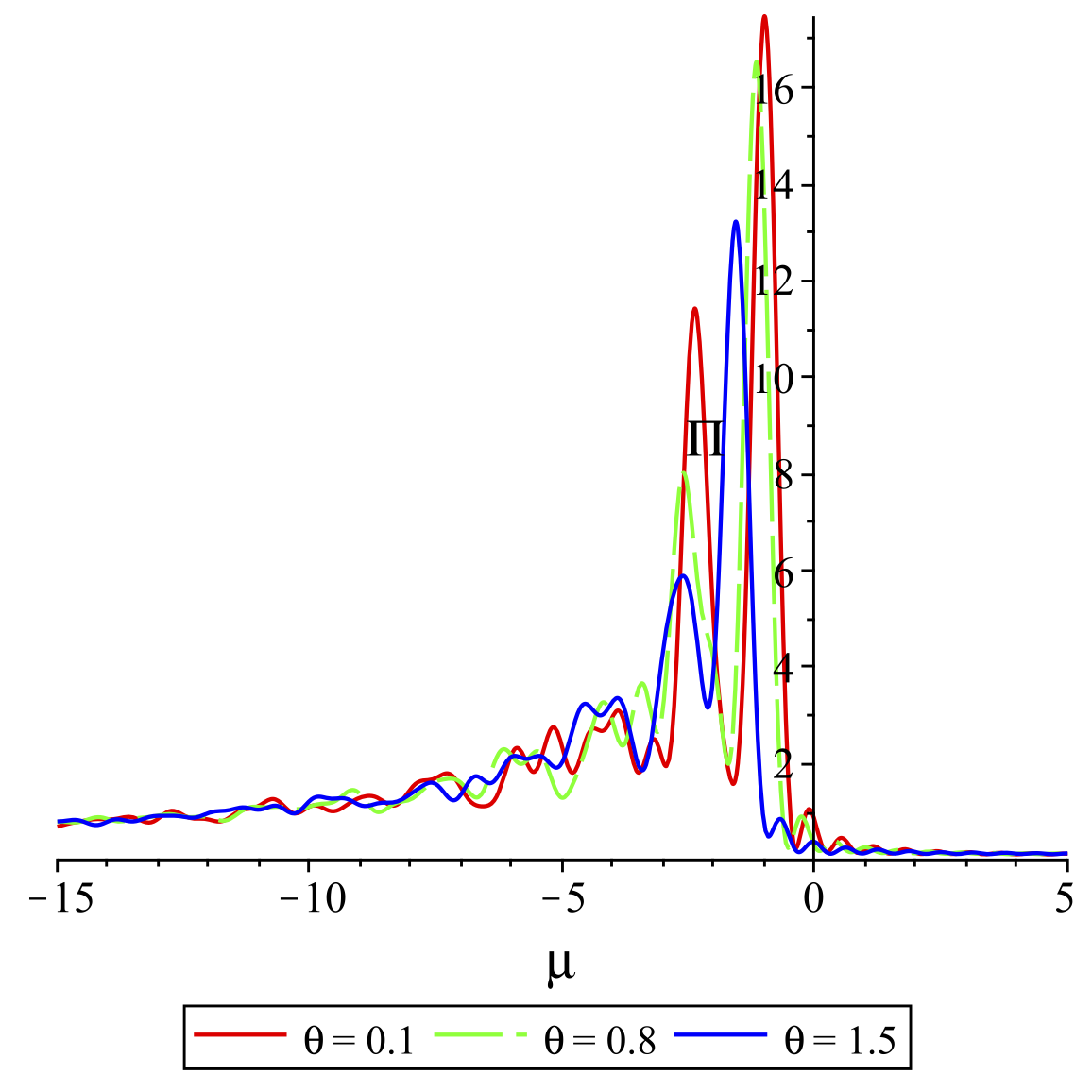} 
\label{Pers:MassiveResFunH}}
\subfigure[$\Pi_u^{in}(\mu, 90,100)$ for $aL = 0.2$ and selected $\theta$.]{%
\includegraphics[width=0.48\textwidth]{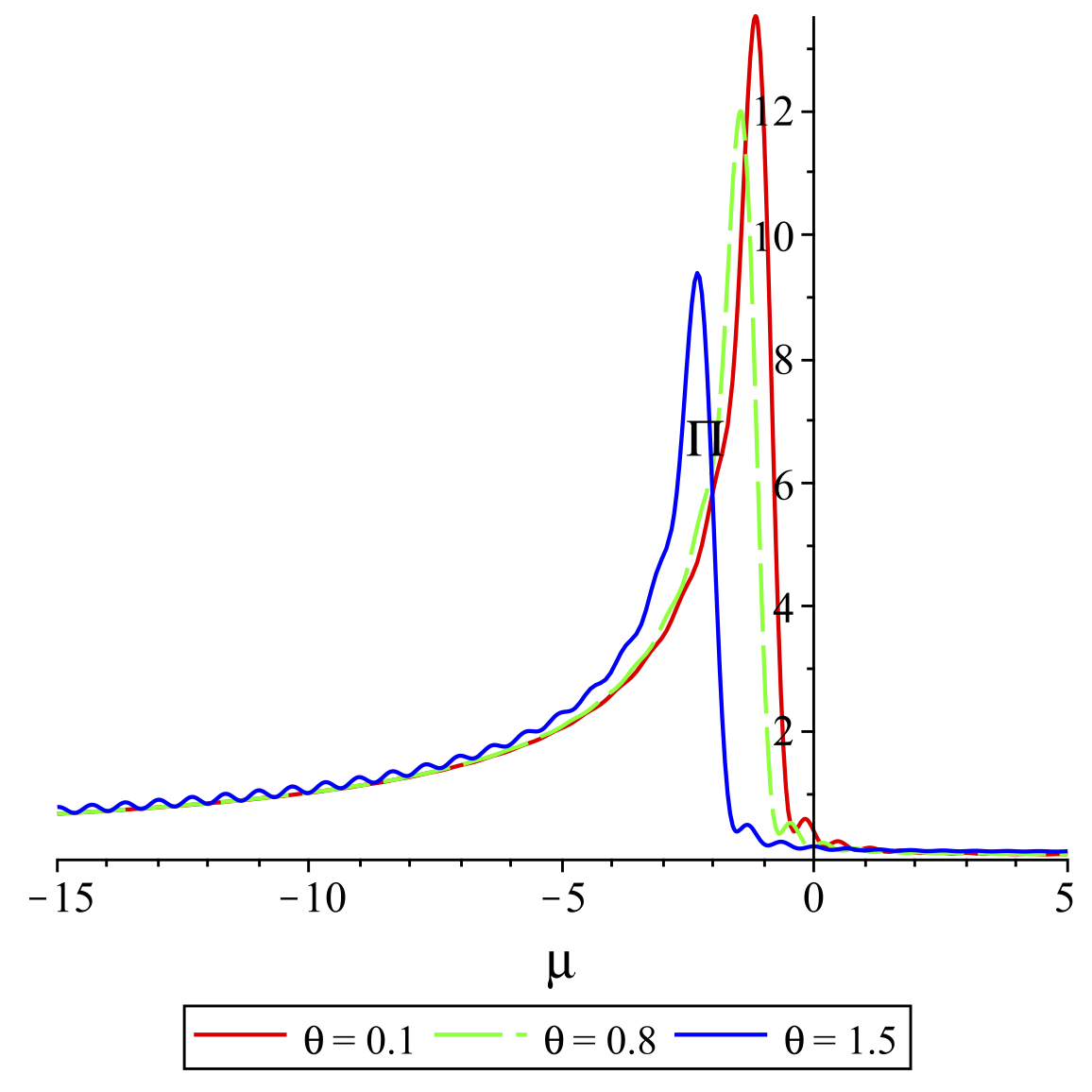} 
\label{CrossSection:MassiveResFun2H}}
\subfigure[$\Pi_u^{in}(\mu, 10,20)$ for $aL = 1$ and selected $\theta$.]{%
\includegraphics[width=0.48\textwidth]{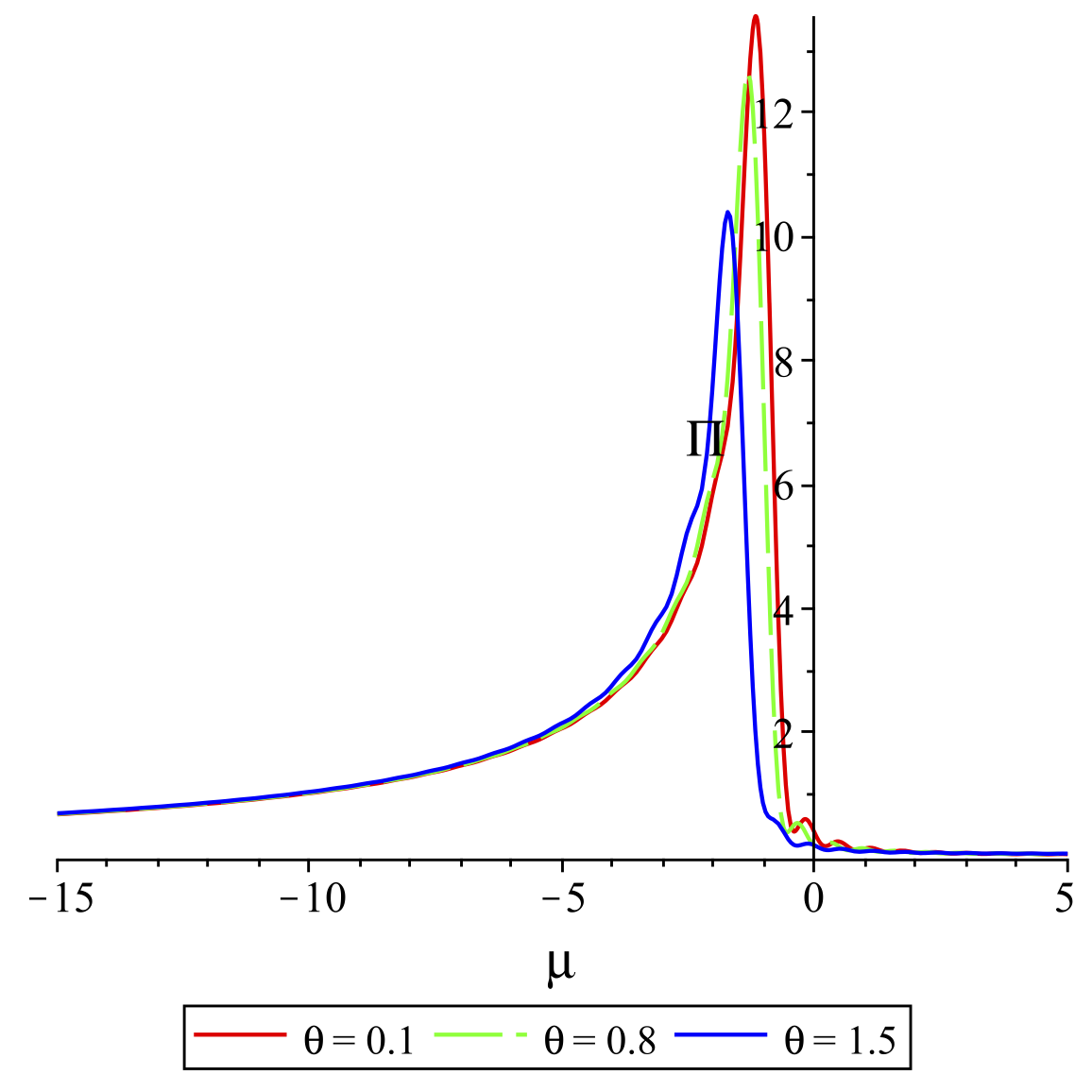} 
\label{CrossSection:MassiveResFun3H}}
\subfigure[$\Pi_u^{in}(\mu, 90,100)$ for $aL = 1$ and selected $\theta$.]{%
\includegraphics[width=0.48\textwidth]{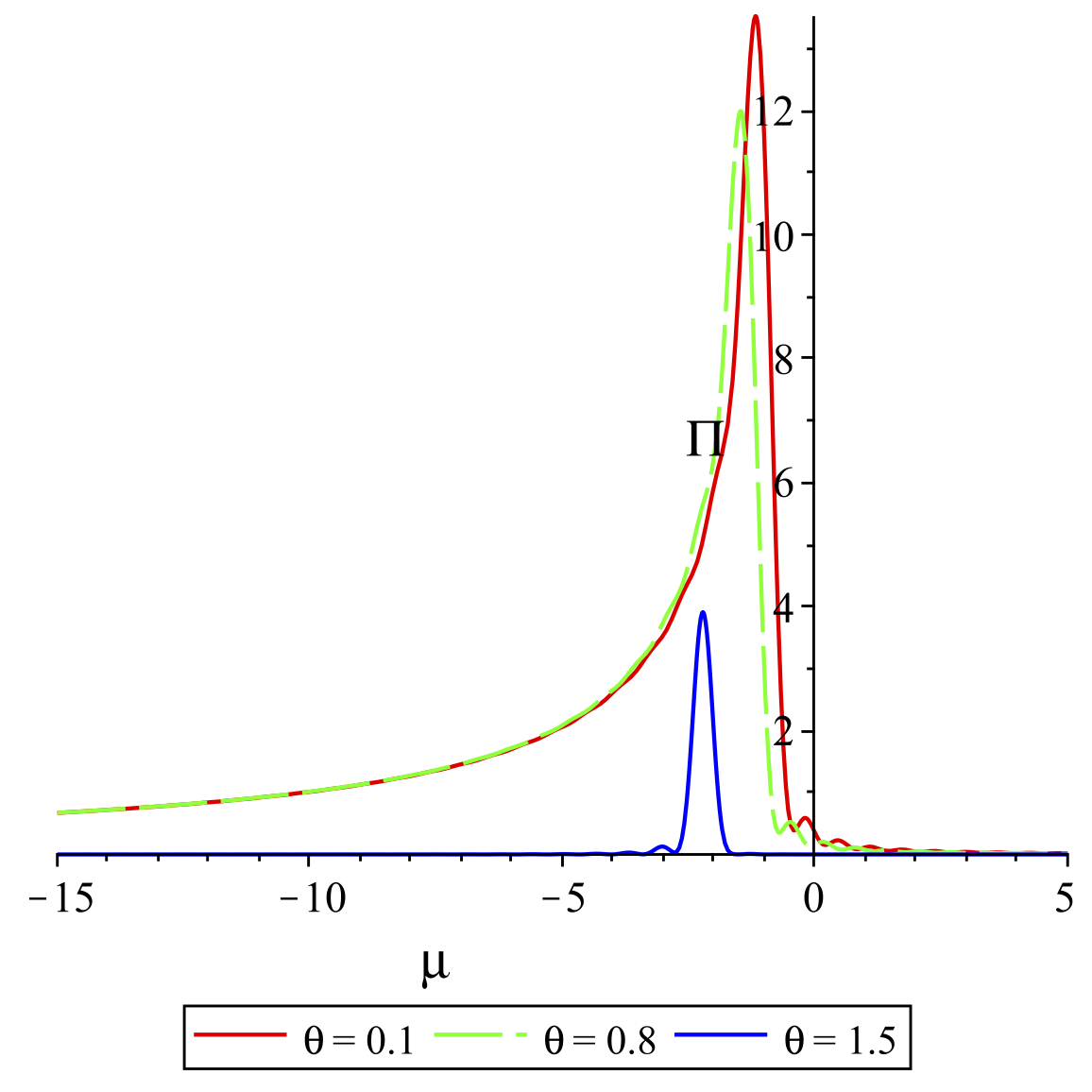} 
\label{CrossSection:MassiveResFun4H}}
\caption{Non-comoving detector's response for the untwisted field in Milne, 
as a function of $\mu = \omega/m$, 
for the ``in'' vacuum, and parameter values as indicated. 
The ``in'' vacuum spatially constant mode parameters are 
$\tilde\alpha=0$ and $\tilde\beta=1$, 
so that this mode coincides with that of the ``out'' vacuum.\label{UTCMD_inVACC}} 
\end{figure}

\begin{figure}[p]
\centering
\subfigure[$\tilde\alpha =0$  and $\tilde\beta = 2/\sqrt\pi$]{%
\includegraphics[width=0.48\textwidth]{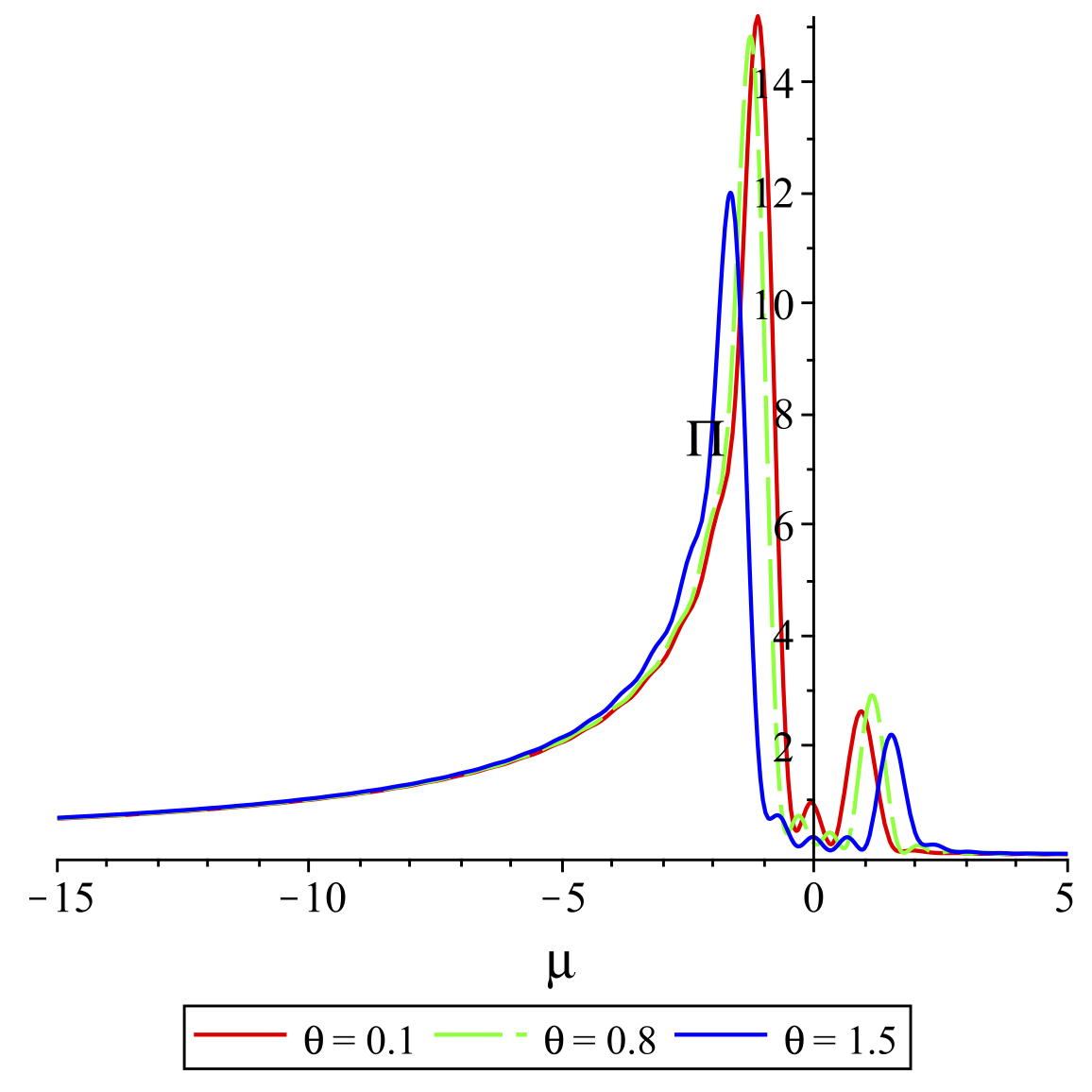} 
\label{Pers:MassiveResFunM}}
\subfigure[$\tilde\alpha =0$ and $\tilde\beta = 10/\sqrt\pi$]{%
\includegraphics[width=0.48\textwidth]{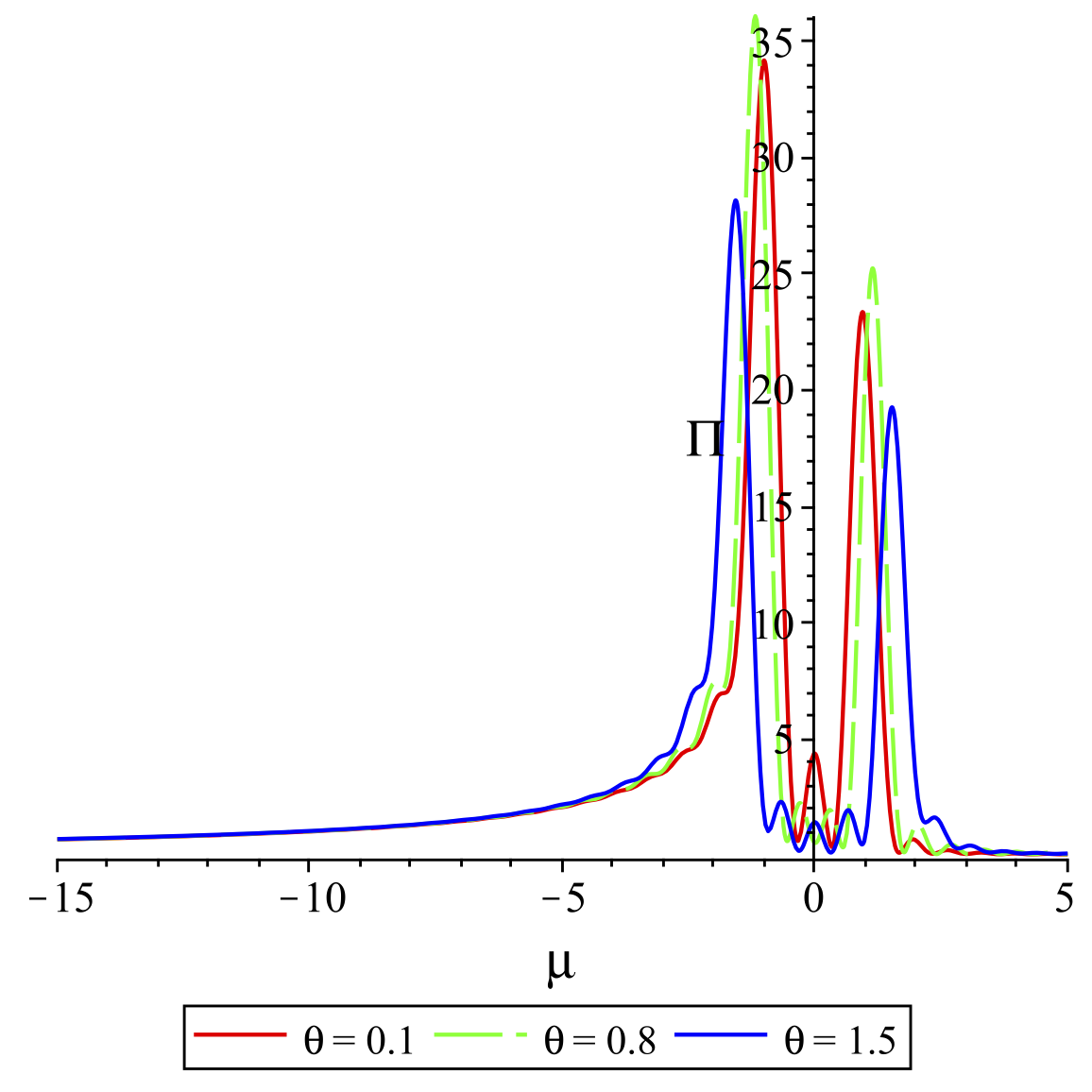} 
\label{CrossSection:MassiveResFun2M}}
\subfigure[$\tilde\alpha =4/\sqrt\pi$ and $\tilde\beta = 1$]{%
\includegraphics[width=0.48\textwidth]{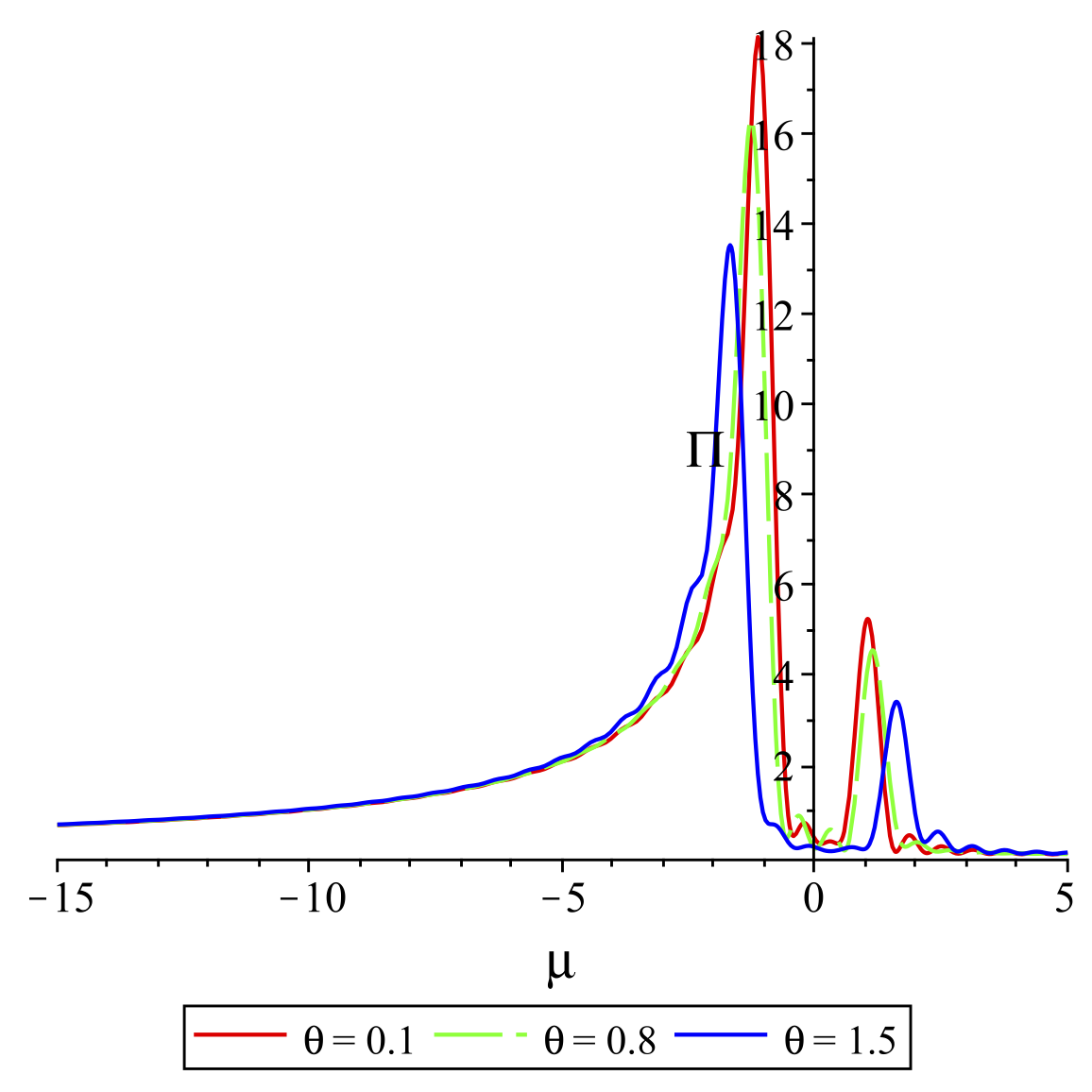} 
\label{Pers:MassiveResFun3M}}
\subfigure[$\tilde\alpha =6/\sqrt\pi$ and $\tilde\beta = 1$]{%
\includegraphics[width=0.48\textwidth]{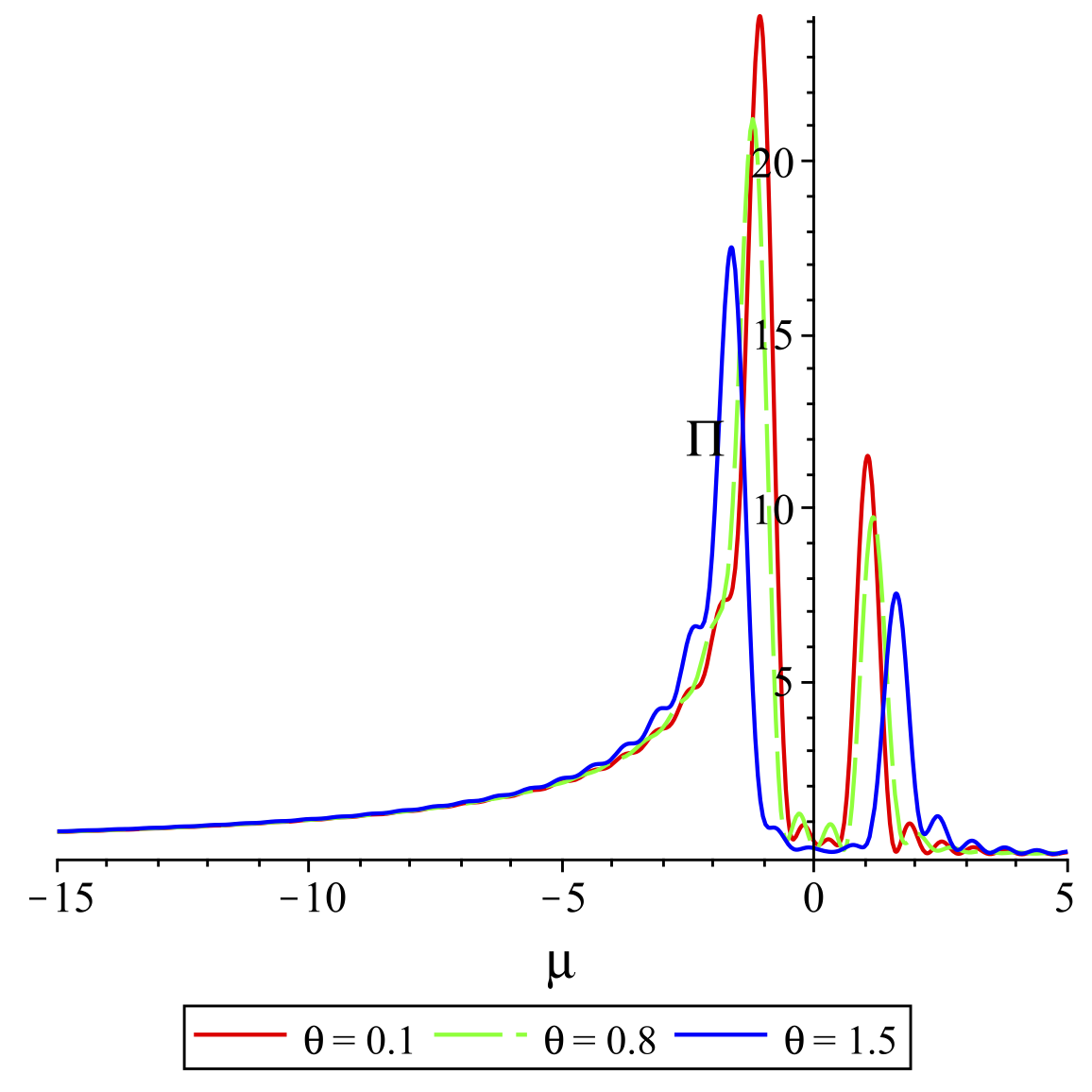} 
\label{CrossSection:MassiveResFun4M}}
\caption{Non-comoving detector's response for the untwisted field in Milne with $aL=1$ 
in the ``in'' vacuum, $\Pi_u^{in}(\mu, 10, 20)$, 
varying the parameters $\tilde\alpha$ and $\tilde\beta$ as indicated.\label{CM_T_in_M}}
\end{figure}

In this appendix we collect 
Figures \ref{Tw_CM_RF}--\ref{CM_T_in_M}, 
which are discussed verbally in Section~\ref{sec:numerical}. 

\end{document}